%% file: clac14.tex
\nonstopmode

\newcommand{\FOSSACS}[1]{#1} \renewcommand{\FOSSACS}[1]{}
\newcommand{\CLAC}[1]{#1} \newcommand{\Paper}[1]{} \newcommand{\MSCS}[1]{}

\CLAC{
\documentclass[copyright,creativecommons]{eptcs}
 \usepackage{qsymbols}
 \usepackage{mathpazo}
 \usepackage{times}

  \newskip \point \point=1pt \setlength {\unitlength} {1\point}

 \newcommand{\Comment}[1]{}

 \newif\ifmycolour \mycolourfalse
 \def\firstitem{\item}
 \def\qed{\QED}

 \def \myitem[#1]{\setbox55=\hbox{#1}%
\myitemmargin\wd55%
		\advance\myitemmargin\labelsep
\item[{#1}]\leavevmode $ \hspace*{1mm} \kern -\myitemmargin
\begin{array}[t]{@{}lclclclclcl}
		\advance\myitemmargin\labelsep
	 \kern \myitemmargin }

 \newenvironment{proof}{\trivlist \item[\hskip \labelsep {\it Proof: }] \begingroup 
}{\endgroup \endtrivlist}

 \newenvironment{Proof}{\begin{proof}}{\end{proof}}

 \def\citet{\cite}

 \def\[{\begin{center}$}
 \def\]{$\end{center}}
}

\usepackage{Xmacros}
\usepackage{PiXmacros}
\def \Weak {\textit{Weak}}

 \CLAC{
 \pagestyle{plain}

 \newtheorem{definition}{Definition}[section]{\bf}{\rm}
 \newtheorem{example}[definition]{Example}{\it}{\rm}
 \newtheorem{remark}[definition]{Remark}{\it}{\rm}
 {\it}{\it}
 {\it}{\it}
 \newtheorem{corollary}[definition]{Corollary}{\it}{\it}
 {\it}{\rm}
 \newtheorem{lemma}[definition]{Lemma}{\it}{\it}
 {\it}{\it}
 \newtheorem{proposition}[definition]{Proposition}{\it}{\it}
 \newtheorem{theorem}[definition]{Theorem}{\bf}{\it}
}

\def\Sink{s}
\def\Red{}
\def \iotn {1 \seq i \seq n}

\begin {document}

\def \Rules{ \begin{figure*}
 \[ \def\TurnPi{\Turn}
 \begin{array}{@{}ccc}
 \begin{array}{@{}rl}
(\Zero): &
 \Inf	{ \Pider {\Zero} : `G |- `D }
 \\ [4mm] %
(!): & %
 \Inf	{ \Pider \proc{P} : `G |- `D }
	{ \Pider \Bang \proc{P} : `G |- `D }
\quad
\\ [4mm]
(`n): &
 \Inf	{ \Pider \proc{P} : `G,a{:}A |- a{:}A,`D }
	{ \Pider \New a \proc{P} : `G |- `D }
 \end{array}
 \kern2mm
 \begin{array}{@{}rl}
(\Par): &
 \Inf	{ \Pider \proc{P}_1 : `G |- `D \quad \dots \quad \Pider \proc{P}_n : `G |- `D }
	{ \Pider \proc{P}_1 \Par \, \dots \, \Par \proc{P}_n : `G |- `D }
\\ [4mm]
(\Weak): &
 \Inf	[`G' \supseteq `G,`D' \supseteq `D]
	{ \Pider \proc{P} : `G |- `D }
	{ \Pider \proc{P} : `G' |- `D' }
\\ [4mm]
(\InRule) : &
 \Inf	{ \Pider \proc{P} : `G, x{:}A |- x{:}A,`D }
	{ \Pider \In a(x) . \proc{P} : `G,a{:}A |- `D }
 \end{array}
 \kern2mm
 \begin{array}{@{}rl}
(\OutRule) : &
 \Inf	[a \not= b]
	{ \Pider \proc{P} : `G,b{:}A |- b{:}A,`D }
	{ \Pider \Out a <b> . \proc{P} : `G,b{:}A |- a{:}A,b{:}A,`D }
\\ [4mm]
(\PairOut) : &
\Inf	[A \notele `D; a,c \notele `G]
	{ \Pider \proc{P} : `G,b{:}A |- c{:}B,`D }
	{ \Pider \Out a <b,c> . \proc{P} : `G,b{:}A |- a{:}A\arr B,c{:}B,`D }
\\ [4mm]
(\Let) : &
 \Inf	[y,z \notele `D; x \notele `G]
	{ \Pider \proc{P} : `G,y{:}B |- x{:}A,`D }
	{ \Pider \Let <x,y> = z in \proc{P} : `G,z{:}A\arr B |- `D }
 \end{array}
 \end{array} \]
\caption{The implicative type assignment rules for the $`p$-calculus.} \label{rules}
 \end{figure*}}

\def\DDinpi{\begin{figure*}[t]
 \[ \def\arraystretch{1.1} \begin{array}{l@{\kern-3mm}cl}
\SPilmuTerm [`D `D] a
	\quad \ByDef \quad
\PilmuTerm [a {l x . xx} `lx.xx] a
	&\redPi (c)& \\
\PilmuTerm [S xx x := `ly.yy] a
	&\ByDef& \\
\SPilmuTerm [S xx x := `ly.yy] a
	&\ByDef& \\
\PilmuTerm [s {A {v x} x} x := {l y . yy}] a
	&\redPi (x,w)& \\
\PilmuTerm [S {a {l y . yy} x} x := `ly.yy] a
	& \ByDef & \\
\SPilmuTerm [S {{}(`ly.yy)x} x := `ly.yy] a
	& \redPi (c)& \\
\NewA x ( \NewA yb ( \PilmuTerm[yy] b \Par \PilmuTerm[y := x] {} \Par \BEq d=a \Par
	\NewA c ( \PiExContsub c := x . a ) ) \Par \PiExSub x := `ly.yy )
	& \wbisim & \\
\PilmuTerm [S {s {A {v y} y} y := x } x := `ly.yy] a
	& \ByDef & \\
\SPilmuTerm [S {S yy y := x } x := `ly.yy] a
	& \redPi (y) & \\
\NewA x ( \NewA cw ( \BEq w=c \Par \PiExContSub c := y . a \Par \PilmuTerm[x] w ) \Par
	\PiExSub y := x \Par \PiExSub x := `ly.yy )
	& \wbisim & \\
\PilmuTerm [S {S {A x y} y := x } x := `ly.yy] a & \ByDef & \\
\SPilmuTerm [S {S xy y := x } x := `ly.yy] a 
& \ldots 
 \end{array} \]
\caption{Running $\SPilmuTerm [{}(`lx.xx) (`lx.xx)] a $} \label{DDinpi}
\end{figure*}}

\def \dDouble{
 \begin{figure*}[t]
 \[ \def\arraystretch{1.2}
\kern-3mm \begin{array}{@{}l@{\kern-1mm}c@{\kern-4mm}l}
\SPilmuTerm[{}(`lx.x)(\muterm`a.[`a](`lq.q)(\muterm`b.[`a]`ly.y))] a
	& \ByDef & \\
\PilmuTerm[a `lx.x \muterm`a.[`a](`lq.q)(\muterm`b.[`a]`ly.y)] a
	& \ByDef & \\
\PilmuTerm[a {l x . x} \muterm`a.[`a](`lq.q)(\muterm`b.[`a]`ly.y)] a
	& \red (c) \\
\New xb ( \PilmuTerm[x] b \Par \PiExSub x := {{}\muterm`a.[`a](`lq.q)(\muterm`b.[`a]`ly.y)} \Par \BEq b=a ) \Par \New c (\Bang \In c (v,d) . \dots)
	& \congruent,\ByDef,\wbisimG & \\
\New xb ({ \PilmuTerm[v x] b \Par \PiExsub x := {{}\muterm`a.[`a](`lq.q)(\muterm`b.[`a]`ly.y)} \Par \BEq b=a })
	& \red (x) \\
\New wb ( \BEq w=b \Par \SPilmuTerm[u `a . `a {}(`lq.q)(\muterm`b.[`a]`ly.y)] w \Par \BEq b=a ) \Par \New x ( \Bang \New w (\Out x <w> . \dots ) )
	& \ByDef,\wbisimG & \\
\New wb ( \BEq w=b \Par \PilmuTerm[{}(`lq.q)(\muterm`b.[w]`ly.y)] w \Par \BEq b=a )
	& \ByDef & \\
\New wb ( \BEq w=b \Par \PilmuTerm[a `lq.q {\muterm`b.[w]`ly.y}] w \Par \BEq b=a )
	& \ByDef,=_{`a} & \\
\setcounter{indb}{1}
\New `ab ( \BEq `a=b \Par \PilmuTerm[a {l q . q} {\muterm `b . [`a]`ly.y}] `a \Par \BEq b=a ) 
	& \red (c), \wbisimG,\ByDef \\
\New `ab ( \BEq `a=b \Par \New qb_1 ( \PilmuTerm[q] b_1 \Par
	\PiExSub q := {{}\muterm`b.[`a]`ly.y} \Par \BEq b_1=`a ) \Par \BEq b=a )
	& \ByDef \\
\New `ab ( \BEq `a=b \Par \New qb_1 ( \PilmuTerm[v q] b_1 \Par 
	\PiExsub q := {{}\muterm`b.[`a]`ly.y} \Par \BEq b_1=`a ) \Par \BEq b=a ) \dquad
	& \red (q), \wbisimG,\ByDef,\congruent \\
\New `ab ( \BEq `a=b \Par
	\New wb_1 ( \Eq w=b_1 \Par \SPilmuTerm[u `b . `a {l y . y}] w \Par \BEq b_1=`a ) \Par \BEq b=a )
	& \ByDef \\
\New `ab ( \BEq `a=b \Par
	\New wb_1 ( \Eq w=b_1 \Par \PilmuTerm[l y . y] `a \Par \BEq b_1=`a ) \Par \BEq b=a )
	& \congruent ,\ByDef \\
\New `ab ( \BEq `a=b \Par
\setcounter{indb}{2}
	\New wb_1 ( \Eq w=b_1 \Par \PilmuTerm[l y . y] `a \Par \BEq b_1=`a ) \Par \BEq b=a )
	& \ByDef,\wbisimR, \wbisimG \\
\New `ab ( \BEq `a=b \Par \PilmuTerm[`l y . y] `a \Par \BEq b=a )
	& \ByDef,\wbisimR, \wbisimG & \\
\PilmuTerm[l y . y] a
	& \congruent \\
\PilmuTerm[u `a . `a {(`l y . y)}] a
	& \ByDef \\
\SPilmuTerm[u `a . `a {l y . y}] a
\\ [-4mm]
 \end{array} \]
\caption{The interpretation of a term with double output} \label{figure double}
 \end{figure*}}

\def \Double{
 \begin{figure*}[t]
 \[ \def\arraystretch{1.2}
\kern-3mm \begin{array}{@{}l@{\kern-1mm}c@{\kern-4mm}l}
\SPilmuTerm[\muterm`a.[`a](`lq.q)(\muterm`b.[`a]`ly.y)] a
	& \ByDef & \\
\PilmuTerm[{}(`lq.q)(\muterm`b.[a]`ly.y)] a
	& \ByDef & \\
\PilmuTerm[a {l q . q} \muterm`b.[a]`ly.y] a
	& \red (c), \wbisimG,\ByDef \\
\New qb ( \PilmuTerm[q] b_1 \Par
	\PiExSub q := {{}\muterm`b.[`a]`ly.y} \Par \BEq b=`a )
	& \ByDef \\
\New qb ( \PilmuTerm[v q] b \Par
	\PiExsub q := {{}\muterm`b.[`a]`ly.y} \Par \BEq b=`a )
	& \red (q), \ByDef,\congruent \\
\New w ({ \New b ({ \BEq w=b \Par \SPilmuTerm[u `b . `a {l y . y}] w \Par \BEq b=`a }) \Par \New q ({ \Out q <w> . (\dots) }) })
	& \wbisimG,\ByDef \\
\New wb ({ \BEq w=b \Par \PilmuTerm[u `b . `a {`l y . y}] w \Par \BEq b=`a })
	& \congruent,\ByDef \\
\New wb ({ \BEq w=b \Par \PilmuTerm[`l y . y] `a \Par \BEq b=`a })
	& \congruent \\
\PilmuTerm[`l y . y] a \Par \New wb ( \BEq w=b \Par \BEq b=`a )
	& \wbisimG \\
\PilmuTerm[u `a . `a {(`l y . y)}] a
	& \ByDef 
\SPilmuTerm[u `a . `a {l y . y}] a
\\ [-4mm]
 \end{array} \]
\caption{The interpretation of a term with double output} \label{figure double}
 \end{figure*}}

\def \DDnorename{\begin{figure*}
 \[ \begin{array}[t]{@{}l@{}cl}
\SPilmuTerm [a `D `D] a \quad = \quad
\SPilmuTerm [a (`lx.xx) `D] a
	& \ByDef & \\
\PilmuTerm [a {l x . xx} `D] a
	& \red (c) \\
\New xb ( \PilmuTerm [xx] b \Par \PilmuTerm[x := `D] {} \Par \BEq b=a ) \Par \New c ( \Bang \In c (v,d) . ( \PilmuTerm[x := `D] {} \Par \BEq d=a )
	& \ByDef,\wbisimG & \\
\New xb ( \PilmuTerm [a {v x} x] b \Par \PiExsub x := `D \Par \BEq b=a )
	& \red (x) \\
\New xbw ( \New c ( \BEq w=c \Par \Bang \In c (v,d) . ( \PiExSub v := x \Par \BEq d=b ) ) \Par
\PilmuTerm[l y . yy] w \Par \PiExSub x := `D \Par \BEq b=a )
	& \red (w) \\
\setcounter{indb}{1}	
\New xb ( \New c ( \PilmuTerm[l y . yy] c \Par \Bang \In c (v,d) . ( \PiExSub v := x \Par \BEq d=b ) ) \Par \PiExSub x := `D \Par \BEq b=a )
	& \red (c),\wbisimG & \\
\New xb ( \New yb_1 ( \PilmuTerm[yy] b_1 \Par \PiExSub y := x \Par \BEq b_1=b ) \Par \PiExSub x := `D \Par \BEq b=a )
	& \ByDef & \\
\NewA xb ( \NewA yb_1 ( \PilmuTerm[a {v y} y] b_1 \Par {} \\ ~ \hfill \PiExsub y := x \Par \BEq b_1=b ) \Par \PiExSub x := `D \Par \BEq b=a )
	& \red (y) \\
\NewA xb ( \NewA yb_1 ( \New c ( \Eq w=c \Par \Bang \In c (v,d) . (\PiExSub v := y \Par \BEq d=b_1 ) ) \Par \PilmuTerm[x] w \Par
{} \\ ~ \hfill
\PiExSub y := x \Par \BEq b_1=b ) \Par \PiExSub x := `D \Par \BEq b=a )
	& \ByDef & \\
\NewA xb ( \NewA yb_1 ( \New c ( \Eq w=c \Par \Bang \In c (v,d) . (\PiExSub v := y \Par \BEq d=b_1 ) ) \Par {} \\ ~ \hfill
\PilmuTerm[v x] w \Par \PiExSub y := x \Par \BEq b_1=b ) \Par \PiExSub x := `D \Par \BEq b=a )
	& \congruent & \\
\NewA xb ( \NewA yb_1 ( \New c ( \Eq w=c \Par \Bang \In c (v,d) . (\PiExSub v := y \Par \BEq d=b_1 ) ) \Par
\PilmuTerm[v x] w \Par \PiExSub y := x \Par \BEq b_1=b ) \Par
			{} \quad \\ \dquad
\PiExsub x := { l z . zz} \Par \PiExSub x := `D \Par \BEq b=a )
	& & \kern -25mm \red (x,w_1,w,c) \\
\New xb ( \New yb_1 ( \New zb_2 ( \PilmuTerm[zz] b_2 \Par \PiExSub z := y \Par \BEq b_2=b_1 ) ) \Par \PiExSub y := x \Par \BEq b_1=b \Par \PiExSub x := `D \Par \BEq b=a ) )
 \end{array} \]
\setbox211=\hbox{$\SPilmuTerm [a `D `D] a $}
\caption{Running {\box211} without renaming, but using garbage collection.} \label{DDnogarb}
 \end{figure*}}

\setbox81=\hbox{$\PilmuTerm[`D`D] a $}

\title{A fully-abstract semantics of $\lmu$ in the $`p$-calculus}

\MSCS{%
\author[S. van Bakel and M.G. Vigliotti]
{S\ls t\ls e\ls f\ls f\ls e\ls n\ns v\ls a\ls n\ns B\ls a\ls k\ls e\ls l$^1$\ns
and\ns
M\ls a\ls r\ls i\ls a\ns G\ls r\ls a\ls z\ls i\ls a\ns V\ls i\ls g\ls l\ls i\ls o\ls t\ls t\ls i$^2$
\\
$^1$ Department of Computing, Imperial College London, 180 Queen's Gate, London SW7 2BZ, UK
\addressbreak
$^2$ Adelard LLP Exmouth House, 3-11 Pine Street London EC1R 0JH, UK \\
}}

\CLAC{
\author{%
Steffen van Bakel
 \institute{Department of Computing, Imperial College London, \\ 180 Queen's Gate, London SW7 2BZ, UK}
 \email{s.vanbakel@imperial.ac.uk}
 \and
Maria Grazia Vigliotti
 \institute{Adelard LLP Exmouth House, \\ 3-11 Pine Street London EC1R 0JH, UK}
 \email{mgv@adelard.com}
}}

 \def\titlerunning{A fully-abstract semantics of $\lmu$ in the $`p$-calculus}
 \def\authorrunning{van Bakel and Vigliotti}

 \def\Abstract{ \begin {abstract}
We study the $\lmu$-calculus, extended with explicit substitution, and define a compositional output-based interpretation into a variant of the $`p$-calculus with pairing that preserves single-step explicit head reduction with respect to weak bisimilarity.
We define four notions of weak equivalence for $\lmu$ -- one based on weak reduction $\equivwbmu$, two modelling weak head-reduction and weak explicit head reduction, $\equivwh$ and $\equivwxh$ respectively (all considering terms without weak head-normal form equivalent as well), and one based on weak approximation $\equivA$ -- and show they all coincide.
\Comment{
We define notions of weak equivalence for $\lmu$, based on weak reduction and on weak explicit head reduction (all considering terms without weak head-normal form equivalent as well) and show they coincide.
}
We will then show full abstraction results for our interpretation for the weak equivalences with respect to weak bisimilarity on processes.
 \end {abstract}
}

\MSCS{\bibliographystyle {harvard}}
\CLAC{\bibliographystyle {eptcs}}

\date{}

\maketitle

\Abstract


 \section*{Introduction}
The research presented in this paper is part of an ongoing investigation into the suitability of classical logic in the context of programming languages with control.
Rather than looking at how to encode known control features into calculi like the $`l$-calculus \cite{Church'36,Barendregt'84}, Parigot's $\lmu$-calculus \cite{Parigot'92}, or $\Lmu$\Paper{\footnote{The name $\Lmu$ was first introduced in \cite{Saurin'05}, that also introduced a different notation for terms, in placing names \emph{behind} terms, rather than in front, as done by Parigot and de Groote; we use their notation here.}} \cite{deGroote'94}, as has been done in great detail by others, we focus on trying to understand what is exactly the notion of computation that is embedded in calculi like $\lmu$; we approach that problem here by presenting a fully abstract interpretation for that calculus into the (perhaps better understood) $`p$-calculus \cite{Milner'92}.

\CLAC{

In the past, many }
researchers \Paper{to }investigate\CLAC{d} 
interpretations into the $`p$-calculus of various calculi that have their foundation in classical logic\Paper{, as done in, for example, \cite{Honda-Yoshida-Berger'04,vBCV-CLaC'08,CiminiCS'10,Beffara-Mogbil'12}}.
From these papers it might seem that the interpretation of such `classical' calculi comes at a great expense; for example, to encode \emph{typed} $\lmu$, \cite{Honda-Yoshida-Berger'04} defines an extension of Milner's encoding and considers a \Paper{version of the $`p$-calculus that is strongly typed}\CLAC{strongly typed $`p$-calculus}; \cite{vBCV-CLaC'08} shows preservation of reduction in $\X$ \cite{Bakel-Lescanne-MSCS'08} only with respect to $\Ctg$, the contextual ordering (so not with respect to $\equivC$, contextual equivalence, nor with respect to weak bisimilarity); \cite{CiminiCS'10} defines a non-compositional interpretation of $\lmmt$ \cite{Curien-Herbelin'00} that strongly depends on recursion, and does not regard the logical aspect\Paper{ at all}.

\CLAC{In \cite{Bakel-Vigliotti-IFIPTCS'12} we started our investigations by presenting an interpretation for de Groote's variant $\Lmu$ into the $`p$-calculus \cite{Milner'92} and proved a soundness result; here we show that this interpretation is fully abstract, but have to limit the interpretation to $\lmu$ terms. }
We study an output-based encoding of $\lmu$ into the $`p$-calculus that is an extension of the one we defined for the $`l$-calculus \cite{Bakel-Vigliotti-CONCUR'09} and is a natural variant of that for $\Lmu$ in \cite{Bakel-Vigliotti-IFIPTCS'12}.
In those papers, we have shown that our encoding respects \emph{single-step explicit head reduction} (which only ever replaces the head variable of a term) modulo $\equivC$\Paper{; here we restate those properties with respect to $\wbisimilar$, weak bisimilation}.
\CLAC{

We will here address the natural question that arises next: are two terms that are equal under the interpretation also operational equivalent, \emph{i.e.}: is the interpretation \emph{fully abstract}?
We answer that question positively, using a new approach to showing full abstraction, for our interpretation of $\lmu$-terms (rather than $\Lmu$ as used in \cite{Bakel-Vigliotti-IFIPTCS'12}) and thereby also for the standard $`l$-calculus. }
\Paper{Using a new technique, we will show that our interpretation is fully abstract, \emph{i.e.}~two terms that are equal under the interpretation are also operational equivalent; our approach will be to first show that our interpretation respects single-step explicit head reduction $\redxh$ modulo \emph{weak bi-simularity} $\wbisim$ and to }
\CLAC{Following the approach of \cite{Bakel-Vigliotti-IFIPTCS'12} we can show that our interpretation respects single-step explicit head reduction $\redxh$ modulo \emph{weak bisimularity} $\wbisim$ (rather than $\equivC$ as used in \cite{Bakel-Vigliotti-IFIPTCS'12}; we omit the details here).
We }
extend this result to $\equivwxh$, the equivalence relation generated by $\redxh$ that equates also terms without (weak) normal form with respect to $\redxh$.
The main proof of the full abstraction result is then achieved through showing that $\equivwxh$ equates to $\equivwbmu$, the equivalence relation generated by standard reduction that also equates terms without weak head normal form\Paper{: this latter result is stated entirely within $\lmu$}.

This technique is considerably different from the one used by Sangiorgi, who has shown a full abstraction result \cite{Sangiorgi'94,Sangiorgi-Walker-Book'01} for Milner's encoding $\MilSem[M] a $ of the lazy $`l$-calculus \cite{Milner'92}.
\Paper{The characterisation of $\MilSem[M] a \wbisim \MilSem[N] a $, left as open problem in \cite{Milner'92}, was achieved through showing that the equivalence classes under weak-bisimilarity of Milner's encoding form a model for the lazy $`l$-calculus in the sense that it provides a denotational semantics, similar to Cor.~\ref{semantics} below. }
To achieve full abstraction, Sangiorgi proves that $\MilSem[M] a \wbisim \MilSem[N] a $ if and only if $M \AppBis N$, where $\AppBis$ is the \emph{applicative bisimularity} on $`l$-terms \cite{Abramsky-Ong'93b}\CLAC{. }\Paper{, an operational notion of equivalence on terms of the lazy $`l$-calculus as defined by Abramsky and Ong, rather than $`b$-equality.

}
However, this result comes at a price, since applicative bisimulation equates \Paper{the }terms \Paper{$ x (x \Theta `D`D) \Theta $ and $ x (`l y . x \Theta `D`D y ) \Theta $ (with $`D = `lx.xx$, and $\Theta$ is such that, for every $N$, $\Theta N$ is reducible to an abstraction) whereas these terms}\CLAC{that} are not weakly bisimilar under \Paper{the interpretation }$\MilSem[`.] `. $: in order to achieve full abstraction, Sangiorgi had to extend Milner's encoding to $`L_c$, a $`l$-calculus enriched with constants and by exploiting a more abstract encoding into the \emph{Higher Order} $`p$-calculus, a variant of the $`p$-calculus with higher-order communications.
Sangiorgi's result then essentially states that the interpretations of closed $`L_c$-terms $M$ and $N$ are contextually equivalent if and only if $M$ and $N$ are applicatively bisimilar; in \cite{Sangiorgi'94} he shows that the interpretation of terms in $`L_c$ in the standard $`p$-calculus is weakly bisimilar if and only if they have the same L\'evy-Longo tree\Paper{ \cite{Levy'76,Longo'83} (a lazy variant of B\"ohm trees \cite{Barendregt'84})}.


We would like to stress that in order to achieve full abstraction for our interpretation, \emph{we did not need to extend the interpreted calculus, and use a first order $`p$-calculus}.
In fact, \CLAC{the main contribution of this paper and }novelty of our proof is the \emph{structure of the proof} of the fact that our interpretation gives a fully abstract semantics.
To wit, we define a choice of operational equivalences for the $\lmu$-calculus, both with and without explicit substitution.
We define the \emph{weak explicit head equivalence} $\equivwxh$ and show that this is exactly the relation that is naturally representable in the $`p$-calculus; we define \emph{weak head equivalence} $\equivxh$ and show that for $\lmu$-terms without explicit substitution, $\equivwxh$ corresponds to $\equivxh$.
The relation $\equivwxh$ essentially equates terms that have the same L\'evy-Longo tree, but of course defined for $\lmu$, which gets shown through a notion of weak approximation.
We then show that the relation $\equivwA$, which expresses that terms have the same set of weak approximants, $\equivwxh$, and $\equivwbmu$ all correspond.

\CLAC{The combined results of \cite{Bakel-Vigliotti-CONCUR'09,Bakel-Vigliotti-IFIPTCS'12} and the full abstraction}\Paper{The} results we present here stress that the $`p$-calculus constitutes a very powerful abstract machine indeed: although the notion of structural reduction in $\lmu$ is very different from normal $`b$-reduction, no special measures had to be taken in order to be able to express it through our interpretation.
In fact, the distributive character of application in $\lmu$, and of both term and context substitution is dealt with \Paper{entirely }by congruence in $`p$\Paper{ (see also Example~\ref{redex example})}, and both naming and $`m$-binding are dealt with entirely statically by the interpretation.

\Comment{
The results of \cite{Bakel-Vigliotti-IFIPTCS'12} concentrated on the relation between $\Lmux$ and the $`p$-calculus, and in particular on Soundness and Completeness (Theorem~\ref{rtc soundness} here); in this paper, we will focus on full abstraction for $\lmu$.
We will define a notion of equivalence $\equivwxh$ between terms of $\lmux$ that equates also terms that have no weak head-normal form, and show that terms are equivalent with respect to $\equivwxh$ if and only if their images under $ \PilmuTerm[`.] `. $ are contextually equivalent.
We will then generalise this to equivalences generated by head reduction, standard reduction, and weak-approximation, respectively, that all equate terms that have no weak head-normal form, and show that all coincide: this will lead to our main result: $M \equivwbmu N \Iff \PilmuTerm[M] a \wbisimilar \PilmuTerm[N] a $.
}

{
 \vspace{2mm}
\noindent
\textbf{Organisation of this paper:}
We start with revisiting the $\lmu$-calculus in Section~\ref{lambda mu calculus} and define a notion of \emph{head-reduction} $\redh$.
In Section~\ref{pi with pairing} we revisit the $`p$-calculus, enriched with \emph{pairing}.
In Section~\ref{lmux section} we define $\lmux$, a version of $\lmu$ with \emph{explicit substitution}, as well as a notion of \emph{explicit head reduction} and in Section~\ref{lambda interpretation} define our \emph{logical interpretation} of $\lmux$ in to $`p$.

Working towards our full abstraction result, in Section~\ref{weak reduction} we will define notions of weak reduction, in particular \emph{weak head reduction} and \emph{weak explicit head reduction}.
We then define the two notions of equivalence these induce, also equating terms without weak head-normal form and show that these notions coincide on pure $\lmu$ terms (\emph{i.e.}~without explicit substitutions).
We also define the equivalence $\equivwbmu$ induced by $\redbmu$ on pure $\lmu$ terms, that also equates terms without weak head-normal form.
In Section~\ref{approximation section}, we define a notion of \emph{weak approximation} for $\lmu$, and show the semantics this induces, $\equivwA$, is fully abstract with respect to both $\equivwh$ and $\equivwbmu$.
We show that our logical interpretation is fully abstract with respect to weak bisimilarity $\wbisim$ on processes and $\equivwxh$, $\equivwh$, $\equivwA$, and $\equivwbmu$ on pure $`l`m$-terms.
}

\vspace*{2mm}
\noindent
\textbf{Notation:}
We will use a vector notation $\Vect{`.}$ as abbreviation for any sequence: for example, $\Vect{x_i}$ stands for $x_1, \ldots, x_n$, for some $n$, or for $\Set{x_1, \ldots, x_n}$, \CLAC{and
}$\langle\Vect{ `a_i := N_i `. `b_i }\rangle$ for $ \excontsub`a_1 := N_1 . `b_1 \,$ $\dots\, \excontsub `a_n := N_n . `b_n $,\Paper{ $\Vect{M_i \eqh N_i}$ for $\Forall \iotn \Pred [ M_i \eqh N_i ] $,} etc.
When possible, we will drop the indices.


 \section{The $\lmu$~calculus\CLAC{ and explicit substitution}} \label{lambda mu calculus}

In this section, we will briefly discuss Parigot's $\lmu$-calculus~\cite{Parigot'92}; we assume the reader to be familiar with the $`l$-calculus and its notion of reduction $\bred$ and equality $=_{`b}$.

$\lmu$ is a proof-term syntax for classical logic, expressed in Natural Deduction, defined as an extension of the Curry type assignment system for the {\LC} by adding the concept of \emph{named} terms, and adding the functionality of a \emph{context switch}, allowing arguments to be fed to subterms.

\Paper{In the next section we will define explicit head reduction for $\lmux$, a variant of $\lmu$ with explicit substitution \emph{\`a la} $\Lx$ \cite{Bloo-Rose'95}, and will show full abstraction results for $\lmux$; since $\lmux$ implements $\lmu$-reduction, this implies that, automatically, our main results are also shown for standard reduction (with implicit substitution).}

 \begin{definition}[Syntax of $\lmu$] \label{lm-terms}
\Paper{The $\lmu$-\emph{terms} we consider are defined over the set of \emph{variables} represented by Roman characters, and \emph{names}, or \emph{context} variables, represented by Greek characters, through the grammar:}%
\CLAC{The $\lmu$-\emph{terms} we consider are defined over the set of \emph{variables} (Roman characters) and \emph{names}, or \emph{context} variables (Greek characters), through:}
\Paper{
 \[ \begin{array}{@{}rrl@{\quad}l}
M,N &::=& x & \textit{variable} \\
& \mid & `l x.M & \textit{abstraction} \\
& \mid & MN & \textit{application} \\
& \mid & \muterm`a.[`b]M & \textit{context switch}
 \end{array} \]
}
\CLAC{
 \[ \begin{array}{rcl}
M,N &::=& x \mid `l x.M \mid MN \mid \muterm`a.[`b]M
 \end{array} \]
}
We will occasionally write $\Cmd$ for the pseudo-term $[`a] M$.
 \end{definition}
\Paper{In fact, the main difference between $\Lmu$ and $\lmu$ is that in the former, $[`a] M$ is considered to be a term.}

As usual, $`l x.M$ binds $x$ in $M$, and $\muterm `a .\Cmd$ binds $`a $ in $\Cmd$, and the notions of free variables $\fv(M)$ and names $\fn(M)$ are defined accordingly; the notion of $`a $-conversion extends naturally to bound names and we assume Barendregt's convention in that we assume that free and bound variables and names are always distinct, using $`a$-conversion when necessary.
\Paper{
As usual, we call a term \emph{closed} if it has no free variables.

Simple type assignment for $\lmu$ is defined as follows:
 \begin {definition}[Types, Contexts, and Typing] \label{types} \label {typing for lmu}
 \begin {enumerate}

 \item
Types are defined by:
 \[ \begin{array}{rcl}
A,B & ::= & \tvar \mid A \arrow B
 \end{array} \]
where $\tvar$ is a basic type of which there are infinitely many.

 \item
A \emph{context of inputs} $`G$ is a mapping from term variables to types, denoted as a finite set of \emph{statements} $\stat{x}{A}$, such that the \emph{subject} of the statements ($x$) are distinct.
We write $\G_1,\G_2$ for the \emph{compatible} union of $\G_1$ and $\G_2$ (if $\stat{x}{A_1} \ele \G_1$ and $\stat{x}{A_2} \ele \G_2$, then $A_1 = A_2$), and write $`G, \stat{x}{A}$ for $`G, \{\stat{x}{A}\}$.

 \item
Contexts of \emph{outputs} $`D$ as mappings from names to types, and the notions $`D_1,`D_2$ and $`a{:}A,`D$ are defined similarly.

 \item
Type assignment for $\lmu$ is defined by the following natural deduction system.
 \[ \kern-5mm 
 \begin{array}{@{}rl@{\quad}rl}
(\Ax) : &
\Inf	{ \derLmu `G,x{:}A |- x : A | `D }
& 
 (`m) : &
 \Inf	[`a \notele `D]
	{ \derLmu `G |- M : B | `a{:}A,`b{:}B,`D
	}{ \derLmu `G |- \muterm`a.[`b]M : A | `b{:}B,`D }
\quad
 \Inf	[`a \notele `D]
	{ \derLmu `G |- M : A | `a{:}A,`D
	}{ \derLmu `G |- \muterm`a.[`a]M : A | `D }
\end{array} \] \[ \begin{array}{@{}rl@{\dquad}rl}
(\arrI) : &
\Inf	[x \notele `G]
	{ \derLmu `G,x{:}A |- M : B | `D
	}{ \derLmu `G |- `l x.M : A\arrow B | `D }
& 
(\arrE) : &
\Inf	{ \derLmu `G |- M : A\arrow B | `D
	 \quad
	 \derLmu `G |- N : A | `D
	}{ \derLmu `G |- MN : B | `D }
 \end{array} \]

 \end {enumerate}
 \end{definition}
So, for the context $`G, \stat{x}{A}$, we have either $\stat{x}{A} \in `G$, or $`G$ is not defined on $x$. 

In $\lmu$, reduction of terms is expressed via implicit substitution; a%
}
\CLAC{A}s usual, $M[N \For x]$ stands for the substitution of all occurrences of $x$ in $M$ by $N$, and $M [ N{`.}`g \For `a ]$, the \emph{structural substitution}, for the term obtained from $M$ when every (pseudo) sub-term of the form $[`a]M'$ is replaced by $[`g]M' N$.\Paper{\footnote{This notion is often defined as $M [ N \For `a ]$, where every (pseudo) sub-term of the form $[`a]M'$ is replaced by $[`a]M' N$; in our opinion, this creates confusion, since $`a$ gets `reintroduced'; it is in fact a new name, which is illustrated by the fact that, in the typed version $`a$ then changes type.}}
\CLAC{(We omit the formal definition here; see Def.~\ref{definition lmux} for the variant with \emph{explicit} structural substitution.)}
\Paper{
For reasons of clarity, and because below we will present a version of $\lmu$ that makes the substitution explicit, we define the $`m$-substitution formally.

 \begin{definition} [Structural substitution] \CLAC{\hfill}
We define $M [ N{`.}`g \For `a ]$ \Paper{(where every sub-term $[`a]L$ of $M$ is replaced by $[`g]LN$) }by induction over the structure of (pseudo-)terms by:
 \[ \begin{array}{@{}r@{\,}lcl@{\quad}l}
([`a]M) & [ N{`.}`g \For `a ] & \ByDef & [`g] ( M \, [ N{`.}`g \For `a ] ) N
	\\
([`b]M)&[ N{`.}`g \For `a ] & \ByDef & [`b] ( M \, [ N{`.}`g \For `a ] ) & (`a\not=`b)
	\\
(\muterm`b.\Cmd) & [ N{`.}`g \For `a ] & \ByDef & \muterm`b. ( \Cmd \, [ N{`.}`g \For `a ] )
	\\
x & [ N{`.}`g \For `a ] & \ByDef & x
	\\
(`lx.M) & [ N{`.}`g \For `a ] & \ByDef & `lx . (M \, [ N{`.}`g \For `a ] )
	\\
(M_1M_2) & [ N{`.}`g \For `a ] & \ByDef & M_1 \, [ N{`.}`g \For `a ]~M_2 \, [ N{`.}`g \For `a ]
 \end{array}\]
 \end{definition}

We have the following rules of computation in $\lmu$:
}

 \begin{definition}[$\lmu$ reduction] \label{lmu reduction}
\CLAC{Reduction on $\lmu$-terms is defined as the contextual closure of the rules:}
\Paper{$\lmu$ has a number of reduction rules: two \emph{computational rules}: }
 \[ \begin{array}{@{}rrcll}
\textit{logical } (`b): & (`l x . M ) N & \red & M [ N \For x ]
	\\
\textit{structural } (\mu): & (\muterm `a . \Cmd ) N & \red & \muterm`g. ( \Cmd [N{`.}`g \For `a] )
\Paper{ \end{array} \]
as well as the \emph{simplification rules}:
 \[ \begin{array}{@{}rrcll}
}\CLAC{\\ }
\textit{renaming} : & \muterm `d . [`b](\muterm`g.[`a]M) & \red & \muterm `d . [`a] M[`b \For `g]
	\\
\textit{erasing} : & \muterm `a . [`a] M & \red & M & (`a \notele \fn(M))
 \end{array} \]
\Paper{which are added mainly to simplify the presentation of results.
We use the contextual rules:\footnote{Normally the contextual rules are not mentioned but are left implicit; here we need to state them, since we will below consider notions of reduction that do not permit all contextual rules.}
 \[ \begin {array}[t]{@{}r@{~~}c@{~~}l}
M \red N &\Then&
 \begin{cases}
ML &\red& NL \\
LM &\red& LN \\
`lx.M &\red& `lx.N \\
\muterm`a.[`b]M & \red & \muterm`a.[`b]N
 \end{cases}
 \end {array} \]}%
%
%
We use ${}\rtcredbmu{}$ for the pre-congruence\Paper{\footnote{A \emph{pre-congruence} is a reflexive and transitive relation that is preserved in all contexts; a \emph{congruence} is symmetric pre-congruence.}} based on these rules, ${}\eqbmu{}$ for the congruence, write $M \rtcredbmu[\nf] N$ if $M \rtcredbmu N$ and $N$ is in normal form, $M \rtcredbmu[\hnf] N$ if $M \rtcredbmu N$ and $N$ is in head-normal form, $M \Shows $ if there exists a finite reduction path starting from $M$,\Paper{\footnote{Note that this does not imply that \emph{all} paths are finite.}} and $M \Diverges$ if this is not the case; we will use these notations for other notions of reduction as well.


 \end{definition}

That this notion of reduction is confluent was shown in \cite{Py-PhD'98}; so we have:

 \begin{proposition} \label{confluence eq}
If $M \eqbmu N $ and $M \rtcredbmu P $, then there exists $Q$ such that $ P \rtcredbmu Q $ and $N \rtcredbmu Q $.
 \end{proposition}

\Paper{
For convenience, Parigot also considers $[`a]M$ and $\muterm`a.M$ as \emph{pseudo}-terms, that represent \emph{deactivation} and \emph{activation}.\footnote{In fact, \cite{Parigot'92} formulates the renaming rule as $ [`b](\muterm`g.M) \red M[`b \For `g] $.}

The intuition behind the structural rule is given by \citet{deGroote'94}: ``\emph{in a $\lmu$-term $\muterm`a.M$ of type $A \arr B$, only the subterms named by $`a$ are \emph{really} of type $A \arr B$ (\ldots); hence, when such a $`m$-abstraction is applied to an argument, this argument must be passed over to the sub-terms named by $`a$.}''
We can think of $[`a]M$ as storing the type of $M$ amongst the alternative conclusions by giving it the name $`a $.

\cite{Parigot'93} has shown that typeable terms are strongly normalisable.
It also defines the extensional rules:
\[ \begin{array}{@{}r@{\quad}rcl@{\quad}l}
(\eta): & `lx.Mx & \red & M & (x\not\in \fv(M))
	\\
(`h`m) : & \muterm`a . [`b] M & \red & `lx\muterm`g . [`b] M[x{`.}`g \For `a]
 \end{array} \]
Here we do not consider these rules: the model we present through our interpretation is not extensional, and we can therefore not show that those rules are preserved by the interpretation (see Rem.~\ref{not extensional}).

 \begin{example} \label{lmu double negation}
As an example illustrating the fact that this system is more powerful than the system for the \LC, here is a proof of it is possible to inhabit Peirce's Law (due to \cite{Ong-Stewart'97}):
 \[ \begin{array}{@{}c}
 \Inf	[\arrI]
	{\Inf	[`m]
		{\Inf	[\arrE]
			{\Inf	[\Ax]
				{ \derLmu x{:}(A\arrow B)\arrow A |- x : (A\arrow B)\arrow A | `a{:}A }
			 \Inf	[\arrI]
				{\Inf	[`m]
					{\Inf	[\Ax]
						{ \derLmu x{:}(A\arrow B)\arrow A,y{:}A |- y : A | `b{:}B }
					}{ \derLmu x{:}(A\arrow B)\arrow A,y{:}A |- \muterm`b.[`a]y : B | `a{:}A }
				}{ \derLmu x{:}(A\arrow B)\arrow A |- `ly.\muterm`b.[`a]y : A\arrow B | `a{:}A }
			}{ \derLmu x{:}(A\arrow B)\arrow A |- x(`ly.\muterm`b.[`a]y) : A | `a{:}A }
		}{ \derLmu x{:}(A\arrow B)\arrow A |- \muterm`a.[`a](x(`ly.\muterm`b.[`a]y)) : A | {} }
	}{ \derLmu { } |- `lx.\muterm`a.[`a](x(`ly.\muterm`b.[`a]y)) : ((A\arrow B)\arrow A)\arrow A | {} }
 \end{array} \]
The underlying logic of the system of Def.~\ref{typing for lmu} corresponds to \emph{minimal classical logic} \cite{Ariola-Herbelin'03}.
 \end{example}

We also consider the notion of head reduction; it is defined in \cite{Wadsworth'76} for the $`l$-calculus by first defining the head-redex of a term as the subterm $(`lx.M)N$ in a term of the form
 \[ `lx_1x_2\dots x_n.((`lx.M)N)L_1L_2\dots L_m \quad (n \geq 0, m \geq 0) \]
Head reduction is then that notion in which each step is determined by contraction of the head redex (when it exists), and head-normal forms (the normal forms with respect to head reduction) are of the generic shape
 \[ `lx_1x_2\dots x_n.yL_1L_2\dots L_m \quad (n \geq 0, m \geq 0) \]
and $y$ in this term is called the head variable.
In $\lmu$, given the naming and $`m$-binding features, the notion of head redex is not this easily defined; rather here we define head reduction by not allowing reductions to take place in the right-hand side of applications (in the context of the $`l$-calculus, this gives the original notion); we also define a notion of head-normal form for $\lmu$.
}

 \begin{definition} [Head reduction for $\lmu$ (cf.~\cite{Lassen'06})] \label{head reduction definition}
 \begin{enumerate}

 \firstitem
We define \emph{head reduction} $\redh$ as the restriction of $\redbmu$ by removing the contextual rule:
 \CLAC{$}\Paper{\[} \begin{array}{rcl@{}}
M \red N &\Then& LM \red LN
 \end{array} \CLAC{$}\Paper{\]}

 \item
The $\lmu$ \emph{head-normal forms} (\HNF) are defined through the grammar:
 \[ \begin {array}{@{}rrl@{\quad}l}
\lmuHNF & ::= &
`lx. \lmuHNF 
	\\ &\mid &
xM_1\dots M_n & (n \geq 0)
	\\ &\mid &
\muterm`a.[`b] \lmuHNF
	& (`b \not= `a \textrm{ or } `a \in \lmuHNF, \textrm{ and }
	\lmuHNF \not= \muterm`g.[`d] \lmuHNF' )
 \end {array} \]

 \end{enumerate}
 \end{definition}
\Paper{Notice that the $\redbmu$ \HNF s are $\redh$-normal forms.}

The following is straightforward: 
 \begin{proposition} [$\redh$ implements $\lmu$'s head reduction]
\label{head reduction} 
%
%
If $ M \rtcredbmu N $ with $N$ in {\HNF} (so $ M \rtcredbmu[\hnf] N $), then there exists $\lmuHNF$ such that $ M \rtcredh[\nf] \lmuHNF $ (so $\lmuHNF$ is in $\redh$-normal form) and $ \lmuHNF \rtcredbmu N $ without using $\redh$. 

 \end{proposition}

\Paper{
Notice that
 $ \begin{array}[t]{@{}lcl}
`lf.(`lx.f(xx))(`lx.f(xx))
	& \redh &
`lf.f((`lx.f(xx))(`lx.f(xx)))
 \end{array} $
and this last term is in \HNF, and in $\redh$-normal form.
}

{

 \section{The synchronous \texorpdfstring{$`p$}{}-calculus with pairing} \label{pi with pairing}

The notion of $`p$-calculus that we consider in this paper was already considered in \cite{Bakel-Vigliotti-CONCUR'09} and is different from other systems studied in the literature \cite{Honda-Tokoro'91} in that it adds \emph{pairing} and uses a {\proc{let}}-construct to deal with inputs of pairs of names that get distributed, similar to that defined in \cite{Abadi-Gordon'97}; in contrast to \cite{vBCV-CLaC'08,Bakel-Vigliotti-CONCUR'09}, we do not consider the asynchronous version of this calculus.


 \begin{definition}[Processes] \label{pi calculus}
\emph{Channel names} and \emph{data} are defined by:
 \[ \begin{array}{rcl@{\quad}l}
& & a,b,c,\dchan,x,y,z &\textsl{names}
 \end{array} \hspace*{1cm} \begin{array}{rcl@{\quad}l}
\proc{p} & ::= & a\mid \PiPair(a,b) ~ & \textsl{data}
 \end{array} \]
\Paper{(the pairing in data is \emph{not} recursive.)}
Processes are defined by:
\Paper{
 \[ \begin{array}{@{}rrl@{\quad}l}
\proc{P}, \proc{Q} & ::= & \Zero & \textsl{nil}
	\\
&\mid& \proc{P} \Par \proc{Q} & \textsl{composition}
	\\
&\mid& \Bang\proc{P} & \textsl{replication}
	\\ %
&\mid& \New{a} \proc{P} & \textsl{restriction}
	\\
&\mid& \In a(x) . \proc{P} & \textsl{input}
	\\
&\mid& \Out a <\procp> . \proc{P} & \textsl{output}
	\\
&\mid& \Let <x,y> = \proc{p} in \proc{P}	& \textsl{let construct}
 \end{array} \]
}\CLAC{\[ \begin{array}{@{}rrl@{\quad}l}
\proc{P}, \proc{Q} & ::= & \Zero \mid \proc{P} {\Par} \proc{Q} \mid \Bang\proc{P} \mid \New{a} \proc{P} \mid
\In a(x) . \proc{P} \mid \Out a <\procp> . \proc{P} \mid \Let <x,y> = \proc{p} in \proc{P}
 \end{array} \]
}
We see, as usual, $`n$ as a binder, and call the name $n$ \emph{bound} in $\New n \proc{P} $, $x$ bound in $\In a(x) . \proc{P} $ and $x,y$ bound in $\Let <x,y> = \proc{p} in \proc{P} $; we write $\bn(\proc{P})$ for the set of bound names in \proc{P}; $n$ is \emph{free} in $\proc{P}$ if it occurs in $\proc{P}$ but is not bound, and we write $\fn(\proc{P})$ for the set of free names in $\proc{P}$.
\Paper{We call $a$ in $\New a \proc{P}$ a \emph{hidden} channel.
A {\it context} $ \Cont[ \cdot]$ is a process with a hole $[ \,]$; we call $\In a (x) $ and $\Out a <\procp> $ \emph{guards}, and call \proc{P} in $\In a (x) . \proc{P} $ and $\Out a <\procp> . \proc{P} $ a process \emph{under guard}.}

 \end{definition}

Notice that data occurs only in two cases: $\Out a <{\proc{p}}> $ and $\Let <x,y> = \proc{p} in \proc{P} $, and that then $\proc{p}$ is either a single name, or a pair of names; we therefore do not allow $\In a ({ \PiPair<x,y>}) . \proc{P} $, nor $\Out a <{ \PiPair<{ \PiPair<b,c>},d>}> . \proc{P} $, nor $\Out {\PiPair<b,c>} <{\proc{p}}> . \proc{P} $, nor $\New \PiPair<a,b> \proc{P} $, nor $\Let <\mbox{$\PiPair<a,b>$},y> = \proc{p} in \proc{P} $, etc.
\Paper{So substitution $\proc{P}[p\For x]$ is a partial operation, which depends on the places in $\proc{P}$ where $x$ occurs; whenever we use $\proc{P}[p\For x]$, we will assume it is well defined.
It is worthwhile to point out that using pairing is not the same as working with the polyadic (or even dyadic) $`p$-calculus, because there each channel has a fixed arity, whereas we allow data to be sent, which is either a name or a pair of names.}

We abbreviate $\In {a} (x) . \Let <y,z> = x in \proc{P} $ by $\In {a} (y,z) . \proc{P} $, \Paper{as well as }$\New m \New n \proc{P} $ by $\New mn \proc{P} $, and write $\Out a <\procp> $ for $\Out a <\procp> . \Zero $. 
As in \cite{Sangiorgi-Walker-Book'01}, we write $\Eq a=b $ for the \emph{forwarder} $ \In a (x) . \Out b <x> $ \Paper{and $\BOut x (w) . \proc{P} $ for $ \New w ( \Out x <w> . \proc{P} ) $}.

 \begin{definition}[Structural Congruence]
The \emph{structural congruence} is the smallest congruence generated by the rules:
\noindent
 \[ \begin{array}{c@{\quad}c} \begin{array}{rcll}
\proc{P} \Par \Zero &\StrCon& \proc{P}
	\\
\proc{P} \Par \proc{Q} &\StrCon & \proc{Q} \Par \proc{P}
	\\
\Bang \proc{P} &\StrCon& \proc{P} \Par \Bang \proc{P}
	\\
\New n \Zero &\StrCon& \Zero
\CLAC{	\end{array} & \begin{array}{rcll} }
\Paper{\\ }
(\proc{P} \Par \proc{Q} ) \Par \proc{R} &\StrCon& \proc{P} \Par (\proc{Q} \Par \proc{R})
	\\
\New m \New n \proc{P} &\StrCon& \New n \New m \proc{P}
	\\
\New n (\proc{P} \Par \proc{Q}) &\StrCon& \proc{P} \Par \New n \proc{Q} & (n \notin \fn(\proc{P}))
	\\
\Let <x,y> = \PiPair(a,b) in \proc{P} &\StrCon& \proc{P} [a \For x,b \For y]
 \end{array} \end{array}
\]

 \end{definition}

As usual, we will consider processes modulo congruence and $`a$-conversion: this implies that we will not deal explicitly with the process $ \Let <x,y> = \PiPair(a,b) in \proc{P} $, but rather with $ \proc{P} [a \For x,b \For y] $.
Because of rule $(\proc{P} \Par \proc{Q} ) \Par \proc{R} \StrCon \proc{P} \Par (\proc{Q} \Par \proc{R}) $, we will \Paper{normally }not write brackets in a parallel composition of more than two processes.

Computation in the $`p$-calculus with pairing is expressed via the exchange of \textsl{data}.
 \begin{definition}[Reduction] \label{pi reduction}
\Paper{ \begin{enumerate}

 \firstitem}
The \emph{reduction relation} over the processes of the $`p$-calculus is defined by the following (elementary) rules:
 \[ \begin{array}{rcl@{~}l}
\Out a <\procp> . \proc{P} \Par \In a (x) . \proc{Q} &\redPi& \proc{P} \Par \proc{Q} [\proc{p} \For x]
	\Paper{ & (\textsl{synchronisation}) }
	\\
\proc{P} \redPi \proc{P'} &\Then& \New n \proc{P} \redPi \New n \proc{P}' \quad
	\Paper{ & (\textsl{hiding}) }
	\\
\proc{P} \redPi \proc{P}' &\Then& \proc{P} \Par \proc{Q} \redPi \proc{P}' \Par \proc{Q}
	\Paper{ & (\textsl{composition}) }
	\\
 \proc{P} \StrCon \proc{Q} \And \proc{Q} \redPi \proc{Q}' \And \proc{Q}' \StrCon \proc{P}'
	& \Then& \proc{P} \redPi \proc{P}'
	\Paper{ & (\textsl{congruence}) }
 \end{array} \]
\Paper{We write $ \proc{P} \redPi(c) \proc{Q} $ if $\proc{P}$ reduces to $\proc{Q}$ in a single step via a synchronisation over channel $c$, and write $\redPi(=_{`a})$ if we want to point out that $`a$-conversion has taken place during the synchronisation.
We say that $ \proc{P} \redPi(c) \proc{Q} $ takes place \emph{over a hidden channel} if $c$ is hidden in $\proc{P}$.}

\Paper{ \item
We say that a $\proc{P}$ is \emph{irreducible} (is in \emph{normal form}) if it does not contain a possible synchronisation. 

 \end{enumerate}}
 \end{definition}
Notice that the first rule is only allowed if $\proc{Q} [\proc{p} \For x]$ is a well-defined process.
\Comment{As usual, we write $\rtcredPi[+]$ for the transitive closure of $\redPi$, and $\rtcredPi$ for its reflexive and transitive closure; we write $\redPi(a)$ if we want to point out that a synchronisation took place over channel $a$}
\Paper{
Also,
 \[ \begin{array}[t]{@{}lclclcl}
\Out a <b,c> \Par \In a(x,y) . \proc{Q}
	& \ByDef &
\Out a <{\PiPair(b,c)}> \Par \In a(z) . \Let <x,y> = z in \proc{Q} \\
	& \redPi &
\Let <x,y> = \PiPair(b,c) in \proc{Q} \\
	& \StrCon &
\proc{Q} [b \For x,c \For y]
 \end{array} \]
}

There are several notions of equivalence defined for the $`p$-calculus: the one we consider here, and will show is related to our encoding, is that of weak-bisimilarity.

 \begin{definition} [Weak-bisimilarity] \label{bisimilarity} \label{other reduction}
 \begin{enumerate}

 \firstitem
We write $\proc{P} \outson n $ and say that $\proc{P}$ \emph{outputs on} $n$ (or $\proc{P}$ exhibits an output barb on $n$) if $\proc{P} \StrCon \New \Vect{b} ( \Out n <\procp> . \proc{Q} \Par \proc{R} ) $, where $ n \notele \Vect{b}$, and $\proc{P} \inson n $ ($\proc{P}$ \emph{inputs on} $n$) if $\proc{P} \StrCon \New \Vect{b} ( \In n (x) . \proc{Q} \Par \proc{R} ) $, where $ n\notele \Vect{b}$.

 \item
We write $\proc{P} \Outson n $ ($\proc{P}$ \emph{will output on} $n$) if there exists $\proc{Q}$ such that $\proc{P} \rtcredPi \proc{Q}$ and $\proc{Q} \outson n $, and $\proc{P} \Inson n $ ($\proc{P}$ \emph{will input on} $n$) if there exists $\proc{Q}$ such that $\proc{P} \rtcredPi \proc{Q}$ and $\proc{Q} \inson n $.

 \item A \emph{barbed bisimilarity} $\bbisimilar$ is the largest symmetric relation such that $\proc{P} \bbisimilar \proc{Q}$ satisfies: 

 \begin{itemize}

 \item for every name $n$: if $\proc{P} \outson n $ then $\proc{Q} \Outson n $, and if $\proc{P} \inson n $ then $\proc{Q} \Inson n $;

 \item for all $\proc{P}'$, if $\proc{P} \rtcredPi \proc{P}'$, then there exists $\proc{Q}'$ such that $\proc{Q} \rtcredPi \proc{Q}'$ and $\proc{P}' \bbisimilar \proc{Q}'$;


 \end{itemize}

 \item \emph{Weak-bisimilarity} is the largest relation $\wbisimilar$ defined by: $ \proc{P} \wbisimilar \proc{Q}$ if and only if $\Cont[ \proc{P}] \bbisimilar \Cont[\proc{Q}]$ for any context $\Cont[`.]$.

 \end{enumerate}
 \end{definition}

\Paper{
The following is easy to show.\footnote{This property was stated in \cite{Bakel-Vigliotti-IFIPTCS'12} using contextual equivalence rather than weak bisimilarity.}
 \begin{proposition} \label{reduction hidden}
Let $\proc{P}, \proc{Q}$ not contain $a$ and $a \not= b$, then
 \[ \begin{array}{rcl}
\New a ( \Out a <{\proc{p}}> . \proc{P} \Par \In a (x) . \proc{Q} )
	& \wbisim &
\proc{P} \Par \proc{Q}[\proc{p} \For x]
	\\ [1mm]
\New a ( \Bang \Out a <{\proc{p}}> . \proc{P} \Par \In a (x) . \proc{Q} )
	& \wbisim &
\proc{P} \Par \proc{Q}[\proc{p} \For x]
 \end{array} \]
 \end{proposition}
This expresses that synchronisation over hidden (internal) channels is unobservable.

The following property is needed in the proof of Theorem~\ref{soundness}.

 \begin{lemma} [\cite{Bakel-Vigliotti-JLC'14}] \label {replication lemma}
Let $x$ only be used as input channel in $\proc{P}$ and $\proc{Q}$, and not appear in \proc{R}, then:

 \[ \begin{array}{@{}rcl@{}}
\New x (\proc{P} \Par \proc{Q} \Par \Bang \Out x <w> . \proc{R})
	&\wbisimilar&
\New x (\proc{P} \Par \Bang \Out x <w> . \proc{R}) \Par \New x (\proc{Q} \Par \Bang \Out x <w> . \proc{R})
	\\
 \New x ( \Bang \proc{P} \Par \Bang \Out x <w> . \proc{R})
	&\wbisimilar&
\Bang \New x ( \proc{P} \Par \Bang \Out x <w> . \proc{R})
	\\
\New x ({ \In c (y) . \proc{P} \Par \Bang \Out x <w> . \proc{R} })
	&\wbisimilar&
\In c (y) . ({ \New x ({ \proc{P} \Par \Bang \Out x <w> . \proc{R} }) })
	\\
\New x ({ \Out c <y> . \proc{P} \Par \Bang \Out x <w> . \proc{R} })
	&\wbisimilar&
 \Out c <y> . ({ \New x ({ \proc{P} \Par \Bang \Out x <w> . \proc{R} }) })
 \end{array} \]
 \end{lemma}

 \begin{lemma} \label {replication lemma}
 \begin{enumerate}

 \firstitem \label{out distributes}
Assume that $a$ does not occur in $\proc{p}$, \proc{P}, and is only used for input in \proc{R} and \proc{Q}. Then:
 \[ \begin{array}{lcl}
\New a (\Bang \Out a <\procp> . \proc{P} \Par \proc{Q} \Par \proc{R} )
	& \wbisim &
{ \New a (\Bang \Out a <\procp> . \proc{P} \Par \proc{Q} ) \Par \New a (\Bang \Out a <\procp> . \proc{P} \Par \proc{R} ) }
 \end{array} \]


%

 \item \label{out splits}
Assume \proc{P}, \proc{Q} only output on $a$ and $a$ does not appear in \proc{R} or \proc{p}. Then:
 \[ \begin{array}{lcl}
\New a ( \Out a <\procp> . \proc{P} \Par \proc{Q} \Par \Bang \In a (x) . \proc{R} )
	& \wbisim & 
{ \New a ( \New b ({ \Out b <\procp> . \proc{P} \Par \proc{Q} \Par \Bang \In b (x) . \proc{R} }) \Par \Bang \In a (x) . \proc{R} ) }
 \end{array} \]

 \item \label{in distributes}
Assume that $a$ does not appear in \proc{P} and is only used for output in \proc{R} and \proc{Q}. Then:
 \[ \begin{array}{lcl}
\New a (\Bang \In a (x) . \proc{P} \Par \proc{Q} \Par \proc{R} )
	& \wbisim &
{ \New a (\Bang \In a (x) . \proc{P} \Par \proc{Q} ) \Par \New a (\Bang \In a (x) . \proc{P} \Par \proc{R} ) }
 \end{array} \]

 \end{enumerate}

 \end{lemma}

 \begin{Proof} Straightforward. \qed
 \end{Proof}

The following properties follow directly from the definition of $\wbisim$ and Proposition~\ref{reduction hidden}:

 \begin{proposition} \label{wbisim lemma}
 \begin{enumerate}

 \firstitem If for all $\proc{P}'$, $\proc{Q}'$ such that $ \proc{P} \rtcredPi \proc{P}' $, $ \proc{Q} \rtcredPi \proc{Q}' $ over hidden channels, we have $ \proc{P}' \wbisim \proc{Q}' $, then $ \proc{P} \wbisim \proc{Q} $.

 \item Let \proc{P}, \proc{Q} be such that no interaction is possible between them, then: $ \proc{P} \wbisim \proc{P}' $ and $ \proc{Q} \wbisim \proc{Q}'$ if and only if $ \proc{P} \Par \proc{Q} \wbisim \proc{P}' \Par \proc{Q}' $.


 \end{enumerate}
 \end{proposition}

The $`p$-calculus is equipped with a rich type theory \cite{Sangiorgi-Walker-Book'01}, from the basic type system for counting the arity of channels \cite{Pierce-Sangiorgi'96} to session types \cite{Honda'93} and sophisticated linear types in \cite{Honda-Yoshida-Berger'04}.
The notion of type assignment we use here is the one first defined in \cite{vBCV-CLaC'08} and differs from systems presented in the past in that types do not contain channel information, and in that it expresses \emph{implication}, \emph{i.e.}~has functional types and describes the `\emph{input-output interface}' of a process.

 \begin {definition}[Context assignment for $`p$ \cite{vBCV-CLaC'08}] \label{type assignment for pi}
Functional type assignment for the {\PiC} is defined by the following sequent system:\footnote{Type assignment is classical in nature (\emph{i.e.}~not intuitionistic), since we can have more than one conclusion.}
{\def\TurnPi{\Turn}
 \[ \begin{array}{rlrlrl}
(\Zero): &
 \Inf	{ \Pider {\Zero} : `G |- `D }
 &
(\InRule) : &
 \Inf	{ \Pider \proc{P} : `G, x{:}A |- x{:}A,`D }
	{ \Pider \In a(x) . \proc{P} : `G,a{:}A |- `D }
 &
~ (\OutRule) : &
 \Inf	[a \not= b]
	{ \Pider \proc{P} : `G,b{:}A |- b{:}A,`D }
	{ \Pider \Out a <b> . \proc{P} : `G,b{:}A |- a{:}A,b{:}A,`D }
 \\ [5mm]
(!): &
 \Inf	{ \Pider \proc{P} : `G |- `D }
	{ \Pider \Bang \proc{P} : `G |- `D }
 &
(`n): &
 \Inf	{ \Pider \proc{P} : `G,a{:}A |- a{:}A,`D }
	{ \Pider \New a \proc{P} : `G |- `D }
 &
(\Par): &
 \Inf	{ \Pider \proc{P} : `G |- `D \quad \Pider \proc{Q} : `G |- `D }
	{ \Pider \proc{P} \Par \proc{Q} : `G |- `D }
\\ [5mm]
&&(\Let) : &
 \multicolumn{3}{l}{
 \Inf	[y,z \notele `D; x \notele `G]
	{ \Pider \proc{P} : `G,y{:}B |- x{:}A,`D }
	{ \Pider \Let <x,y> = z in \proc{P} : `G,z{:}A\arr B |- `D }
}%
\\ [5mm]
&  \multicolumn{2}{r}{(\PairOut)} : &
 \multicolumn{3}{l}{
\Inf	[a \notele `D; a,c \notele `G]
	{ \Pider \proc{P} : `G,b{:}A |- c{:}B,`D }
	{ \Pider \Out a <b,c> . \proc{P} : `G,b{:}A |- a{:}A\arr B,c{:}B,`D }
}%
 \end{array} \] }

%
We write $\Pider \proc{P} : `G |- `D $ if there exists a derivation using these rules that has this expression in the conclusion.
 \end {definition}
We should perhaps stress that it is not known if this system has a relation with logic, other than the one established in \cite{vBCV-CLaC'08}.

Notice that the `\emph{input-output interface of a $`p$-process}' property is nicely preserved by all the rules; handling of arrow types is restricted by the type system to the rules $(\Let)$ and $(\PairOut)$.

 \begin {example}
The inference rules
 \[ \begin{array}{@{~}rlrl}
(\Weak): &
\Inf	[`G' \supseteq `G,`D' \supseteq `D]
	{ \Pider \proc{P} : `G |- `D }
	{ \Pider \proc{P} : `G' |- `D' }
 &
(\PairIn) : &
 \Inf	[y,a \not\in `D, \, x \not\in `G]
	{ \Pider \proc{P} : `G,y{:}B |- x{:}A,`D
	}{ \Pider \In a (x,y) . \proc{P} : `G,a{:}A\arr B |- `D }
 \\ [5mm]
(\OutRule\,') : &
\Inf	{ \Pider \Out a <b> : `G,b{:}A |- a{:}A,b{:}A,`D }
 &
(\PairOut\,') : &
\Inf	{ \Pider \Out a <b,c> : `G,b{:}A |- a{:}A\arr B,c{:}B,`D }
 \\ [5mm]
({\forwarder}) : &
\Inf	{ \Pider \BEq u=a : u{:}A |- a{:}A }
 &
(\PiOverline{`n}) : &
\Inf	{ \Pider \PilmuTerm [Q] w : `G |- w{:}B,`D
	}{ \Pider { \BOut b (w) . { \PilmuTerm [Q] w } } : `G |- b{:}B,`D }
 \end{array} \]
are admissible.
That weakening is admissible follows by a straightforward reasoning over the structure of derivations; for the other rules, consider:
 \[ \begin{array}{ccc}
\Inf	[\InRule]
	{\Inf	[\Let]
		{\InfBox{ \Pider \proc{P} : `G,y{:}B |- x{:}A,`D } }
		{ \Pider \Let <x,y> = z in \proc{P} : `G,z{:}A\arr B |- `D }
	}
	{ \Pider \In a (z) . \Let <x,y> = z in \proc{P} : `G,a{:}A\arr B |- `D }
 \end{array} \]
 \[ \begin{array}{ccc}
 \Inf	[\OutRule]
	{\Inf	[\Zero]
		{\Pider {\Zero} : `G,b{:}A |- b{:}A,`D }
	}{ \Pider \Out a <b> . \proc{P} : `G,b{:}A |- a{:}A,b{:}A,`D }
 &&
\Inf	[\PairOut]
	{\Inf	[\Zero]
		{ \Pider {\Zero} : `G,b{:}A |- c{:}B,`D }
	}{ \Pider \Out a <b,c> . {\Zero} : `G,b{:}A |- a{:}A\arr B,c{:}B,`D }
 \end{array} \]
 \[ \begin{array}{ccc}
\Inf	[\InRule]
	{\Inf	[\OutRule\,']
		{ \Pider \Out a <w> : `G,w{:}A |- a{:}A,w{:}A }
	}{ \Pider \In u (w) . \Out a <w> : `G,u{:}A |- a{:}A }
 &&
\Inf	[`n]
	{\Inf	[\OutRule]
		{\InfBox{ \Pider \PilmuTerm [Q] w : `G |- w{:}B,`D }
		}{ \Pider \Out b <w> . \PilmuTerm [Q] w : `G |- b{:}B,w{:}B,`D }
	}{ \Pider \New w ({ \Out b <w> . \PilmuTerm [Q] w }) : `G |- b{:}B,`D }
 \end{array} \]
 \end {example}


Since weakening is included, we allow ourselves to be a little less precise when we construct derivations, and freely switch to multiplicative style where rules join contexts whenever convenient, by using, for example, the rule
 \[ \begin{array}{@{}rl}
(\mid): &
 \Inf	{ \Pider \proc{P}_1 : \G_1 |- `D_1 \quad \dots \quad \Pider \proc{P}_n : \G_n |- `D_n }
	{ \Pider \proc{P}_1 \Par\dots \Par \proc{P}_n : \G_1, \ldots, \G_n |- `D_1, \ldots, `D_n }
 \\ [4mm]
 \end{array} \]

\Comment{
The main soundness result for our notion of type assignment for $`p$ is stated as:
 \begin {theorem}[Witness reduction \cite{vBCV-CLaC'08}] \label{Witness reduction}
If $\Pider \proc{P} : `G |- `D $ and $\proc{P} \redPi \proc{Q}$, then $\Pider \proc{Q} : `G |- `D $.
 \end {theorem}
}

 \section{Context and background of this paper} \label{context and background}

In the past, there have been several investigations of interpretation from the $`l$-calculus into the $`p$-calculus.
Research in this direction started by Milner's interpretation $\MilTerm [`.] `. $ of $`l$-terms \cite{Milner'92}; Milner's interpretation is input based and the interpretation of closed $`l$-terms respects large-step \emph{lazy} reduction $\redlazy$ \cite{Abramsky'90} to normal form up to substitution; this was later generalised to $`b$-equality, but using weak bisimilarity \cite{Sangiorgi-Walker-Book'01}.

For many years, it seemed that the first and final word on the interpretation of the $`l$-calculus has been said by Milner; in fact, input-based interpretations of the $`l$-calculus into the $`p$-calculus
have become the \emph{de facto} standard, and most published systems are based on Milner's interpretation.
The various interpretations studied in \cite{Sangiorgi-Walker-Book'01} constitute examples, also in the context of the higher-order $`p$-calculus; \cite{Honda-Yoshida-Berger'04} used Milner's approach with a typed version of the $`p$-calculus; \cite{Thielecke'97} used it in the context of continuation-passing style languages.

Milner's interpretation of the $`l$-calculus into the (synchronous, monadic) $`p$-calculus is defined by:

 \begin {definition}[Milner's interpretation \cite{Milner'92}] \label{Milner's interpretation}
The \emph{input-based interpretation} of $`l$-terms into the $`p$-calculus
It is defined by:
 \[ \begin{array}{rcll}
\MilSem [v x] a & \ByDef & \Milvar x a & (x \not= a)
 \\
\MilSem [l x . M] a & \ByDef & \MilTerm [l x . M] a & (b \textit{ fresh})
\\
\MilSem [a M N] a & \ByDef & \MilTerm [a M N] a & (c,z \textit{ fresh})
\\
\MilSem [x := M] & \ByDef & \MilSubX [M] x & (w \textit{ fresh})
 \end{array} \]
Milner calls $\MilSem [x := M] $ an ``\emph{environment entry}''; it could be omitted from the definition above, but is of use separately.
 \end {definition}

Notice that, in $\MilSem [a M N] a $, the interpretation of the operand $N$ is placed under output, and thereby blocked from running; this comes at a price: now $`b$-reductions that occur in the operand can no longer be mimicked.
Combined with using input to model abstraction, this makes that a redex can only be contracted if it occurs on the outside of a term (is the \emph{top} redex): the modelled reduction is \emph{lazy}.

Milner states an Operational Soundness result:
 \begin{theorem}[\cite{Milner'92}] \label{Milner's result}
For closed 
$M$, either $M \diverges$ ($M$ diverges) and $\MilSem[M] u \diverges$, or $M \redlazy `l y . R \Vect{[N \For x ]} $, and
 \[ \begin{array}{rcl}
\MilSem[M] u & \rtcredPi & (\Vect{`nx}) \, ( \MilSem[`ly.R] u \Par \Vect{\MilSem[x:=N ]} ).
 \end{array} \]
 \end{theorem}


Although obviously the intention is that the substitutions $\Vect{[N \For x ]}$ are generated by the lazy reduction, the way the result is stated this need not be the case.
This `glitch' was fixed in \cite{Bakel-Vigliotti-CONCUR'09} by reformulating Milner's result using \emph{explicit} substitution.
That paper also
%
presented a \emph{logical}, \emph{output-based} interpretation $\PiLSem [`.] `. $ that interprets abstraction $`lx.M$ not using \emph{input}, but via an asynchronous \emph{output} which leaves the interpretation of the body $M$ free to reduce.
That interpretation is defined as:

 \begin{definition} [Spine interpretation \cite{Bakel-Vigliotti-CONCUR'09}] \label{head interpretation} ~
\def\SPiHTerm{\PiSTerm}
 \[ \begin {array}{rcll}
\PiLSem [v x] a & \ByDef & \PiHTerm [v x] a
 \\
\PiLSem [l x . M] a & \ByDef & \PiHTerm [l x . M] a 
 \\
\PiLSem [a M N] a & \ByDef & \PiHTerm [a M N] a
 \\
\PiLSem [s M x := N] a & \ByDef & \PiHTerm [s M x := N] a
 \end {array} \]
 \end{definition}
For this interpretation, \cite{Bakel-Vigliotti-CONCUR'09} showed Operational Soundness and Type Preservation, but with respect to the notion of \emph{explicit head reduction} $\redxh$, similar to the notion defined in Def.~\ref{explicit head reduction}, and the notion of type assignment defined in Def.~\ref{type assignment for pi}.
The main results shown are (using $\Diverges$ to denote non-termination):
 \begin{theorem} [\cite{Bakel-Vigliotti-CONCUR'09}] \label{head reduction preservation}
 \begin{enumerate}

 \item If $M \Diverges $ then $\PiLSem[M] a \Diverges$, and if $M \redxh N $ then $ \PiLSem[M] a \rtcredPi\equivC \PiHTerm[N] a $.

 \item If $\der `G |- M : A $ then $ \derPi \PiLSem[M] a : `G |- a{:}A $.

 \end{enumerate}
 \end{theorem}

As argued in \cite{Bakel-Vigliotti-CONCUR'09}, to show the above result, which formulates a direct \emph{step-by-step} relation between $`b$-reduction and the synchronisation in the $`p$-calculus, it is necessary to make the substitution explicit.
This is a direct result of the fact that, in the $`p$-calculus, $`l$'s implicit substitution gets `implemented' \emph{one variable at the time}, rather than all in one fell swoop.
Since we aim to show a similar result for $\lmu$, we will therefore define a notion of explicit substitution.

Although termination is not studied in that paper, it is easily achieved through restricting the notion of reduction in the $`p$-calculus by not allowing reduction to take place inside processes whose output cannot be received, or by placing a guard on the replication as we do in this paper. \\

\noindent
A natural extension of this line of research is to see if the $`p$-calculus can be used to interpret more complex calculi as well, as for example calculi that relate not to intuitionistic logic, but to classical logic, as $\lmu$, $\lmmt$, or $\X$.
There are, to date, a number of papers on this topic.

In \cite{Honda-Yoshida-Berger'04} an interpretation of Call-by-Value $\lmu$ is defined that is based on Milner's.
Contrary to what is claimed in that paper, this interpretation itself is not based on Milner's \CBV-interpretation, but rather is a variant of $\MilTerm [`.] `. $; this implies that, \emph{a priori}, a process like $\sem{(`lx.x)((`lx.x)(`lx.x))} \, a $ cannot be reduced: the top redex should not be contractable since the interpretation should respect the \CBV-reduction, and the contraction of the right-hand redex cannot be modelled, since that term is placed under a guard.
The authors address this problem by considering \emph{typed processes only}, and using a much more liberal notion of reduction on processes. The syntax of processes there considered is
 \[ \begin{array}{rcl}
P &::=& \Bang \In x (\Vect{y}) . \proc{P} ~\mid~ \New \Vect{y} ( \Out x <{\Vect{y}}> \Par \proc{P} ) ~\mid~ \proc{P} \Par \proc{Q} ~\mid~ \New x \proc{P} ~\mid~ \Zero
 \end{array} \]
and the notion of reduction on processes is extended to that of $\searrow$, defined as the least compatible relation over typed processes (\emph{i.e.}~closed under typed contexts), taken modulo $\congruent$, that includes:
 \[ \begin{array}{rcl}
\Bang \In x (\Vect{y}) . \proc{P} \Par \New \Vect{a} ( \Out x <{\Vect{a}}> \Par \proc{Q} ) & \red & \Bang \In x (\Vect{y}) . \proc{P} \Par \New \Vect{a} ( \proc{P}[\Vect{a \For y}] \Par \proc{Q} )
 \end{array} \]
as the basic synchronisation rule, as well as
\[ \begin{array}{rcl}
\Cont[{ \New \Vect{a} ( \Out x <{\Vect{a}}> \Par \proc{P} ) }] \Par \Bang \In x (\Vect{y}) . \proc{Q} &\searrow_r&
\Cont[{ \New \Vect{a} ( \proc{P}[\Vect{a \For y}] \Par \proc{Q} ) }] \Par \Bang \In x (\Vect{y}) . \proc{Q} 
	\\
\New x ( \Bang \In x (\Vect{y}) . \proc{Q} ) & \searrow_g & \Zero \qquad
 \end{array} \]
where $\Cont[`.]$ is an arbitrary (typed) context; note that $\searrow$ synchronises with any occurrence of $\Out x <{\Vect{a}}> $, no matter what guards they may be placed under.
The resulting calculus is thereby very different from the original $`p$-calculus.
Types for processes prescribe usage of names, and name passing is restricted to \emph{bound (private) name passing}.\footnote{This is a feature of all related interpretations into the $`p$-calculus.}

On the relation between Girard's linear logic \cite{Girard'87} and the $`p$-calculus, 
\citet{Bellin-Scott'94} give a treatment of information flow in proof-nets; only a small fragment of Linear Logic was considered, and the translation between proofs and $`p$-calculus was left rather implicit as also noted by \citet{Caires-Pfenning'10}.

To illustrate this, notice that Bellin and Scott use the standard syntax for the polyadic $`p$-calculus
 \[ \begin{array}{rrl@{\quad}l}
\proc{P},\proc{Q} & ::= &
\Zero
\mid \proc{P} \Par \proc{Q}
\mid \Bang \proc{P}
\mid \New a \proc{P}
\mid \In a ({ \Vect{x} }) . \proc{P}
\mid \Out a < {\Vect{p} }> . \proc{P}
 \end{array} \]
similar to the one we use here (see Def.~\ref{pi calculus}) but for the fact that for us output is not synchronous, and there the $\Let$-construct is not used.
However, the encoding of a `cut' in linear logic
\[ \def \TurnLK{\Turn}
\Inf	{\derLK {} |- x{:}A\otimes B,y{:}{(A\otimes B)^\perp}
	 \quad
	 \Inf	{\derLK {} |- n{:}A,m{:}{A^\perp}
		 \quad
		 \derLK {} |- z{:}B,w{:}{B^\perp}
		}{\derLK {} |- m{:}A^\perp,w{:}B^\perp,v{:}A\otimes B }
	}{ \derLK {} |- x{:}A\otimes B,m{:}A^\perp,w{:}{B^\perp} }%
\]
\emph{i.e.}~the `term' $ x{:}A\otimes B,m{:}A^\perp, w{:}B^\perp$, gets translated into a `language of proofs' which looks like:
 \[
Cut^k(I,\bigotimes^{n,z}_ v(I,I)m w z )x,(m,w) = (\nu k) \big( I[k/y ] \mid \bigotimes^{n,z}_ v(I,I)m w z[k/v ] \big)
 \]
where the terms $Cut$ and $I$ are (rather loosely) defined in.
Notice the use of arbitrary application of processes to channel names, and the operation of pairing; the authors do not specify how to relate this notation to the above syntax of processes they consider.

However, even if this relationship is made explicit, also then a different $`p$-calculus is needed to make the encoding work.
To clarify this point, consider the translation in the $`p$-calculus of the term above, which according to the definition given in \cite{Bellin-Scott'94} becomes:
 \[
(\nu k) \big ( x(a).\underbrace{k(a)} \mid (\nu nz) (\underbrace{\Vect k(n,z).}\big( n(b).m(b) \mid z(b).w(b) \big) ) ) .
 \]
Although intended, no communication is possible in this term.
We have `underbraced' the desired communication which is impossible, as the arity of the channel $k$ does not match.
To overcome this kind of problem, Bellin and Scott would need the \Let-construct with use of pairs of names as we have introduced in this paper in Def.~\ref{pi calculus}.
Moreover, there is no relation between the interpreted terms and proofs stated in \cite{Bellin-Scott'94} in terms of logic, types, or provable statements; here, we make a clear link between interpreted proofs and the logic through our notion of type assignment for the $`p$-calculus.

In \cite{vBCV-CLaC'08} an interpretation into $`p$ of the sequent calculus $\X$ is defined that enjoys the Curry-Howard isomorphism for Gentzen's {\LK} \cite{Gentzen'35}, which is shown to respect reduction.
However, this result is only partial, as it is formulated as ``\emph{if $ P \redX Q $, then $\PiSem[P] \gtC \PiSem[Q] $}'', allowing $\PiSem[P] $ to have more observable behaviour than $\PiSem[Q] $; the main reason for this is that reduction in $\X$ is non-confluent.
Although in \cite{vBCV-CLaC'08} it is reasoned that this is natural in the context of non-confluent, symmetric sequent calculi, and is shown that the interpretation preserves types, it is a weaker result than could perhaps be expected.

An interpretation of $\lmmt$ is studied in \cite{CiminiCS'10}; the interpretation defined there strongly depends on recursion, is not compositional, and preserves only outermost reduction; no relation with types is shown.
}

\section{$\lmux$: $\lmu$ with explicit substitution}
\label{lmux section}

One of the main achievements of \cite{Bakel-Vigliotti-CONCUR'09} is that it establishes a strong link between reduction in the $`p$-calculus and step-by-step \emph{explicit substitution} \cite{Bloo-Rose'95} for the $`l$-calculus, by formulating a result 
with respect to explicit head reduction and the spine interpretation defined there. 

In view of this, for the purpose of our interpretation it was natural to study a variant of $\Lmu$ in \cite{Bakel-Vigliotti-IFIPTCS'12} with explicit substitution as well; since here we work with $\lmu$, we present here $\lmux$, as a variant of $\Lmux$ as presented in that paper.
Explicit substitution treats substitution as a first-class operator, both for the logical and the structural substitution, and describes all the necessary steps to effectuate both.
 \begin {definition}[$\lmux$] \label{definition lmux}

 \begin {enumerate}

 \firstitem
The syntax of the \emph{explicit $\lmu$ calculus}, $\lmux$, is defined by:
 \[ \begin {array}{@{}rrl}
M,N & ::= & x \mid `lx.M \mid MN \mid \Sub M x := N \mid
\muterm`a.[`b]M \mid \ContSub M `a := N . `g
 \end {array} \]
We consider the occurrences of $x$ in $M$ bound in $ \Sub M x := N $, and those of $`a$ in $M$ in $ \Sub M `a := N{`.}`g $; by Barendregt's convention, $x$ and $`a$ do not appear outside $M$.


 \item
The reduction relation $\redx$ on \Paper{terms in }$\lmux$ is defined \CLAC{as the contextual closure of}\Paper{through} the following rules\Paper{ (for the sake of completeness, we list all)}:

 \begin{enumerate} \itemsep 0pt

 \item Main reduction rules:
 \[ \begin {array}{rcl@{\quad}l}
 (`l x.M) N &\red & \Sub M x := N \\
(\muterm `a . {\Cmd} ) N & \red & \muterm`g . \Sub {\Cmd} `a := N{`.}`g & (`g \textit{ fresh}) \\
\muterm`b.[`b]M &\red& M & (`b \not\in \fn(M)) \\ {}
[`b]\muterm`g.\Cmd &\red& \Cmd[`b \For `g]
 \end {array} \]

 \item Term substitution rules:
 \[ \begin {array}{rcl@{\quad}l}
\Sub x x := N &\red & N \\
\Sub M x := N &\red & M & (x \not\in \fv(M)) \\
\Sub (`ly.M) x := N &\red & `l y.(\Sub M x := N ) \\
\Sub (PQ) x := N &\red & (\Sub P x := N )(\Sub Q x := N ) \\
\Sub (\muterm`a.[`b]M) x := N & \red & \muterm`a.[`b](\Sub M x := N)
 \end {array} \]

 \item Structural rules:
 \[ \begin {array}{rcl@{\quad}l}
\Sub (\muterm`d.\Cmd) `a := N{`.}`g & \red & \muterm`d.(\Sub {\Cmd} `a := N{`.}`g ) \\
\Sub ([`a]M) `a := N{`.}`g & \red & [`g](\Sub M `a := N{`.}`g ) N \\
\Sub ([`b]M) `a := N{`.}`g & \red & [`b](\Sub M `a := N{`.}`g ) & (`a \not= `b) \\
\Sub M `a := N{`.}`g & \red & M & (`a \not\in \fn(M)) \\
\Sub (`lx.M) `a := N{`.}`g & \red & `lx.\Sub M `a := N{`.}`g \\
\Sub (PQ) `a := N{`.}`g & \red & (\Sub P `a := N{`.}`g )( \Sub Q `a := N{`.}`g )
 \end {array} \]

\Paper{
 \item Contextual rules\CLAC{ ~ \vspace*{-10pt} }
 \[ \begin {array}[t]{rcl}
M \red N &~\Then ~~ &
 \begin{cases}
`lx.M &\red& `lx.N \\
ML &\red& NL \\
LM &\red& LN \\
\muterm`a.[`b]M & \red & \muterm`a.[`b]N \\
\Sub M x := L &\red& \Sub N x := L \\
\Sub L x := M &\red& \Sub L x := N \\
\Sub M `a := L{`.}`g &\red& \Sub N `a := L{`.}`g \\
\Sub L `a := M{`.}`g &\red& \Sub L `a := N{`.}`g \\ [1mm]
 \end{cases}
 \end {array} \]
}

 \end {enumerate}

 \item \label{redxsub definition}
We use $\redxsub$ for the notion of reduction where only term substitution and structural rules are used (so not the main reduction rules)\Paper{, and $\eqx$ for the congruence generated by $\redx$}.

 \end {enumerate}

 \end{definition}

\Paper{
This is a system different from that of \cite{Audebaud'94}, where a version with explicit substitution is defined for a variant of $\lmu$ that uses de Bruijn indices \cite{deBruijn'72}.
}
Notice that since reduction in $\lmux$ \Paper{actually }is formulated via term rewriting rules \cite{Klop'92}, reduction is allowed to take place also inside the substitution term.
\Paper{

Explicit substitution describes explicitly the process of executing a $`b`m$-reduction, \emph{i.e.}~expresses syntactically the details of the computation as a succession of atomic steps (like in a first-order rewriting system), where the implicit substitution of each $`b`m$-reduction step is split up into reduction steps.
Thereby t}\CLAC{T}he following is straightforward:

 \begin{proposition} [$\lmux$ implements $\lmu$-reduction]
\label{lmu vs lmux}
 \begin{enumerate}
 \firstitem $ M \redbmu N \Implies M \rtcredx N $.
 \item $M \ele \lmu \And M \redx N \Implies \Exists L \ele \lmu \Pred [ N \rtcredxsub L ] $.
 \end{enumerate}
 \end{proposition}

\Paper{
The notion of type assignment on $\lmux$ is a natural extension of the system for the $\lmu$-calculus of Def.~\ref{typing for lmu} by adding rules $(\TCut)$ and $(\CCut)$.
 \begin {definition}
Using the notion of types in Def.~\ref{types}, type assignment for $\lmux$ is defined by:
 \[ \begin {array}{@{}r@{~~}l@{\dquad}r@{~~}l}
(\Ax) : &
\Inf	{ \derLmu `G,x{:}A |- x : A | `D }
&
 (`m) : &
 \Inf	{ \derLmu `G |- M : B | `a{:}A,`D
	}{ \derLmu `G |- \muterm`a.[`b]M : A | `b{:}B,`D }
\quad
 \Inf	{ \derLmu `G |- M : A | `a{:}A,`D
	}{ \derLmu `G |- \muterm`a.[`a]M : A | `D }
\\ [5mm]
(\arrI) : &
\Inf	{ \derLmu `G,x{:}A |- M : B | `D
	}{ \derLmu `G |- `l x.M : A\arrow B | `D }
&
(\arrE) : &
\Inf	{ \derLmu `G |- M : A\arrow B | `D
	 \quad
	 \derLmu `G |- N : A | `D
	}{ \derLmu `G |- MN : B | `D }
 \end{array} \]
 \[ \begin {array}{@{}r@{~~}l@{\dquad}r@{~~}l}
(\TCut): &
\Inf	{ \derLmu `G,x{:}A |- M : B | `D \quad \derLmu `G |- N : A | `D }
	{ \derLmu `G |- {\Sub M x := N } : B | `D }		
& 
(\CCut): &
\Inf	{ \derLmu `G |- M : C | `a{:}A\arr B,`D \quad \derLmu `G |- N : A | `D }
	{ \derLmu `G |- {\Sub M `a := N{`.}`g } : C | `g{:}B,`D }
 \end {array} \]
We write $\derLmux `G |- M : A $ for judgements derivable in this system.

 \end {definition}
}

\Comment{
 \begin{lemma}
If $M \redx N$, and $\derLmu `G |- M : A $ then $\derLmu `G |- N : A $.
 \end{lemma}

 \begin{Proof}
{\Red To do.}

}

In the context of head reduction and explicit substitution, we can economise further on how substitution is executed, and perform only those that are essential for the continuation of reduction.
We will therefore limit substitution to allow it to \emph{only replace} the head variable of a term.
(This principle is also found in Krivine's machine \cite{Krivine'07}.)\,
The results of \cite{Bakel-Vigliotti-CONCUR'09} show that this is exactly the kind of reduction that the $`p$-calculus naturally encodes.

%
 \begin{definition}[Explicit head reduction]
 \label{explicit head reduction}

The 
\emph{head variable} of $M$, $\hv(M)$, is defined as expected, adding $ hv(\Sub M x := N ) = \hv(M) $ if $ \hv(M) \not= x $, and the \emph{head name} $\hn(M)$ is defined by $ \hn(\muterm`a.[`b] \lmuHNF) = `b $, $ \hn(\Sub M x := N ) = \hn(M) $, and $ \hn(\ContSub M `a := N . `g ) = \hn(M) $ if $ \hn(M) \not= `a $.

We define \emph{explicit head reduction} $\redxh$ on $\lmux$ as $\redx$, but change and add a few rules
{ (we only give the changes)}:

\Comment{
 \begin{enumerate}
 \item \Paper{Main reduction rules: 
 \[ \begin {array}{rcl@{\quad}l}
 (`l x.M) N &\red & \Sub M x := N & \\
(\muterm `a . M ) N & \red & \muterm`g . \Sub M `a := N{`.}`g & (`g \textit{ fresh}) \\
\muterm`a.[`a]M &\red& M & (`a \not\in \fn(M)) \\
\muterm`a.[`b]\muterm`g.\Cmd &\red& \muterm`a.\Cmd[`b \For `g]
 \end {array} \]

 \item \label{term rules} }
\label{changed rules}
Term substitution rules: 
 \Paper{ \[ }\CLAC{$} \begin {array}{rcl@{\quad}l}
\Paper{\Sub x x := N &\red & N \\
\Sub M x := N &\red & M & (x \not\in \fv(M)) \\
\Sub (`ly.M) x := N &\red & `l y.(\Sub M x := N ) & (x = \hv(M)) \\ }
\Sub (PQ) x := N &\red & \Sub {( \Sub P x := N \,Q )} x := N & (x = \hv(P)) \Paper{ \\
\Sub (\muterm`a.[`b]M) x := N & \red & \muterm`a.[`b](\Sub M x := N) & (x = \hv(M)) \\ }
 \end {array} \Paper{ \] }\CLAC{$}

 \item\label{structural rules}
There are only two structural rules: 
 \[ \begin {array}{rcl@{\quad}l}
\Sub (\muterm`b.[`a]M) `a := N{`.}`g & \red & \muterm`b.[`g](\ContSub M `a := N . `g ) N & (`a = \hn(\muterm`b.[`a]M)) \\
\Sub M `a := N{`.}`g & \red & M & (`a \not\in \fn(M)) \\
 \end {array} \]

 \item
We remove the following contextual rules:
 \[ \begin {array}[t]{rcl}
M \red N &\Then&
 \begin{cases}
 LM &\red& LN \\
 \Sub L x := M &\red& \Sub L x := N \\
 \Sub L `a := M{`.}`g &\red& \Sub L `a := N{`.}`g \\
 \end{cases}
 \end {array} \]
so no longer allow reduction inside the substitution or inside the right-hand side of an application.

 \item \label{substitution rules} \label{new rules}
We add two substitution rules:
 \[ \kern-3mm \begin {array}{@{}rcll}
\Sub {\Sub M x := N } y := P &\red&
	{ \Sub { \Sub { \Sub M y := P } x := N } y := P }
	& (y = \hv(M))
	\\
\Sub {\Sub M `a := N{`.}`g } `b := L{`.}`d &\red&
	{ 
	\Sub { \Sub { \Sub M `b := L{`.}`d } `a := N{`.}`g } `b := L{`.}`d }
	& (`b = \hn(M))
 \end {array} \]

 \end {enumerate}
}

 \begin{enumerate}
 \item \label{changed rules}  
%
We replace the term substitution rule for application and add side-conditions:%
 \[ \begin {array}{rcl@{\quad}l}
\Sub (`ly.M) x := N &\red & `l y.(\Sub M x := N ) & (x = \hv(`ly.M)) \\
\Sub (PQ) x := N &\red & \Sub {( \Sub P x := N \,Q )} x := N & (x = \hv(PQ)) \\
\Sub (\muterm`a.[`b]M) x := N & \red & \muterm`a.[`b](\Sub M x := N) & (x = \hv(\muterm`a.[`b]M))
 \end {array} \]

 \item\label{structural rules}
There are only two structural rules: 
 \[ \begin {array}{rcl@{\quad}l}
\Sub (\muterm`b.[`a]M) `a := N{`.}`g & \red & \muterm`b.[`g](\ContSub M `a := N . `g ) N 
\\
\Sub M `a := N{`.}`g & \red & M & (`a \not\in \fn(M)) \\
 \end {array} \]

 \item
We remove the following contextual rules:
 \[ \begin {array}[t]{rcl}
M \red N &\Then&
 \begin{cases}
 LM &\red& LN \\
 \Sub L x := M &\red& \Sub L x := N \\
 \Sub L `a := M{`.}`g &\red& \Sub L `a := N{`.}`g \\
 \end{cases}
 \end {array} \]

 \item \label{substitution rules} \label{new rules}
We add four substitution rules:
 \[ \kern-3mm \begin {array}{@{}rcll}
\Sub {\Sub M x := N } y := L &\red& \Sub { \Sub { \Sub M y := L } x := N } y := L
	& (y = \hv(M)) \\
\Sub {\ContSub M `a := N . `b } y := L &\red& \Sub { \ContSub { \Sub M y := L } `a := N . `b } y := L
	& (y = \hv(M)) \\
\ContSub {\ContSub M `a := N . `g } `b := L . `d &\red& \ContSub { \ContSub { \ContSub M `b := L . `d } `a := N . `g } `b := L . `d
	& (`b = \hn(M)) \\
\ContSub {\Sub M x := N } `b := L . `d &\red& \ContSub { \Sub { \ContSub M `b := L . `d } x := N } `b := L . `d
	& (`b = \hn(M))
 \end {array} \]

 \end {enumerate}
 \end{definition}
Notice that, for example, in case\CLAC{~}\ref{changed rules}, \Comment{the fourth of }the clause\Comment{s} postpones the substitution $ \exsub x := N $ on $Q$ until such time that an occurrence of the variable $x$ in $Q$ becomes the head-variable of the full term, and that we no longer allow reduction inside the substitution or inside the right-hand side of an application.

\Paper{
Notice that we do not add rules like
 $ \Sub { \Sub M x := N } y := L 
	\red
\Sub { \Sub M y := L } x := {\Sub N y := L }
$;
as in \cite{Bloo-Rose'95}, this would introduce undesired non-termination.

Remark that we need to add the rules of case~\ref{new rules}: if we take the term $\Sub { \Sub P x := Q } y := R $, under `normal' reduction $\redx$, the innermost substitution has to complete first before the outermost can run.
When moving towards explicit head reduction, this would mean that we cannot reduce, for example, $\Sub { \Sub yx x := Q } y := R $; the innermost substitution cannot advance, since $x$ is not the head variable, and the outermost cannot advance since it is not defined for the substitution term it has to work on, $\Sub yx x := Q $ in this case.
To allow outermost substitution to `jump' the innermost, we need to add the extra rules.
This, potentially, introduces non-termination, but by demanding that the variable concerned is actually the head-variable of the term, we avoid this.

Normal forms of explicit head reduction are naturally defined as follows:
 \begin{definition} [cf.~\cite{Lassen'06}]
The normal forms with respect to $\redxh$ are defined through the grammar:
 \[ \begin {array}{@{}rrll}
\lmuNF & ::= &
`lx. \lmuNF
	\\ &\mid &
xM_1\dots M_n & (n \geq 0)
	\\ &\mid &
\muterm `a . [`b] \lmuNF & (`a \not= `b \Or `a \in \lmuNF', \lmuNF \not= \muterm`g.[`d] \lmuNF' )
	\\ &\mid &
\Sub {\lmuNF} x := M & (\hv(\lmuNF) \not= x)
	\\ &\mid &
\ContSub {\lmuNF} `a := M . `g & (\hn(\lmuNF) \not= `a)
 \end {array} \]
 \end{definition}
}


The following proposition states the relation between explicit head reduction, head reduction, and explicit reduction.
 \begin{proposition} \label{explicit versus head}
 \begin{enumerate}

 \firstitem
If $ M \rtcredh N $, then there exists $L \ele \lmux$ such that $ M \rtcredxh L $ and $ L \rtcredxsub N $.


 \item
If $ M \rtcredxh[\nf] N $ with $M \ele \lmu$, then there exists $ L \ele \lmu$ such that $ N \rtcredxsub[\nf] L $, and $ M \rtcredh[\nf] L $.

 \item
$M \rtcredbmu[\nf] N $ if and only if there exists $L \ele \lmux$ such that $ M \rtcredxh[\nf] L $ and $ L \rtcredx[\nf] N $.

 \end{enumerate}
 \end{proposition}
This result gives that we can show our main results for $\lmux$ for reductions that reduce to \HNF.

\Paper{
We will give some examples that illustrate $\lmux$ and $\redxh$.
 \begin{example} \label{redxh examples} \label{reductions example}
 \begin{enumerate} \itemsep1mm

 \item
$ 
\begin{array}[t]{lclcl} 
(`lx.xx)(`ly.y)
	& \redxh &
\Sub xx x := `ly.y
	& \redxh & \\
\Sub {(\Sub x x := `ly.y x)} x := `ly.y
	& \redxh &
\Sub {}(`ly.y)x x := `ly.y
	& \redxh & \\
\Sub {\Sub y y := x } x := `ly.y
	 & \redxh &
\Sub x x := `ly.y
	& \redxh & \\
\Sub `ly.y x := `ly.y
	& \redxh &
`ly.y
 \end{array} $

 \item
Reduction in $\redxh$ is not deterministic; notice that the term $\LTerm<S ({a ({l x . {a y x}}) N}) y := P>$ can reduce in two ways:
 \[ \begin{array}{rcccl}
\LTerm<S {a ({l x . {a y x}}) N} y := P>
	&\redxh&
 \left \{
\begin{array}{l}
\LTerm<S {S {a y x} x := N} y := P> 
	\\
\LTerm<S {a ({l x . {a P x}}) N} y := P> 
 \end{array}
\right \}
	& \redxh &
\LTerm<S {S {a P x} x := N} y := P>
 \end{array} \]
(the last step is not the only one possible).

 \item \label{mu reduction}
$ \begin{array}[t]{@{}lclcl}
(\muterm`a.[`b]\muterm`d.[`a](`ly.y))(`lz.z)
	& \redxh & 
\Sub \muterm`g.[`b]\muterm`d.[`a](`ly.y) `a := `lz.z{`.}`g
	& \redxh & \\
\Sub \muterm`g.[`a](`ly.y)[`b\For `d] `a := `lz.z{`.}`g
	& = & 
\Sub \muterm`g.[`a](`ly.y) `a := `lz.z{`.}`g
	& \redxh & \\
\muterm`g.[`g](`ly.y)(`lz.z)
	& \redxh & 
\muterm`g.[`g]\Sub y y := `lz.z
	& \redxh & \\
\muterm`g.[`g]`lz.z
	& \redxh &
`lz.z
 \end{array} $

 \item
Some reductions leave substitutions in place:

\[ \begin{array}[t]{@{}lcl}
`lf.(`lx.f(xx))(`lx.f(xx))
	& \redxh &
`lf.(\Sub f(xx) x := `lx.f(xx) )
 \end{array} \]

and the last term is in $\redxh$-normal form.

\Comment{
 \item
 $ \begin{array}[t]{@{}lclcl}
\muterm`a.[`a](`lq.q)(\muterm`b.[`a]`ly.y)
	& \redxh &
\muterm`a.[`a]\Sub q q := \muterm`b.[`a]`ly.y
	& \redxh & \quad \\
\muterm`a.[`a]\muterm`b.[`a]`ly.y
	& \redxh &
\LTerm<u `a . `a {l y . y}>
 \end{array} $
}

 \item \label{non-terminating reduction}
Of course in $\redxh$ we can have non-terminating reductions.
We know that in $\redbmu$ and $\redh$, $ (`lx.xx) (`lx.xx) $ reduces to itself; this is not the case for $\redxh$, as is illustrated by:
 \[ \begin{array}[t]{@{}lclcl}
`D`D ~=~ (`lx.xx) (`lx.xx)
	& \redxh &
\Sub xx x := `D
	& \rtcredxh & \\
\Sub {{}(\Sub x x := `D x)} x := `D
	& \redxh 
	&
\Sub {{}(`ly.yy) x} x := `D
	& \redxh & \\
\Sub {\Sub yy y := x } x := `D
	& \rtcredxh &
\Sub {\Sub {(\Sub y y := x y)} y := x } x := `D
	& \redxh & \\
\Sub {\Sub xy y := x } x := `D
	& \redxh &
\Sub {\Sub {\Sub xy x := `D } y := x } x := `D
	& \redxh & \\
\Sub {\Sub {\Sub {(\Sub x x := `D y)} x := `D } y := x } x := `D
	& \rtcredxh 
	&
\Sub {\Sub {{}(`lz.zz)y} y := x } x := `D
	& \rtcredxh & \\
\Sub {\Sub {\Sub {{}(`lw.ww)z} z := y } y := x } x := `D
	& \rtcredxh &
			\dots
 \end{array} \]
(notice the $`a$-conversions, needed to adhere to Barendregt's convention).
This reduction is deterministic and clearly loops.
Notice that $`D`D$ does not run to itself; however,
 \[ \begin{array}[t]{@{}lclcl}
\Sub {\Sub xy y := x } x := `D
	& \rtcredxsub & 
\Sub xx x := `D
	& \rtcredxsub & `D`D
 \end{array} \]
so, as stated by Proposition~\ref{explicit versus head}, the standard reduction result can be achieved by reduction in $\redxsub$ (we will use $`D$ again below).

 \end{enumerate}
 \end{example}

 \Comment{
 \begin{example} \label{reductions example}
To illustrate the 
notions of reduction, take $ M = (`lf.(`lx.f(xx))(`lx.f(xx)))(`lab.a) $, and $N = `lxb.xx $; then
 \begin{itemize}

 \item $\begin{array}[t]{@{}lclclclclc}
M &\rtcredbmu& NN
	&=& (`lxb.xx)N
	&\rtcredbmu& `lb.NN
	&\rtcredbmu& `lbb'.NN & \dots
 \end{array} $

 \item $\begin{array}[t]{@{}lclclclclc}
M &\rtcredh& NN
	&=& (`lxb.xx)N
	&\rtcredh& `lb.NN
	&\rtcredh& `lbb'.NN & \dots
 \end{array} $

 \item $\begin{array}[t]{@{}lclc}
M &\rtcredxh& \Sub (`lx.f(xx))((`lx.f(xx))) f := `lab.a \\
	&\rtcredxh& \Sub {\Sub f(yy) y := `lx.f(xx) } f := `lab.a \\
	&\rtcredxh& \Sub {\Sub (`lab.a)(yy) y := `lx.f(xx) } f := `lab.a \\
	&\rtcredxh& \Sub {\Sub {\Sub `lb.a a := yy } y := `lx.f(xx) } f := `lab.a \\
	&\rtcredxh& \Sub {\Sub `lb.yy y := `lx.f(xx) } f := `lab.a \\
	&\rtcredxh& \Sub {\Sub `lb.(`lz.f(zz))y y := `lx.f(xx) } f := `lab.a \\
	&\rtcredxh& \Sub {\Sub { \Sub `lb.f(zz) z := y } y := `lx.f(xx) } f := `lab.a \\
	&\rtcredxh& \Sub {\Sub { \Sub `lb.(`lab'.a)(zz) z := y } y := `lx.f(xx) } f := `lab.a \\
	&\rtcredxh& \Sub {\Sub { \Sub {\Sub `lbb'.a a := zz } z := y } y := `lx.f(xx) } f := `lab.a \\
	&\rtcredxh& \Sub {\Sub { \Sub `lbb'.zz z := y } y := `lx.f(xx) } f := `lab.a
	& \dots
 \end{array} $

 \end{itemize}
 \end{example}
 }

}

 \section{A logical interpretation of \texorpdfstring{$\lmux$}{}-terms to \texorpdfstring{$`p$}{}-processes}
\label{lambda interpretation}

We will now define our logical,\footnote{It is called \emph{logical} because it has its foundation in the relation between natural deduction and Gentzen's sequent calculus\CLAC{.}\Paper{ \LK; in particular, the case for application is based on the representation of \emph{modus ponens}
 \[ \def \TurnLK {\Turn} \begin{array}[t]{@{}c@{\quad}c@{\quad}c}
\Inf{ \derLK `G |- A\Arr B \quad \derLK `G |- A }{ \derLK `G |- B }
& \textrm{ by } &
 \Inf{ \derLK `G |- A\Arr B,`D \quad \Inf { \derLK `G |- A,`D \quad \derLK `G,B |- `D }{ \derLK `G,A\Arr B |- `D }
}{ \derLK `G |- B,`D }
 \end{array} \]}}
output-based interpretation $\PilmuTerm[M] a $ of the $\lmux$-calculus into the $`p$-calculus\Paper{. }\CLAC{ (where $M$ is a $\lmu$-term, and $a$ is the name given to its (anonymous) output), which is essentially the one presented in \cite{Bakel-Vigliotti-IFIPTCS'12}, but no longer considers $[`a]M$ to be a term.
The reason for this change is the following: using the interpretation of \cite{Bakel-Vigliotti-IFIPTCS'12},
 \[ \begin{array}{rcl}
\SPilmuTerm[\muterm`a.`lx.x] a
	&=&
\PilmuTerm[U `a . {l x . {v x}}] a
 \end{array} \]
is in normal form, and all inputs and outputs are restricted; thereby, it is weakly bisimilar to $\Zero$ and to $\SPilmuTerm[a {(l x . xx)} {(l x . xx)}] a $.
So using that interpretation, we cannot distinguish between \emph{blocked} and \emph{looping} computations, which clearly affects any full-abstraction result.
When restricting our interpretation to $\lmu$, this problem disappears: since naming has to follow $`m$-abstraction, $\muterm`a.`lx.x$ is not a term in $\lmu$.
Since $\lmu$ is a subcalculus of $\Lmu$, this change clearly does not affect the results shown in \cite{Bakel-Vigliotti-IFIPTCS'12} that all hold for the interpretation we consider here as well.

}
The main idea behind the interpretation, as in \cite{Bakel-Vigliotti-CONCUR'09}, is to give a name to the anonymous output of terms; it combines this with the inherent naming mechanism of $\lmu$.
\Paper{As we will show in Thm.~\ref{soundness},}\CLAC{As shown in \cite{Bakel-Vigliotti-IFIPTCS'12},} this encoding naturally represents explicit head reduction; we will need to consider weak reduction later for the full abstraction result, but not for soundness, completeness, or termination.

 \begin{definition}[Logical interpretation\Paper{ of $\lmux$ terms}\CLAC{ \cite{Bakel-Vigliotti-IFIPTCS'12}}] \label{lmu interpretation}
The interpretation of $\lmux$ terms into the $`p$-calculus is defined by:
 \[ \begin {array}{rcl@{\quad}l}
\SPilmuTerm [x] a & \ByDef & \PilmuTerm [v x] a & (u \textit{ fresh})
	\\
\SPilmuTerm [`lx.M] a & \ByDef & \PilmuTerm [l x . M] a & (b \textit{ fresh})
	\\
\SPilmuTerm [MN] a & \ByDef & \PilmuTerm [a M N] a & (c,v,d \textit{ fresh})
	\\
\SPilmuTerm [\Sub M x := N ] a & \ByDef & \PilmuTerm [S M x := N] a
	\\
\SPilmuTerm [x := N] {\textcolor{white}{a}} & \ByDef & \PiExsub x := N & (w \textit{ fresh})
	\\
\SPilmuTerm [\muterm`g.\Cmd] a & \ByDef & \PilmuTerm [m `g . \Cmd] a & (\Sink \textit{ fresh})
	\\
\SPilmuTerm [{[`b]M}] a & \ByDef & \PilmuTerm [n `b M] a
	\\
\SPilmuTerm [\Sub M `b := N{`.}`g ] a & \ByDef & \PilmuTerm [C M `b := N . `g] a
	\\
\SPilmuTerm [`a := M{`.}`g] {\textcolor{white}{a}} & \ByDef & \PiExContsub `a := N . `g & (v,d \textit{ fresh})
 \end {array} \]
 \end{definition}

 \Comment{
As we will show below, this interpretation improves on the results of \cite{Bakel-Vigliotti-CONCUR'09} in that we encode not only explicit head reduction for the $`l$-calculus, but also explicit head reduction for the $\lmu$-calculus.

In \cite{Bakel-Vigliotti-IFIPTCS'12}, this interpretation was defined for $\Lmu$, where naming and $`m$-abstraction are separated; in fact, there the definition above contained:
 \[ \begin {array}{rcll}
\SPilmuTerm [\muterm`g.\Cmd] a & \ByDef & \PilmuTerm [m `g . \Cmd] a , & \Sink \textit{ fresh}
	\\
\SPilmuTerm [[`b]M] a & \ByDef & \PilmuTerm [n `b M] a
 \end {array} \]}
\Paper{
Note that the this encoding very elegantly expresses that the main computation in $\muterm`g.\Cmd$ is blocked: the name $\Sink$ is fresh and bound, so the main output of $\PilmuTerm [m `g . \Cmd] a $ cannot be received.
Moreover, n}%
\CLAC{\noindent N}otice that
 \CLAC{$ \begin{array}{rclclcl}
\SPilmuTerm [\muterm`g.[`b]M] a
	&\ByDef&
\PilmuTerm [m `g . {{}[`b]M}] a
	&\ByDef&
\PilmuTerm [m `g . n `b M] a 
	& \congruent &
\PilmuTerm [u `g . `b M] a
 \end{array} $}%
\Paper{ \[ \begin{array}{rclclcl}
\SPilmuTerm [\muterm`g.[`b]M] a
	&\ByDef&
\PilmuTerm [m `g . {{}[`b]M}] a
	&\ByDef&
\PilmuTerm [m `g . n `b M] a 
	& \congruent &
\PilmuTerm [u `g . `b M] a
 \end{array} \]}
%
which implies that we can add $ \SPilmuTerm [\muterm`g.[`b]M] a
	\ByDef
\PilmuTerm [u `g . `b M] a $ to our encoding.

\Paper{
Note that we could have avoided the implicit renaming in the case for $`m$-abstraction by defining:
 $ \SPilmuTerm [\muterm`g.\Cmd] a \ByDef \New {\Sink} ({ \PilmuTerm[\Cmd] {\Sink} \Par \BEq `g=a }) $
which is operationally the same as $ \PilmuTerm [m `g . \Cmd] a $ (they are, in fact, weakly bisimilar) but then we could not show that terms in {\HNF} are translated to processes in normal form (Lem.~\ref{head pi normal form}), a property that is needed in the proof of termination.

There is a strong relation between this interpretation and the abstract machine defined in \cite{Crolard'99}, but for the fact that that only represents lazy reduction.

As in \cite{Bakel-Vigliotti-JLC'14}, we can make the following observations:

 \begin{remark} \label{natural observations}
 \begin{itemize}

 \item
The synchronisations generated by the encoding only involve processes of the shape:%
 \[ \begin{array}{c@{\dquad}c@{\dquad}c}
\Picaps<x,`a> & \Out `b <x,`a> & \In z (`b,y) . (\proc{P} \Par \proc{Q})
 \end{array} \]
so in particular, substitution is always well defined.
These synchronisations are of the shape:
 \[ \begin{array}{rcl}
\New c ({\New y`b ({\proc{P} \Par \Out c <y,b> }) \Par \In c (v,d) . ({ \proc{R} \Par \In d (w) . \Out a <w> }) })
	&\redPi&
\New y`b ({\proc{P} \Par \proc{R}[y\For v] \Par \In b (w) . \Out a <w> })
 \end{array} \]
and after the synchronisation over $c$, $\proc{P}$ can receive over $y$ from $\proc{R}[y\For `a]$ and send over $b$ to $\In b (w) . \Out a <w> $; or of the shape
 \[ \begin{array}[t]{@{}lcl@{}}
\New c ({\New yb ({\proc{P} \Par \Out c <y,b> }) \Par \Picaps<c,a> })
	&\redPi&
\New yb ({\proc{P} \Par \Out a <y,b> })
 \end{array} \]

 \item
All synchronisation takes place \emph{only} over channels whose names are bound connectors in the terms that are interpreted.

 \item
To underline the significance of our results, notice that the encoding is not trivial, since%
 \[ \begin{array}{rcl}
\SPilmuTerm [`lyz.y] a & = & \PilmuTerm [l y . {l z . {v y}}] a
\\
\SPilmuTerm [`lx.x] a &=& \PilmuTerm [l x . {v x}] a
 \end{array} \]
processes that differ under $\wbisimilar$.

 \end{itemize}
 \end{remark}

Notice that the context switches do not really influence the structure of the process that is created by the interpretation since they have no representation in $`p$, but are statically encoded through renaming. 
This strengthens our view that, as far as our interpretation is concerned, $`m$-reduction is not a separate computational step, but is essentially static administration, a reorganisation of the applicative structure of a term, which is dealt with by our interpretation statically rather than by synchronisation between processes.
In other words, the point of view from the $`p$-calculus is that modelling $`b$-reduction involves a computational step, but context switches are dealt with by congruence; this is only possible, of course, because the interpretation of the operand in application uses replication.

In \cite{Bakel-Vigliotti-CONCUR'09} the case for application in the interpretation for $`l$-terms was defined as:
 \[ \begin {array}{rcll}
\PiLSem [a M N] a & \ByDef & \PiHTerm [a M N] a
 \end {array} \]
In particular, there the input on name $c$ is \emph{not replicated}: this corresponds to the fact that for $`l$-terms, in $\PiHTerm[M] c $, the output $c$ is used \emph{exactly once}, which is not the case for the interpretation of $\lmu$-terms: for example, $`a$ might appear many times in $M$, and since $\SPilmuTerm[u `a . `a M] a = \PilmuTerm[u `a . `a M] a = \PilmuTerm[{M[a/`a]}] a $, then the name $a$ appears many times in the latter.

We would like to stress that, although inspired by logic, our interpretation does not depend on types \emph{at all}; in fact, we can treat untypeable terms as well, and can show that $ \SPilmuTerm [{}(`lx.xx)(`lx.xx)] a $ (perhaps the prototype of a non-typeable term) runs forever without generating output (see Example~\ref{DD is a zero process}
; this already holds for the interpretation of \cite{Bakel-Vigliotti-CONCUR'09}).

Notice that, as is the case for Milner's interpretation and in contrast to the interpretation of \cite{Bakel-Vigliotti-CONCUR'09}, a guard is placed on the replicated terms.
This is not only done with an eye on proving preservation of termination, but more importantly, to make sure that $ \New x ( \PiExSub x := N ) \wbisim \Zero $, a property we need for our full abstraction result: since a term can have named sub-terms, the interpretation will generate output not only for the term itself, but also for those named terms, so $ \New x ( \PilmuTerm[N] x ) $ -- as used in \cite{Bakel-Vigliotti-CONCUR'09} -- \emph{can have} observable behaviour, in contrast to here, where $ \New x ( \PiExSub x := N ) = \New x ( \PiExsub x := N ) $ \emph{is} weakly bisimilar to $\Zero$.
}

Observe the similarity between
 \[ \begin{array}{rcll}
\Paper{\hspace*{13mm} }\SPilmuTerm[a M N] a &\ByDef& \PilmuTerm[a M N] a & \textrm{and} \\
\SPilmuTerm [\Sub M c := N{`.}`g ] a &\ByDef&
\PilmuTerm [C M c := N . `g ] a \\ &\ByDef&
\PilmuTerm [c M c := N . `g ] a
 \end{array} \]
The first communicates $N$ via the output channel $c$ of $M$\CLAC{ (which might occur more than once inside $\PilmuTerm[M] c $, so replication is needed)}, whereas the second communicates with all the sub-terms that have $c$ as output name, and changes the output name of the process to $`g$.\Paper{\footnote{A similar observation can be made for the interpretation of $\lmu$ in $\X$ \cite{Bakel-Lescanne-MSCS'08}.}}
In other words, application is just a special case of explicit structural substitution;
this allows us to write $ \PilmuTerm[A M N] a 
$ for $\SPilmuTerm[a M N] a $.
\Paper{This very elegantly expresses exactly what the structural substitution does: it `connects' arguments with the correct position in a term.}
This stresses that the $`p$-calculus constitutes a very powerful abstract machine indeed: although the notion of structural reduction in $\lmu$ is very different from normal $`b$-reduction, no special measures had to be taken in order to be able to express it; the component of our interpretation that deals with pure $`l$-terms is almost exactly that of \cite{Bakel-Vigliotti-CONCUR'09} (ignoring for the moment that substitution is modelled using a guard, which affects also the interpretation of variables), but for the use of replication in the case for application.

\Paper{
The operation of \emph{renaming} we will use below is defined and justified via the following lemma, which states that we can safely rename the output of an interpreted $\lmu$-term.
First we need to show:

 \begin{proposition} [\cite{Bakel-Vigliotti-JLC'14}] \label{aux-renaming lemma}
 \[ \begin{array}{rcl}
\New xb ({ \In c(v,d) . ( \proc{P} \Par \BEq d=e ) }) &\wbisim& \CLAC{ \\ }
\New a ({ \BEq a=e \Par \New xb ({ \In c(v,d) . ( \proc{P} \Par \BEq d=a ) }) })
 \end{array} \]

 \end{proposition}

We use this result to show the following:

 \begin{lemma} [Renaming lemma] \label{renaming lemma}
 \begin{enumerate}

 \firstitem \label{renaming outside}
Let $ a \not\in \fv(M)$, then
 $ \begin{array}[t]{@{}lcl@{}}
\New a ( \BEq a=e \Par \SPilmuTerm [M] a ) & \wbisim & \SPilmuTerm [M] e .
 \end{array} $

 \item \label{renaming inside}
 $ \begin{array}[t]{@{}lcl@{\quad}l@{}}
\New a ( \BEq a=e \Par \SPilmuTerm [M] b ) & \wbisim & \SPilmuTerm [{M[e \For a]}] b & (b \not= a)
 \end{array} $.

 \end{enumerate}
 \end{lemma}

 \begin{Proof}
By induction on the structure of $\lmux$-terms.

 \begin{description} \itemsep 4pt

 \item [$ M = x $]
$ \begin{array}[t]{@{}lclclcl}
\New a ( \BEq a=e \Par \SPilmuTerm [v x] a )
	& \ByDef & 
\New a ( \BEq a=e \Par \PilmuTerm [v x] a )
	& \wbisim &
\PilmuTerm [v x] e
	& \ByDef & 
\SPilmuTerm [v x] e
 \end{array} $

 \myitem [$ M = `lx.N $]
\New a ( \BEq a=e \Par \SPilmuTerm [l x . N] a )
	& \ByDef & 
\New a ( \BEq a=e \Par \PilmuTerm [l x . N] a )
	& \redPi (a) \\
\New axb ( \BEq a=e \Par \PilmuTerm [N] b \Par \Out e <x,b> )
	& \wbisim & (a \not= x \Then a \notele \fv(N)) \\ 
\New a ( \BEq a=e ) \Par \PilmuTerm [l x . N] e
	& \wbisim &
\SPilmuTerm [l x . N] e
 \end{array} $

 \myitem [$ M = PQ $]
\New a ( \BEq a=e \Par \SPilmuTerm [a P Q] a )
	& \ByDef \\
\New a ( \BEq a=e \Par \PilmuTerm [X P Q] a ) 
	& \wbisim & (\Ref{aux-renaming lemma}, a \not=c, a \notele \fv(P,Q)) \\
\PilmuTerm [X P Q] e
	& \ByDef & 
\SPilmuTerm [PQ] e
 \end{array} $

 \myitem [$ M = \Sub P x := Q $]
\New a ( \BEq a=e \Par \SPilmuTerm [s P x := Q] a )
	& \ByDef & \\
\New a ( \BEq a=e \Par \PilmuTerm [s P x := Q] a )
	& \wbisim & (\IH\Ref{renaming outside}, x \not= a, a \notele \fv(P,Q)) & \\
\PilmuTerm [s P x := Q] e
	& \ByDef & 
\SPilmuTerm [s P x := Q] e
 \end{array} $

 \myitem [{$ M = \muterm`b. [`g] N $}]
\New a ( \BEq a=e \Par \SPilmuTerm [{\muterm`b. [`g] N}] a )
	\kern-2cm &&
	& \ByDef & \\
\New a ( \BEq a=e \Par \PilmuTerm [u `b . `g N] a )
	& \wbisim & \multicolumn{3}{l}{ (\IH\Ref{renaming inside}, a \not=`g, a \notele \fv(N)) } \\
\PilmuTerm [u `b . `g N] a \, [e \For a]
	& \wbisim & 
\PilmuTerm [u `b . `g N] e
	& \ByDef &
\SPilmuTerm [u `b . `g N] e
 \end{array} $

 \myitem [$ M = \Sub P `b := Q{`.}`g $]
\New a ( \BEq a=e \Par \SPilmuTerm [\Sub P `b := Q{`.}`g ] a )
	& \ByDef & \\
\New a ( \BEq a=e \Par \PilmuTerm [C P `b := Q . `g] a )
	& \wbisim & (\IH\Ref{renaming outside}, a\not= `b, a \notele \fv(P,Q)) \\
\PilmuTerm [C {P[e \For a]} `b := {Q[e \For a]} . `g] e
	& \ByDef & \\
\PilmuTerm [C P `b := Q . `g] e
	& \ByDef & 
\SPilmuTerm [C P `b := Q . `g] e \qed
 \end{array} $

 \end{description}
 \end{Proof}

For reasons of clarity, we use some auxiliary notions of equivalence, that are used in Thm~\ref{soundness}.

 \begin{definition} \label{sbired definition}
 \begin{enumerate}

 \firstitem
We define a \emph{garbage collection} bisimilarity by: $ \proc{P} \redGC \proc{Q} $ 
if and only if there exists $\proc{R}$ such that $ \proc{P} = \proc{Q} \Par \proc{R} $ and $ \proc{R} \wbisimilar \Zero $.

We call a process that is weakly bisimilar to $\Zero$ \emph{garbage}.

 \item
We define $\redR$ as the largest, symmetric equivalence such that:

 \begin{enumerate}

 \item for all $\proc{P}$ such that the name $a$ is the only free output of $\proc{P}$ and is only used to output and only once: $ \New a ( \proc{P} \Par \Eq a=e ) \redR \proc{P}[e \For a]$,

 \item for all contexts $\Cont$, if $ \proc{P} \redR \proc{Q} $ then $ \Cont [\proc{P}] \redR \Cont [\proc{Q}] $.

 \end{enumerate}

We will use $\redR$ if we want to emphasise that two processes are equivalent just using renaming, \emph{i.e.}~define: $ \New a ( \proc{P} \Par \Eq a=e ) \redR \proc{P}[e \For a]$.
 \item
We define $ \conredPi $ as $ \lpired $.

 \end{enumerate}
 \end{definition}
Notice that ${\redGC} \subset {\wbisimilar}$ and ${\redR} \subset {\wbisimilar}$.


Using this lemma, we can show the following:
 \begin{example} \label{redex example}
The interpretation of the $`b$-redex $(`lx.P)Q$ reduces as follows:
 \[ \def\arraystretch{1.1} \begin{array}[t]{@{}l\CLAC{@{~}}c@{~}l}
\SPilmuTerm [a {(l x . P )} Q] a
	& \ByDef & \\
\PilmuTerm [a {l x . P} Q] a
	& \redPi & (c) \\ 
\New c ( \New bx ({ \PilmuTerm[P] b \Par \BEq b=a \Par \PilmuTerm[x := Q] {} }) \Par
\PiExContsub c := Q . a )
	& \congruent & (c \notele \fn(P,Q)) \\
\New bx ({ \PilmuTerm[P] b \Par \BEq b=a \Par \PiExSub x := Q }) \Par \New c ( \PiExContSub c := Q . a )
	& \wbisimG & (\ast) \\
\New bx ({ \PilmuTerm[P] b \Par \BEq b=a \Par \PiExSub x := Q })
	& \wbisimR & (\Ref{renaming lemma}) \\
\PilmuTerm[S P x := Q] a
	& \ByDef &
\SPilmuTerm[s P x := Q] a
 \end{array} \]

This shows that $`b$-reduction is implemented in $`p$ by at least one synchronisation.
Notice that, in step $(\ast)$, the process
 $ \New c ( \PiExContSub c := Q . a ) \ByDef \New c ( \PiExContsub c := Q . a ) $
is weakly bisimilar to $\Zero$.

\Comment{
Renaming is not always needed; for a concrete example, take
 \[ \begin{array}[t]{@{}l@{~}c@{~}l}
\SPilmuTerm [a {(l x . x )} (`ly.y)] a
	& \ByDef & \\
 \PilmuTerm [a {l x . x} `ly.y] a
	& \redPi & (c) \\ 
\New bx ({ \PilmuTerm[x] b \Par \BEq b=a \Par \PilmuTerm[x := `ly.y ] {} }) \Par {}
\New c ( \PiExContsub c := `ly.y . a )
	& \wbisimG & \\
\New bx ( \PilmuTerm[x] b \Par \BEq b=a \Par \PilmuTerm[x := `ly.y] {} )
	& \ByDef & \\
\New bx ( \PilmuTerm[v x] b \Par \BEq b=a \Par \PiExsub x := `ly.y )
	& \redPi & (x) \\
\New bw ({ \BEq w=b \Par \BEq b=a \Par \PilmuTerm[`ly.y ] w \Par \New x ( \PiExsub x := `ly.y ) })
	& \wbisimG & \\
\New bw ( \BEq w=b \Par \BEq b=a \Par \PilmuTerm[`ly.y ] w )
	& \ByDef & \\
\New bw ( \BEq w=b \Par \BEq b=a \Par \PilmuTerm[l y . y] w )
	& \redPi & (w) \\
\New bw ( \BEq w=b \Par \BEq b=a ) \Par \PilmuTerm[l y . y] a
	& \wbisimG & 
\SPilmuTerm[l y . y] a
 \end{array} \]
which shows that renaming is not needed here; however, we need it for cases like
 \[ \begin{array}[t]{@{}l@{~}c@{~}l}
\SPilmuTerm [a {(l x . x )} y] a
	& \ByDef & \\
 \PilmuTerm [a {l x . x} y] a
	& \redPi & (c) \\ 
\New bx ({ \PilmuTerm[x] b \Par \BEq b=a \Par \PilmuTerm[x := y ] {} }) \Par 
\New c ( \PiExContsub c := y . a )
	& \wbisimG & \\
\New bx ( \PilmuTerm[x] b \Par \BEq b=a \Par \PilmuTerm[x := y] {} )
	& \ByDef & \\
\New bx ( \PilmuTerm[v x] b \Par \BEq b=a \Par \PiExsub x := y )
	& \redPi & (x) \\
\New bw ({ \BEq w=b \Par \BEq b=a \Par \PilmuTerm[v y] w \Par \New x ( \PiExsub x := y ) })
	& \wbisimG & \\
\New bw ( \BEq w=b \Par \BEq b=a \Par \PilmuTerm[v y] w )
	& \wbisimR & (\Ref{renaming lemma}) \\
\PilmuTerm[v y] a
	& \ByDef &
\SPilmuTerm[y] a
 \end{array} \]

On the other hand, as mentioned above, $`m$-reduction consists of a reorganisation of the structure of a term by changing its applicative structure.
Since application is modelled through parallel composition, this implies that the interpretation of a $`m$-redex is essentially dealt with by congruence and renaming.
For example,
 \[ \begin{array}[t]{@{}lll}
\SPilmuTerm [a {(u `b . `b P)} Q] a
	& \,\ByDef & \\
\PilmuTerm [a {(u `b . `b P)} Q] a
	& =_{`a} & \\
\New `b ({ \PilmuTerm [P] `b \Par \PiExContsub `b := Q . a })
 \end{array} \]
We can show, using Lem.~\ref{replication lemma}, this last process is weakly bisimilar to
 \[ \begin{array}[t]{@{}lllll}
\New `g ( \New `b ({ \PilmuTerm [P] `g \Par \PiExContSub `b := Q . a }) \Par
\PiExContSub `g := Q . a )
	& \ByDef &
\SPilmuTerm [a {c P `b := Q . a} Q] a
 \end{array} \]
(notice that
we have separated out the outside name of the term $P$, being $`b$, which we renamed to $`g$; this leaves two context substitutions, one dealing with the occurrences of $`b$ inside $P$, and one with $`g$.\footnote{This corresponds to the behaviour of rule $(\Riii)$ in $\X$.})
This illustrates that structural substitution is dealt with by structural congruence and $`a$-conversion.
}

 \end{example}
This example stresses that all synchronisations in the image of $\PilmuTerm[`.] `. $ are over hidden channels, so by Proposition~\ref{reduction hidden} are in $\wbisim$.

We can also show that typeability is preserved: first, we need to prove a substitution lemma.

 \begin{proposition}[Substitution] \label{substitution lemma}
If $\Pider \proc{P} : `G,x{:}A |- x{:}A,`D $, then $\Pider \proc{P}[b/x] : `G,b{:}A |- b{:}A,`D $ for $b$ fresh or $b{:}A \ele `G \union `D$.
 \end{proposition}

 \begin{Proof} Easy.
\qed
 \end{Proof}

 \begin {theorem} [{$\SPilmuTerm [`.] `. $} preserves $\lmux$ types]
 \label{typeability is preserved}
If $\derLmux `G |- M : A | `D $, then $\Pider \SPilmuTerm [M] a : `G |- a{:}A,`D $.
 \end {theorem}

 \begin{Proof}
By induction on the structure of derivations in $\TurnLmux$:

 \begin {description} \itemsep3mm

 \item [$(\Ax)$]
Then $M= x$, and $ `G = `G ', x{:}A$.
Notice that $\SPiHTerm[v x] a = \PilmuTerm [v x] a $, and that
 \[ \begin {array}{@{}c}
 \Inf	[\InRule]
	{\Inf	[\Bang]
		{\Comment{%
		\Inf	[\InRule]
			{\Inf	[\OutRule]
				{\Inf	[\Zero]
					{\Pider {\Zero} : `G',w{:}A |- w{:}A }
				}{ \Pider \Out a <w> : `G',w{:}A |- a{:}A,w{:}A }
			}{ \Pider \PiHTerm[v u] a : `G',u{:}A |- a{:}A }}
		 \Inf	[\forwarder]
			{ \Pider \Eq u=a : `G',u{:}A |- a{:}A }	
		}{ \Pider \Bang \PiHTerm[v u] a : `G',u{:}A |- a{:}A }
	}{ \Pider \PilmuTerm[v x] a : `G',x{:}A |- a{:}A }
 \end {array} \]

 \item [$(\arrI)$]
Then $M = `lx.N$, $A = C \arr D$, and $\derLmux `G,x{:}C |- N : D | `D $; by definition, $\SPilmuTerm[l x . N ] a = \PilmuTerm [l x . N ] a $.
Then, by induction, $ \D \dcol \Pider \SPilmuTerm [N] b : `G,x{:}C |- b{:}D,`D $ exists, and we can construct:
 \[ \def\TurnPi{\Turn}
 \Inf	[`n]
	{\Inf	[`n]
		{\Inf	[\mid]
			{ \InfBox{\D}{ \Pider \SPilmuTerm [N] b : `G,x{:}C |- b{:}D,`D }
			 \quad
			 \Inf	[\PairOut\,']
				{ \Pider \Out a <x,b> : x{:}C |- a{:}C\arr D,b{:}D }
			}{ \Pider \SPilmuTerm [N] b \Par \Out a <x,b> : x{:}C |- a{:}C\arr D,b{:}D,`D }
		}{ \Pider {\New b ( \SPilmuTerm [N] b \Par \Out a <x,b> )} : `G,x{:}C |- a{:}C\arr D,`D }
	}{ \Pider {\PilmuTerm[l x . N ] a } : `G |- a{:}C\arr D,`D }
 \]
Notice that $\PilmuTerm[l x . N ] a = \SPilmuTerm [`lx . N ] a $.

 \item [$(`m)$]
Then $M = \muterm `a.[`b]N$, and either:

 \begin{description}

 \item[$`a \not= `b $]
Then $\derLmux `G |- N : A | `a{:}A,`b{:}B,`D $.
By induction, there exist a derivation for $ \Pider \SPilmuTerm [N] `b : `G |- `a{:}A,`b{:}B,`D $; then by Lem.~\ref{substitution lemma}, also $ \Pider \PilmuTerm [u `a . `b N] a : `G |- a{:}A,`b{:}B,`D $ as well, and $\PilmuTerm [u `a . `b N] a = \SPilmuTerm [u `a . `b N] a $.

 \item[$`a = `b $]
Then $\derLmux `G |- N : A | `a{:}A,`D $.
By induction, there exist a derivation for $ \Pider \SPilmuTerm [N] a : `G |- a{:}A,`a{:}A,`D $; then by Lem.~\ref{substitution lemma}, also $ \Pider \PilmuTerm [u `a . `a N] a : `G |- a{:}A,`D $ as well, and $\PilmuTerm [u `a . `a N] a = \SPilmuTerm [u `a . `a N] a $.

 \end{description}

 \item [$(\CCut)$]
Then $M = \ContSub P `a := Q . `g $ and we have $\derLmux `G |- P : C | `a{:}A\arr B,`D $ and $\derLmux `G |- Q : A | `g{:}B,`D $ for some $B$.
By induction, there exist derivations $ \D_1 \dcol \Pider { \SPilmuTerm [P] a } : `G |- a{:}C,`a{:}A\arr B,`D $ and, since $a$ is fresh,
$ \D_2 \dcol \Pider \SPilmuTerm [Q] w : `G |- w{:}B,`D $, and we can construct
the derivation
 \[ \def\TurnPi{\Turn}
\kern-1mm \begin{array}{@{}c}
\Inf	[`n]
	{\Inf	[\Par]
		{\InfBox{\D_1}{ \Pider \PilmuTerm [P] a : `G |- a{:}B\arr A,`D }
		 \Inf	[!]
			{\Inf	[\PairIn]
				{\Inf	[\mid]
					{\Inf	[!]
						{\Inf	[\PiOverline{`n}]
								{\InfBox{\D_2}{ \Pider \PilmuTerm [Q] w : `G |- w{:}B,`D }
								}{ 
								\Pider \BOut b (w) . { \PilmuTerm [Q] w } : `G |- b{:}B,w{:}B,`D }
						}{ \Pider \Bang { 
							\BOut b (w) . { \PilmuTerm [Q] w } } : `G |- b{:}B,`D }
					 \Inf	[!]
						{\Comment{
						 \Inf	[\InRule]
							{\Inf	[\OutRule]
								{\Inf	[\Zero]
									{\Pider {\Zero} : w{:}A |- w{:}A }
								}{ \Pider \Out `g <w> : w{:}A |- `g{:}A,w{:}A }
							}{ \Pider \Eq d=`g : d{:}A |- `g{:}A }}
						 \Inf	[\forwarder]
							{ \Pider \Eq d=`g : d{:}A |- `g{:}A }
						}{ \Pider \BEq d=`g : d{:}A |- `g{:}A }
					}{ \Pider \PiExSub b := Q \Par \BEq d=`g : `G,d{:}A |- `g{:}A,b{:}B,`D }
				}{ \Pider \In `a (b,d) . ( \PiExSub b := Q \Par \BEq d=`g ) : `G,`a{:}B\arr A |- `g{:}A,`D }
			}{ \Pider \Bang \In `a (b,d) . ( \PiExSub b := Q \Par \BEq d=`g ) : `G,`a{:}B\arr A |- `g{:}A,`D }
		}{ \Pider \PilmuTerm [P] a \Par \Bang \In `a (b,d) . ( \PiExSub b := Q \Par \BEq d=`g ) : `G,`a{:}B\arr A |- `g{:}A,`D }
	}{ \Pider \PilmuTerm [c P `a := Q . `g] a : `G |- `g{:}A,`D }
 \end{array} \]
and $ \PilmuTerm [c P `a := Q . `g] a = \SPilmuTerm[c P `a := Q . `g] a = \PilmuTerm [C P `a := Q . `g] a $.

 \item [\CLAC{$\arrE)$, $(\TCut$}\Paper{$(\arrE)$, $(\TCut)$}]
Notice that since $ \SPilmuTerm[a P Q] a = \PilmuTerm[A P Q] a $ and $\SPilmuTerm[\Sub M x := N ] a = \PilmuTerm[S M x := N ] a $, these cases are very similar to that for $(\CCut)$.
\qed
 \end {description}
 \end{Proof}

We can show a witness reduction result for our encoding, for which we need the property:

 \begin{lemma}[Contraction \cite{Bakel-Vigliotti-JLC'14}] \label{Pi witness reduction}
 \begin{enumerate}

 \firstitem If $ \Pider \New bc ( \proc{P} \Par \Out a <b,c> ) \Par \In a (x) . \Out e <x> : `G,a{:}C |- a{:}C,`D $, $a$ does not occur in $\proc{P}$, and $a \not= e$, then $ \Pider \New bc ( \proc{P} \Par \Out e <b,c> ) : `G |- `D $.

 \item If $a$ does not occur in $\proc{P}$ and $\proc{Q}$ and $ \Pider \New bc ( \proc{P} \Par \Out a <b,c> ) \Par \In a (x,y) . \proc{Q} : `G,a{:}C |- a{:}C,`D $, then \\ $ \Pider \New bc ( \proc{P} \Par \proc{Q} [b \For x,c \For y] ) : `G |- `D $.

 \end{enumerate}
 \end{lemma}

Using this result, we can also show a witness reduction result:

 \begin{theorem}
If $ \Pider \PilmuTerm[P] a : `G |- `D $, and $ \PilmuTerm[P] a \rtcredPi \proc{Q} $, then $ \Pider \proc{Q} : `G |- `D $.
 \end{theorem}

 \begin{Proof}
By Remark~\ref{natural observations}, Theorem~\ref{typeability is preserved}, and Lemma~\ref{Pi witness reduction}.
\qed
 \end{Proof}

Notice that, as for Milner's and Sangiorgi's interpretations, ours is not extensional, since
$ \PilmuTerm[`D`D] a \wbisim \Zero $, but not $ \PilmuTerm[`lx.`D`Dx] a \wbisim \Zero $ (see Lem.~\ref{zero lemma} and \ref{bmu reduction characterisation}).

 \begin{remark} \label{not extensional}
We could not have represented the extensional rules: note that
 \[ \begin{array}{@{}lcl}
\PilmuTerm [`lx.yx] a
	& \ByDef & 
\PilmuTerm [l x . {A y x}] a
 \end{array} \]
does not reduce to $\PilmuTerm[y] a $, and neither does
 \[ \begin{array}{@{}lclcl}
\PilmuTerm [{`m`a . [`b] y}] a
	& \ByDef & 
\PilmuTerm [u `a . `b y] a
	& = & 
\PilmuTerm [y] `b
 \end{array} \]
reduce to:
 \[ \begin{array}{@{}lclclcl}
\New xb ( \PilmuTerm[y] `b \Par \Out a <x,b> )
	& = &
\PilmuTerm [l x . {u `g . `b y}] a
	& \ByDef & \\
\PilmuTerm [l x . {u `g . `b y}] a
	& \ByDef &
\PilmuTerm [l x . {`m`g . {[`b]} y }] a
	& \ByDef & \\
\PilmuTerm [{`lx.`m`g . [`b] y }] a
	& = &
\PilmuTerm [{`lx.`m`g . [`b] y [x{`.}`g \For `a]}] a
 \end{array} \]
 \end{remark}

\Comment{ \Double }

To illustrate the expressiveness and elegance of our interpretation, we give some examples:
 \begin{example} \label{our example}
 \begin{enumerate} \itemsep 1mm

 \firstitem
The interpretation of the fixed-point combinator runs as follows:

\[ \begin{array}[t]{lcl}
\PilmuTerm [`lf.(`lx.f(xx))(`ly.f(yy))] a
	& \ByDef & \\
\PilmuTerm [l f . (`lx.f(xx))(`ly.f(yy))] a
	& \ByDef & \\
\PilmuTerm [l f . {a `lx.f(xx) `ly.f(yy)}] a \quad
	& \ByDef & \\
\PilmuTerm [l f . {d {l x . {{}f(xx)}} `ly.f(yy)}] a
	& \redPi(c),\wbisimG & \\
\PilmuTerm [l f . {S {}f(xx) x := `ly.f(yy)}] a
	& \ByDef & \\
\PilmuTerm [l f . {t 
		{a {v f} {(xx)}} x := `ly.f(yy)}] a
	& \ByDef & \\
\SPilmuTerm [l {}f . {t {a {}f {(xx)}} x := `ly.f(yy)}] a
 \end{array} \]

\noindent
This last process is in normal form, \emph{i.e.}~cannot be reduced further: since the only `reachable inputs' are $c$ and $f$, and the only `reachable output' is $x$, which is different from both $c$ and $f$, no synchronisation is possible.
Moreover, this is not a lazy reduction.

\Comment{
 \item
In Fig.~\ref{figure double} we run the interpretation of the term $(`lx.x)(\muterm`a.[`a](`lq.q)(\muterm`b.[`a]`ly.y))$
from Example~\ref{redxh examples}
$\PilmuTerm[{}(`lx.x)(\muterm`a.[`a](`lq.q)(\muterm`b.[`a]`ly.y))] a $,
 \[ \begin{array}{@{}lcl}
(`lx.x)(\muterm`a.[`a](`lq.q)(\muterm`b.[`a]`ly.y)) & \red \\
\sub x x := \muterm`a.[`a](`lq.q)(\muterm`b.[`a]`ly.y) & \red \\
\muterm`a.[`a](`lq.q)(\muterm`b.[`a]`ly.y) & \red \\
\muterm`a.[`a](\sub q q := \muterm`b.[`a]`ly.y ) & \red \\
\muterm`a.[`a]\muterm`b.[`a]`ly.y & \red \\
\muterm`a.[`a]`ly.y
 \end{array} \]
as an example of a term that generates two outputs over $`a$, and highlights the need for the repeated use of replication; notice that the individual reduction steps return (translated) in that figure.

Notice that
 \[ \begin{array}{@{}lcl}
(`lx.x)(\muterm`a.[`a](`lq.q)(\muterm`b.[`a]`ly.y))
	& \redxh & \\
\Sub x x := \muterm`a.[`a](`lq.q)(\muterm`b.[`a]`ly.y)
	& \redxh & \\
\muterm`a.[`a](`lq.q)(\muterm`b.[`a]`ly.y)
	& \redxh & \\
\muterm`a.[`a]\Sub q q := \muterm`b.[`a]`ly.y
	& \redxh & \quad \\
 \multicolumn{3}{l}{
\muterm`a.[`a]\muterm`b.[`a]`ly.y
	\hfill \redxh
\LTerm<u `b . `a {l y . y}>
 }
 \end{array} \]

 \item
 $ \begin{array}[t]{@{}lclcl}
\PilmuTerm[\Sub M x := N L ] a & = \\
\PilmuTerm[A {S M x := N } L ] a & \congruent \\
\PilmuTerm[S {a M L} x := N ] a & = \\
\PilmuTerm[\Sub ML x := N ] a
 \end{array} $
}

 \item
When running the reduction of Example \ref{redxh examples}\ref{mu reduction}, the naming features are taken care of by the interpretation:
 \[ \begin{array}{lcl}
\PilmuTerm[{}(\muterm`a.[`b]\muterm`d.[`a](`ly.y))(`lz.z)] a
	& \ByDef & \\
\PilmuTerm[a {u `a . `b \muterm`d.[`a](`ly.y)} `lz.z] a 
	& = & \\
\PilmuTerm[a {\muterm`d.[c](`ly.y)} `lz.z] a
	& \ByDef & \\
\PilmuTerm[a {u `d . c `ly.y} `lz.z] a
	& = & \\
\PilmuTerm[a `ly.y `lz.z] a
	& \ByDef & \\ 
\SPilmuTerm[a (`ly.y) (`lz.z)] a
	& \ByDef & \\
\PilmuTerm[a {l y . y} `lz.z] a
	& \redPi (c) & \\
\New yb ({ \PilmuTerm[y] b \Par \PiExSub y := `lz.z \Par \BEq b=a })
	& \wbisimilar & (\Ref{renaming lemma}) \\
\PilmuTerm[S y y := `lz.z] a
	& \ByDef & \\
\PilmuTerm[f {v y} y := {l z . z}] a
	& \red (y) \\
\New w ( \BEq w=a \Par \PilmuTerm[l z . z] w ) \Par \New y ( \PiExSub y := `lz.z )
	& \red (w) \\
\PilmuTerm[l z . z] w \Par \New w ( \BEq w=a ) \Par \New y ( \PiExSub y := `lz.z )
	& \wbisim &
\PilmuTerm[`lz.z] a
 \end{array} \]

\Comment{
 \[ \begin{array}{@{}lcl}
\SPilmuTerm[(\muterm`a.[`b]\muterm`d.[`a](`ly.y))(`lz.z)] a
	& \ByDef & \\
\PilmuTerm[A (\muterm`a.[`b]\muterm`d.[`a](`ly.y)) {`lz.z}] a
	& \ByDef & \\
\PilmuTerm[a {u `a . `b {\muterm`d.[`a](`ly.y)}} {`lz.z}] a
	& \ByDef & \\
\PilmuTerm[A {n `b {\muterm`d.[c](`ly.y)}} {`lz.z}] a
	& \ByDef & \\
\PilmuTerm[A {n `b {u `d . c (`ly.y)}} {`lz.z}] a
	& = & \\
\PilmuTerm[A {`ly.y} {`lz.z}] a
	& \ByDef & \\
\PilmuTerm[a {l y . y} {`lz.z}] a
	\quad & \redPi & (c) \\
\New yb ( \PilmuTerm[y] b \Par \PiExSub y := `lz.z \Par \BEq b=a ) \Par
\New c ( \PiExContSub c := {`lz.z} . a )
	& \wbisimG & \\
\New yb ( \PilmuTerm[y] b \Par \PiExSub y := `lz.z \Par \BEq b=a )
	& \ByDef & \\
\New yb ( \PilmuTerm[v y] b \Par \PiExsub y := `lz.z \Par \BEq b=a )
	& \redPi & (y) \\
\New bw ( \BEq w=b \Par \PilmuTerm[`lz.z] w \Par \BEq b=a ) \Par \New y ( \PiExSub y := `lz.z )
	& \wbisimG & \\
\setcounter{indb}{1}	
 \New bw ( \BEq w=b \Par {} \PilmuTerm[l z . z] w \Par \BEq b=a )
	& \redPi & (w,b) \\
\PilmuTerm[l z . z] a \Par {} \New bw ( \BEq w=b \Par \BEq b=a )
	& \wbisimG &
\SPilmuTerm[l z . z] a
 \end{array} \]
}

 \item
%
$ \begin{array}[t]{@{}lcl}
\SPilmuTerm [a {a P Q} R] a
	& \ByDef,\congruent & 
\NewA cc' ( \PilmuTerm [P] c' \Par \PiExContsub c' := Q . c \Par
\PiExContsub c := R . a )
 \end{array} $

\noindent
so components of applications are placed in parallel under the interpretation. 
%
Similarly, we have
 $ 
\SPilmuTerm [C {C M `a := Q . `b } `g := L . `d ] a
= 
\New `g`a ( \PilmuTerm [M] a \Par \PiExContSub `a := Q . `b \Par \PiExContSub `g := L . `d )
 $, so repeated structural substitutions are also placed in parallel under the interpretation and can be applied independently.

 \end{enumerate}
 \end{example}

 \section{Soundness, completeness, and termination}
 \label{Soundness, completeness, and termination}

}

We can now show a reduction-preservation result for explicit head reduction for $\lmux$, by showing that $\PilmuTerm [`.] {`.} $ preserves $\redxh$ up to weak bisimularity, stated using $\equivC$ in \cite{Bakel-Vigliotti-IFIPTCS'12}.
\Paper{
Notice that we prove the result for $\lmux$ terms, do not require the terms to be closed, and that the result is shown for single step reduction.


 \begin{theorem} [Soundness] \label{soundness} \label{head reduction simulation}
If $ M \redxh N $, then $\SPilmuTerm [M] a \conredPi
\SPilmuTerm [N] a $.
 \end {theorem}

 \begin{Proof}
By induction on the definition of explicit head reduction\Comment{; for convenience, we separate naming and $`m$-binding, so actually use the encoding of \cite}.

 \begin{description} \itemsep5pt

 \item [Main reduction rules]

 \myitem[$ (`l x.M) N \red \Sub M x := N $]
\SPilmuTerm [a {(l x . M )} N ] a
	& \ByDef & \\
\multicolumn{3}{@{}l}{ \PilmuTerm [a {l x . M} N] a
	\quad \redPi(c),\wbisimG } \\ 
\New b ({ \PilmuTerm[M] b \Par {\PiExSub x := N } \Par \BEq b=a })
	& \wbisimR (\Ref{renaming lemma}) &
\PilmuTerm[S M x := N] a
 \end{array} $

 \myitem[{$(\muterm `b . [`a]M ) N \red \muterm`g . \Sub {[`a]M} `b := N{`.}`g , ~ `g \textit{ fresh} $}]
\kern-20mm && \kern10mm
\SPilmuTerm [a {(u `b . `a M)} N] a
	& \ByDef & \\
\PilmuTerm [A {(\muterm `b . [`a]M)} N] a
	& \ByDef & 
\PilmuTerm [A {(u `b . `a M)} N] a
	& =_{`a} & \\
\New `b ( \PilmuTerm[M] `a \Par \PiExContSub `b := N . a )
	& = & 
\PilmuTerm[u `g . `a {C M `b := N . `g }] a
	& \ByDef & \\
\PilmuTerm[u `g . `a {\Sub M `b := N{`.}`g }] a
	& \ByDef &
\SPilmuTerm[u `g . `a {s M `b := N{`.}`g }] a
 \end{array} $

 \myitem[{$ \muterm`b.[`b]M \red M \hfill \textit{ if } `b \not\in \fn(M)$}]
\SPilmuTerm [\muterm`b.[`b]M] a
	& \ByDef &
\PilmuTerm [u `b . `b M] a
	& \ByDef (`b \not\in \fn(M) ) &
\PilmuTerm [M] a
 \end{array} $

 \myitem[{$\muterm`a.[`b]\muterm`g.[`d]M \red \muterm`a.[`d]M[`b \For `g], `g \not= `d$}]
\kern-35mm &&&& \kern-10mm
\SPilmuTerm [\muterm`a.[`b]\muterm`g.[`d]M] a
	& \ByDef & \\
\PilmuTerm [u `a . `b {\muterm`g.[`d]M}] a
	& \ByDef &
\PilmuTerm [u `a . `b {u `g . `d M}] a
	& =&
\PilmuTerm [u `a . `d {M[`b \For `g]}] a
	& \ByDef& \\
\SPilmuTerm [u `a . `d {M[`b \For `g]}] a
 \end{array} $

 \myitem[{$\muterm`a.[`b]\muterm`g.[`g]M \red \muterm`a.[`b]M[`b \For `g]$}]
\kern-35mm &&&& \kern-25mm
\SPilmuTerm [\muterm`a.[`b]\muterm`g.[`g]M] a
	& \ByDef & \\
\PilmuTerm [u `a . `b {\muterm`g.[`g]M}] a
	& \ByDef &
\PilmuTerm [u `a . `b {u `g . `g M}] a
	& =&
\PilmuTerm [u `a . `b {M[`b \For `g]}] a
	& \ByDef& \\
\SPilmuTerm [u `a . `b {M[`b \For `g]}] a
 \end{array} $

 \item [Term substitution rules]

 \myitem[$ \Sub x x := N \red N $] 
\SPilmuTerm [S x x := N] a
	& \ByDef & \\
\PilmuTerm [s {v x} x := N] a
	& \redPi (x) & \\
\New w ( \BEq w=a 
\Par \PilmuTerm [N] w ) \Par \New x ( \PiExSub x := N )
	& \wbisimR (\Ref{renaming lemma}) & \\
\PilmuTerm [N] a \Par \New x ( {\PiExSub x := N } )
	& \wbisimG & 
\SPilmuTerm [N] a
 \end{array} $

 \myitem[$ \Sub M x := N \red M, ~ x \not\in \fv(M) $]
\SPilmuTerm [S M x := N] a
	& \ByDef & 
\PilmuTerm [s M x := N] a
	& \congruent & \\
\PilmuTerm[M] a \Par \New x ( \PiExsub x := N )
	& \wbisimG &
\SPilmuTerm [M] a
 \end{array} $

\myitem [$\Sub (PQ) x := N \red \Sub {(\Sub P x := N Q)} x := N , ~ x = \hv(P) $]
\SPilmuTerm [S (PQ) x := N] a
	& \ByDef & \\
\PilmuTerm [S {A P Q} x := N] a
	& \wbisim (\Ref{replication lemma}) & \\
\PilmuTerm [S {A {S P x := N} Q } x := N] a
	& \ByDef & \\
\PilmuTerm [S {A {\Sub P x := N } Q} x := N] a
	& \ByDef & \\
\PilmuTerm [S {\Sub P x := N Q } x := N] a
	& \ByDef & \\
\SPilmuTerm [S {(a {\Sub P x := N } Q)} x := N] a
 \end{array} $

 \myitem[$ \Sub (`ly.M) x := N \red `l y.(\Sub M x := N ) , ~ x = \hv(M) $] 
\kern-16mm && \kern10mm \SPilmuTerm [S {(l y . M)} x := N] a
	& \ByDef & \\
\PilmuTerm [S {l y . M} x := N] a
	& \congruent & \\
\PilmuTerm [l y . {S M x := N}] a
	& \ByDef & 
\SPilmuTerm [l y . {S M x := N}] a
 \end{array} $

 \myitem[{$ \Sub (\muterm`a.[`b]M) x := N \red \muterm`a.[`b](\Sub M x := N) , ~ x = \hv(M)$}]
\kern-35mm && \kern10mm
\SPilmuTerm [S {(u `a . `b M)} x := N] a
	& \ByDef & \\
\PilmuTerm [S {(u `a . `b M)} x := N] a
	& \congruent (`a \notele \fn(N) ) & 
\PilmuTerm [u `a . `b {S M x := N}] a
	& \ByDef & \\
\SPilmuTerm [u `a . `b {S M x := N}] a
 \end{array} $


 \item [Structural rules]

\myitem[{$\Sub (\muterm`d.\Cmd) `a := N{`.}`g \red \muterm`d.(\Sub {\Cmd} `a := N{`.}`g ) N$}] \kern-10mm && 
\SPilmuTerm [S {(m `d . \Cmd)} `a := N{`.}`g ] a
	& \ByDef & \\
\PilmuTerm [C {m `d . \Cmd} `a := N . `g ] a
	& =_{`a} & (`d \not\in \fn(\exsub `a := {N{`.}`g} ) ) \\
\PilmuTerm [m `d . {C {\Cmd} `a := N . `g }] a
	& \ByDef &
\PilmuTerm [m `d . { {\Cmd} \excontsub `a := N . `g }] a
	& \ByDef & \\
\SPilmuTerm [\muterm`d.(\Sub {\Cmd} `a := N{`.}`g ) ] a
 \end{array} $

\myitem[{$\Sub ([`a]M) `a := N{`.}`g \red [`g](\Sub M `a := N{`.}`g ) N$}] 
\kern-10mm && \kern-10mm
\SPilmuTerm [S {(n `a M)} `a := N{`.}`g ] a
	& \ByDef & \\
\PilmuTerm [C {n `a M} `a := N . `g ] a
	& \wbisim (\Ref{replication lemma}) &
	\\ 
\PilmuTerm [n `g {A {C M `a := N . `g } N}] a
	& \ByDef & \\
\PilmuTerm [n `g {A {M \excontsub `a := N . `g } N}] a
	& \ByDef & 
\SPilmuTerm [A {C M `a := N . `g } N] `g
	& \ByDef & \\
\SPilmuTerm [n `g {A {c M `a := N . `g } N}] a
 \end{array} $

\myitem[{$\Sub ([`b]M) `a := N{`.}`g \red [`b]\Sub M `a := N{`.}`g , ~ `b \not= `a$}] 
\kern-30mm && \kern20mm
\SPilmuTerm [C {(n `b M)} `a := N . `g ] a
	& \ByDef & \\
\PilmuTerm [C {[`b]M} `a := N . `g ] a
	& \ByDef & 
\PilmuTerm [C {n `b M} `a := N . `g ] a
	& \ByDef & \\
\PilmuTerm [n `b {M \excontsub `a := N . `g }] a
	& \ByDef & 
\SPilmuTerm [n `b {c M `a := N . `g }] a
 \end{array} $

 \myitem[$ \Sub M `a := N{`.}`g \red M , ~ `a \not\in \fn(M) $]
\kern-7mm &&
\SPilmuTerm [C {M} `a := N . `g ] a
	& \ByDef &
\PilmuTerm [C {M} `a := N . `g ] a
	& \congruent & \\
\SPilmuTerm [M] a \Par \New `a ( \PiExContSub `a := N . `g )
	& \wbisimG &
\SPilmuTerm [M] a
 \end{array} $

 \myitem[$ \Sub (`lx.M) `a := N{`.}`g \red `lx.\Sub M `a := N{`.}`g $]
\kern-7mm && \kern-10mm
\SPilmuTerm [C {(l x . M)} `a := N . `g ] a
	& \ByDef & \\
\PilmuTerm [C {l x . M} `a := N . `g ] a
	& \congruent & \\
\PilmuTerm [l x . {C M `a := N . `g }] a
	& \ByDef & 
\SPilmuTerm [l x . {C M `a := N . `g }] a
 \end{array} $

 \myitem[$ \Sub (PQ) `a := N{`.}`g \red \Sub {(\Sub P `a := N{`.}`g Q )} `a := N{`.}`g $]
\SPilmuTerm [C (PQ) `a := N . `g ] a
	& \ByDef & \\
\PilmuTerm [C {A P Q} `a := N . `g ] a
	& \wbisim (\Ref{replication lemma}) &
	\\
\PilmuTerm [C {A {C P `a := N . `g } Q } `a := N . `g ] a
	& \ByDef & \\
\PilmuTerm [C {C P `a := N . `g Q } `a := N . `g ] a
	& \ByDef & \\
\SPilmuTerm [C {(a {C P `a := N . `g } Q)} `a := N . `g ] a
 \end{array} $

 \item [Substitution rules]

 \myitem[$ \Sub {\Sub M x := N } y := P \red \Sub { \Sub { \Sub M y := P } x := N } y := P $]
&& \kern-15mm
\SPilmuTerm [S {S M x := N } y := P ] a
	& \ByDef & \\
\PilmuTerm [S {S M x := N } y := P ] a
	& \wbisim (\Ref{replication lemma}) &
	\\
\PilmuTerm [S {S {S M y := P } x := N } y := P] a
	& \ByDef & \\
\SPilmuTerm [S {S {S M y := P } x := N } y := P] a
 \end{array} $

\item[$ \Sub {\Sub M `a := N{`.}`g } `b := L{`.}`d \red \Sub { \Sub { \Sub M `b := L{`.}`d } `a := N{`.}`g } `b := L{`.}`d $] ~


$ \begin{array}{lclcl}
\SPilmuTerm [C {C M `a := N . `g } `b := L . `d ] a
	& \ByDef & \\
\PilmuTerm [C {C M `a := N . `g } `b := L . `d ] a
	& \wbisim (\Ref{replication lemma}) &
	\\
\Paper{\PilmuTerm [C {C {C M `b := L . `d } `a := N . `g } `b := L . `d ] a }
\CLAC{\PilmuTerm [C {C {B M `b := L . `d } `a := N . `g } `b := L . `d ] a }
	& \ByDef & \\
\SPilmuTerm [C {C {C M `b := L . `d } `a := N . `g } `b := L . `d ] a
 \end{array} $

 \item [Contextual rules] By induction. \qed

\Comment{
 \myitem[$ M \red N \Then ML \red NL $]
\SPilmuTerm [a M L] a
	& \ByDef &
\PilmuTerm [A M L] a
	& \wbisim (\IH) & \\
\PilmuTerm [A N L] a
	& \ByDef &
\SPilmuTerm [a N L] a
 \end{array} $

 \myitem[$ M \red N \Then `lx.M \red `lx.N $]
\SPilmuTerm [l x . M] a
	& \ByDef &
\PilmuTerm [l x . M] a
	& \wbisim (\IH) & \\
\PilmuTerm [l x . N] a
	& \ByDef &
\SPilmuTerm [l x . N] a
 \end{array} $

 \myitem[{$ M \red N \Then \muterm`a.[`b]M \red \muterm`a.[`b]N$}]
 \kern-10mm &&
\SPilmuTerm [u `a . `b M] a
	& \ByDef &
\PilmuTerm [u `a . `b M] a
	& \wbisim (\IH) & 
	\\
\PilmuTerm [u `a . `b N] a
	& \ByDef &
\SPilmuTerm [u `a . `b N] a
 \end{array} $

 \myitem[$ M \red N \Then \Sub M x := L \red \Sub N x := L $]
\kern-15mm &&
\SPilmuTerm [S M x := L] a
	& \ByDef & \\
\PilmuTerm [S M x := L] a
	& \wbisim (\IH) & 
\PilmuTerm [S N x := L] a
	& \ByDef &
\SPilmuTerm [S N x := L] a
 \end{array} $

 \myitem[$ M \red N \Then \Sub M `a := L{`.}`g \red \Sub N `a := L{`.}`g $]
\kern-15mm &&
\SPilmuTerm [C M `a := L . `g] a
	& \ByDef & \\
\PilmuTerm [C M `a := L . `g] a
	& \wbisim (\IH) &
\PilmuTerm [C N `a := L . `g] a
	& \ByDef & \\
\SPilmuTerm [C N `a := L . `g] a \qed
 \end{array} $
}

 \end {description}
 \end{Proof}

Remark that, by Prop.~\ref{reduction hidden}, all proper reductions in this proof are in $\wbisim$.
Also, the proof \CLAC{in \cite{Bakel-Vigliotti-IFIPTCS'12} }shows that $`b$-reduction is implemented in $`p$ by at least one synchronisation.

We can now easily show:
}

 \begin{theorem} [Operational Soundness\Paper{ for $\redwxh$}\CLAC{ \cite{Bakel-Vigliotti-IFIPTCS'12}}] \label{Operational Soundness redxh} \label{rtc soundness}

 \begin{enumerate}
 \firstitem $M \rtcredxh N \Then \SPilmuTerm [M] a \wbisim \SPilmuTerm [N] a $.

 \item If $M \Divergesxh $ then $\PilmuTerm[M] a \Diverges $.

 \end{enumerate}
 \end {theorem}
\Paper{

 \begin{Proof}
The first is shown by induction using Theorem~\ref{head reduction simulation}, using Proposition~\ref{reduction hidden}; the second follows from Example~\ref{redex example}, and the fact that $`m$-reduction and substitution do not loop \cite{Py-PhD'98} (\emph{i.e.}~non-termination is caused only by $`b$-reduction).%
\qed
 \end{Proof}

}
\CLAC{The proof in \cite{Bakel-Vigliotti-IFIPTCS'12} shows that $`b$-reduction is implemented in $`p$ by at least one synchronisation.}

\Comment{
This result states that our interpretation gives, in fact, a semantics for the explicit head reduction for $\lmu$.
 By Lem.~\ref{explicit versus head} and Prop.~\ref{lmu vs lmux} we can show the same for $\redh$ and $\redbmu$.

By Lem.~\ref{explicit versus head}, we can also show:

%
 \begin{theorem} [Operational Soundness for $\redh$] \label{Operational Soundness redh} \CLAC{~}

 \begin{enumerate}
 \firstitem If $M \rtcredh N $, then $ \SPilmuTerm [M] a \wbisimilar \SPilmuTerm [N] a $.
 \item $M \divergesh \Implies \SPilmuTerm [M] a \divergesPi $.
 \end{enumerate}
 \end {theorem}

 \begin{Proof}
 \begin{enumerate}
 \firstitem

If $M \rtcredh N $, then, by Lem.~\ref{explicit versus head}, there exists $L$ such that $ M \rtcredxh L $ and $ L \rtcredxsub N $; \Paper{then, }by Theorem~\ref{Operational Soundness redxh}, $ \SPilmuTerm [M] a \wbisimilar \SPilmuTerm [L] a $.

 \item
Since every $`b$-reduction step in $\redh$ invokes at least one step in $\redxh$.
\qed

 \end{enumerate}
 \end{Proof}
and similarly, by Prop.~\ref{lmu vs lmux}, we can show:

%
 \begin{theorem} [Operational Soundness for $\redbmu$] \label{Operational Soundness}

 \begin{enumerate}
 \firstitem
$M \rtcredbmu N \Then \SPilmuTerm [M] a \wbisimilar \SPilmuTerm [N] a $.
 \item
$M \divergess \Implies \SPilmuTerm [M] a \divergesPi $.
 \end{enumerate}
 \end {theorem}

 \begin{Proof}
If $M \rtcredbmu N $, then, by Prop.~\ref{lmu vs lmux}, there exists $L$ such that $ M \rtcredx L $ and $ L \rtcredxsub N $; then, by Theorem~\ref{Operational Soundness}, $ \SPilmuTerm [M] a \wbisimilar \SPilmuTerm [L] a $,
\qed
 \end{Proof}

We can also show operational completeness for $\redxh$.
 \begin{theorem} [Operational Completeness for $\redxh$] \label{Operational Completeness for Lmux}
If $ \PilmuTerm [M] a \redPi \proc{P} $, then there exists $N$ such that $ \proc{P} \wbisim \PilmuTerm [N] a $ and $ M \rtcredxh N $.
 \end {theorem}

 \begin{Proof}
By easy induction on the structure of terms, using the fact that all reductions that are possible in $\PilmuTerm [M] a $ are generated by the interpretation, and correspond to $`b$-contractions or are the result of reduction over $y$ in
the process
$ \New y ( \PilmuTerm [l x . M] y \Par \Eq y=a ) $, and is then the result of executing a substitution $\Sub y y := `lx.M $.%
\qed
 \end{Proof}

This in turn can be used to show:
 \begin{lemma} \label{Completeness}
 \begin{enumerate}

 \firstitem
Let $M$ be a term in $\lmux$.
If $ \PilmuTerm[M] a \rtcredPi \PilmuTerm[N] a $ then $ M \rtcredxh N $.

 \item
Let $M \ele \lmu$, \emph{i.e.}~a (pure) $\lmu$-term.
If $ \PilmuTerm[M] a \redPi \proc{P} $ then there exists $N \ele \lmux$ and $L \ele \lmu$ such that $ \proc{P} \wbisimilar \PilmuTerm[N] a $, and $ M \rtcredxh N $ and $ N \rtcredxsub L $.

 \end{enumerate}
 \end {lemma}

 \begin{Proof}
The first is an obvious consequence of Theorem~\ref{Operational Completeness for Lmux}, the second follows from Lem.~\ref{head reduction}, Lem.~\ref{lmu vs lmux}, and Theorem~\ref{Operational Completeness for Lmux}.\qed
 \end{Proof}

Since renaming can be executed for abstractions, we can even generalise this result for lazy reduction $\redlazy$ \cite{Abramsky'90} and explicit lazy reduction $\redxl$ \cite{Bakel-Vigliotti-CONCUR'09} on closed $`l$-terms:
 \begin{theorem}
Let $M$ be closed. Then
 \begin{enumerate}
 \item $M \redxl `ly.M' \, \Vect{\exsub x := N } \Implies \CLAC{$ \\ $} \SPilmuTerm [M] a \rtcredPi
 y := P {(`nx)} \, ( \SPilmuTerm [{`ly.M}'] a \Par \Vect{ \Bang \SPilmuTerm [x := N] {} } ) $.

 \item $M \redlazy `ly.M' \, \Vect{[N \For x]} \Implies \CLAC{$ \\ $} \SPilmuTerm [M] a \rtcredPi
 y := P {(`nx)} \, ( \SPilmuTerm [`ly.M'] a \Par \Vect{ \SPilmuTerm [N] x } ) $.

 \end{enumerate}
 \end{theorem}

}


We can also show that equality with explicit substitution, $\eqx$, is preserved under our encoding by weak bisimulation.

 \begin{theorem} \label{lambda mu x model}
If $M \eqx N$, then $\PilmuTerm[M] a \wbisimilar \PilmuTerm[N] a $.
 \end{theorem}

 \begin{Proof}
By induction on the definition of $\eqx$\Paper{
; we only show the cases that are different from the proof of Theorem~\ref{soundness}.

 \begin{description} \itemsep 4pt

 \myitem [$ \Sub (PQ) `a := N{`.}`g \red (\Sub P `a := N{`.}`g ) \, (\Sub Q `a := N{`.}`g) $]
\PilmuTerm[\sub (PQ) x := N ] a
	& \ByDef & \\
\PilmuTerm[S {A P Q} x := N ] a
	& \wbisimilar (\Ref{replication lemma}) \\
\New c ({ \New x ({ \PilmuTerm[P] c \Par \PiLExSub x := N }) \Par
\quad {} \\ \hfill
	\New x ({ \In c (v,d) . ({ \PiExsub v := Q \Par \Eq d=a }) \Par \PiLExSub x := N })
})
	& \wbisimilar (\Ref{replication lemma}) \\
\New c ({ \New x ({ \PilmuTerm[P] c \Par \PiLExSub x := N }) \Par
	\In c (v,d) . ({ \New x ({ \PiExsub v := Q \Par \PiExSub x := N }) \Par \Eq d=a })
})
	& \wbisimilar (\Ref{replication lemma}) \\
\New c ({ \New x ({ \PilmuTerm[P] c \Par \PiExSub x := N }) \Par
	\In c (v,d) . ({ \PiExsub v := {S Q x := N } \Par \Eq d=a }) })
	& \ByDef & \\
\New c ({ \New x ({ \PilmuTerm[M] c \Par \PiExSub x := L }) \Par
	\In c (v,d) . ({ \PiExsub v := {\sub N x := L } \Par \Eq d=a }) })
	& \ByDef & \\
\PilmuTerm[A {\sub P x := N } {\sub Q x := N }] a
	& \ByDef & \\
\PilmuTerm[{}(\sub P x := N ) (\sub Q x := N )] a
 \end{array} $

 \myitem [$ \Sub (PQ) `a := N{`.}`g \red (\Sub P `a := N{`.}`g ) \, (\Sub Q `a := N{`.}`g) $]
\SPilmuTerm [C (PQ) `a := N . `g ] a
	& \ByDef & \\
\PilmuTerm [C {X P Q} `a := N . `g ] a
	& \wbisim (\Ref{replication lemma}) & \\
\PilmuTerm [Y {(C P `a := N . `g)} {C Q `a := N . `g}] a
	& \ByDef & \\
\PilmuTerm [a {(C P `a := N . `g)} {C Q `a := N . `g}] a
	& \ByDef & \\
\PilmuTerm [A {C P `a := N . `g } {(\ContSub Q `a := N . `g )}] a
	& \ByDef & \\
\SPilmuTerm [a {(C P `a := N . `g)} {(C Q `a := N . `g )}] a
 \end{array} $

 \item [If $ M \red N $ then $ LM \red LN $, $\sub L x := M \red \sub L x := N $, and $ \Sub L `a := M{`.}`g \red \Sub L `a := N{`.}`g $] ~

By induction.
\qed

\Comment{
 \myitem [$ M \red N \Then LM \red LN $]
\PilmuTerm[LM] a
	& \ByDef & \\
\PilmuTerm[X L M] a
	& \wbisimilar & (\IH) \\
\PilmuTerm[X L N] a
	& \ByDef & 
\PilmuTerm[LN] a
 \end{array} $

 \myitem [$ M \red N \Then \sub L x := M \red \sub L x := N $]
\PilmuTerm[\sub L x := M ] a
	& \ByDef & \\
\PilmuTerm[s L x := M] a
	& \wbisimilar & (\IH) \\
\PilmuTerm[s L x := N] a
	& \ByDef & 
\PilmuTerm[\sub L x := N ] a
 \end{array} $

 \myitem [$ M \red N \Then \Sub L `a := M{`.}`g \red \Sub L `a := N{`.}`g $]
\PilmuTerm[\Sub L `a := M{`.}`g ] a
	& \ByDef & \\
\PilmuTerm[F L `a := M . `g ] a
	& \wbisimilar & (\IH) & \\
\PilmuTerm[F L `a := N . `g ] a
	& \ByDef & 
\PilmuTerm[\Sub L `a := N{`.}`g ] a
 \end{array} $

 \myitem [$ \sub x x := L \red L $]
\PilmuTerm[\, \sub x x := L ] a
	& \ByDef & \\
\PilmuTerm[s {v x} x := L] a
	& \wbisimilar (\Ref{reduction hidden}) \\
\New w ({ \Eq w=a \Par \PilmuTerm[L] w }) \Par \New x ({ \PiLExSub x := N })
	& \wbisimilar (\Ref{other reduction},\Ref{renaming lemma}) & 
\PilmuTerm[L] a
 \end{array} $

 \myitem [$ \sub M x := L \red M \hfill (x \notele \fv(M) ) $]
\SPiLTerm [s M x := N] a
	& \ByDef & 
\PilmuTerm [S M x := N] a
	& \wbisimilar & \\
\PilmuTerm[M] a \Par \New x ( \PiLExSub x := N )
	~ \wbisimilar (\Ref{other reduction}) ~
\SPiLTerm[M] a
 \end{array} $

 \myitem [$ M \red N \Then ML \red NL $]
\PilmuTerm[ML] a
	& \ByDef & \\
\PilmuTerm[A M L] a
	& \wbisimilar (\IH) \\
\PilmuTerm[A N L] a
	& \ByDef & 
\PilmuTerm[NL] a
 \end{array} $

 \myitem [$ M \red N \Then LM \red LN $]
\PilmuTerm[LM] a
	& \ByDef & \\
\PilmuTerm[a L M] a
	& \wbisimilar (\IH) \\
\PilmuTerm[a L N] a
	& \ByDef & 
\PilmuTerm[LN] a
 \end{array} $

 \myitem [$ M \red N \Then `lx.M \red `lx.N $]
\PilmuTerm[`lx.ML] a
	& \ByDef & 
\PilmuTerm[l x . M] a
	& \wbisimilar (\IH) & \\
\PilmuTerm[l x . N] a
	& \ByDef & 
\PilmuTerm[`lx.N] a
 \end{array} $

 \myitem [$ M \red N \Then \sub M x := L \red \sub N x := L $]
\kern -11mm &&
\PilmuTerm[\sub M x := L ] a
	& \ByDef & 
\PilmuTerm[S M x := L] a
	& \wbisimilar (\IH) \\
\PilmuTerm[S N x := L] a
	& \ByDef & 
\PilmuTerm[\sub N x := L ] a
 \end{array} $

 \myitem [$ M \red N \Then \sub L x := M \red \sub L x := N $]
\PilmuTerm[\sub L x := M ] a
	& \ByDef & \\
\PilmuTerm[s L x := M] a
	& \wbisimilar (\IH) \\
\PilmuTerm[s L x := N] a
	& \ByDef & 
\PilmuTerm[\sub L x := N ] a
 \end{array} $

 \item [$ M \red N \Then M \rtcred N $]
Immediate.

 \item [$ M \rtcred M $]
Immediate.

 \item [$ M \rtcred N \And N \rtcred L \Then M \rtcred L $]
Since $\PilmuTerm[M] a \wbisimilar \PilmuTerm[N] a $ and $ \PilmuTerm[N] a \wbisimilar \PilmuTerm[L] a $ implies $\PilmuTerm[M] a \wbisimilar \PilmuTerm[L] a $.

 \item [$ M \rtcred N \Then M \eqb N $]
Immediate.

 \item [$ M \eqb N \Then M \eqb N $]
Since $\PilmuTerm[M] a \wbisimilar \PilmuTerm[N] a $ implies $\PilmuTerm[N] a \wbisimilar \PilmuTerm[M] a $.

 \item [$ M \eqb N \And N \eqb L \Then M \eqb L $]
Since $\PilmuTerm[M] a \wbisimilar \PilmuTerm[N] a $ and $ \PilmuTerm[N] a \wbisimilar \PilmuTerm[L] a $ implies $\PilmuTerm[M] a \wbisimilar \PilmuTerm[L] a $.\QED
}

 \end{description}
The steps to a reflexive, transitive closure and equivalence relation follow directly from $\wbisimilar$
}
.\qed

 \end{Proof}

Now the following is an immediate consequence:

 \begin{theorem}[Semantics] \label{lambda model}
If $M \eqbmu N$, then $\PilmuTerm[M] a \wbisimilar \PilmuTerm[N] a $.
 \end{theorem}
 \begin{Proof}
By induction on the definition of $\eqbmu$.
The case $M \rtcredbmu N$ follows from the fact that then, by Proposition~\ref{lmu vs lmux}, also $M \rtcredx N$, so by Theorem~\ref{lambda mu x model}, we have $\PilmuTerm[M] a \wbisimilar \PilmuTerm[N] a $.
The steps to an equivalence relation follow directly from $\wbisimilar$.\QED
 \end{Proof}

Notice that it is clear that we cannot prove the exact reversal of this result, since terms without head-normal form are all interpreted by $\Zero$ (see also Lem.~\ref{zero lemma}), but are not all related through $\eqbmu$.
Using \Paper{our notion of }weak equivalence, we can deal with the reverse part, and will do so in the last sections of this paper.

\Paper{

We can show that interpretation of terms in $\redxh$-normal form are in normal form as well, which is a property that we need here.
 \begin{lemma} \label{head pi normal form}
$\lmuNF$ is a $\redxh$-nf implies $ \PilmuTerm[\lmuNF] a $ is irreducible.
 \end{lemma}

 \begin{Proof}
By induction on the structure of terms in $\redxh$-normal form.

 \begin{description}

 \myitem[$\lmuNF = xM_1\dots M_n ~ (n \geq 0) $]
\textrm{Then } \PilmuTerm [xM_1\dots M_n] a
	& \ByDef & \\
\New c_n (\PilmuTerm [xM_1\dots M_{n-1}] c_n \Par \PiExContSub c_n := M_{n-1} . a )
	& \ByDef & \\
\NewA c_n ({ \NewA c_{n-1} (\PilmuTerm [xM_1\dots M_{n-2}] c_{n-1} \Par \PiExContSub c_{n-1} := M_n . c_n }) \Par \dots \Par \PiExContSub c_n := M_n . a )
	& \congruent & \\
\NewA c_nc_{n-1} (\PilmuTerm [xM_1\dots M_{n-2}] c_{n-1} \Par \PiExContSub c_{n-1} := M_n . c_n \Par \dots \Par \PiExContSub c_n := M_n . a )
	& \ByDef & \\
 \New c_n\dots c_1 (\PilmuTerm [x] c_1 \Par \PiExContSub c_1 := M_1 . c_2 \Par \dots \Par
\PiExContSub c_n := M_n . a ) \\ [1mm]
 \end{array} $

\noindent
Since
 \[ \begin{array}[t]{rcl}
 \SPilmuTerm [x] c_1 &=& \PilmuTerm [v x] c_1 \\
 \PiExContSub c_i := M_i . c_{i+1} &=& \PiExContFsub c_i := M_i . c_{i+1}
 \end{array} \]
all $\PilmuTerm [M_i] w $ appear under input on $c_i$, so no synchronisation inside one of the $\PilmuTerm [M_i] w $ is possible; since all $c_i$ are fresh, all are different from $x$ and no synchronisation is possible over one of the $c_i$.
So this process is in normal form.

 \item[$\lmuNF = `lx. \lmuNF' $]
Then $ \SPilmuTerm [l x . \lmuNF'] a = \PilmuTerm [l x . \lmuNF'] a $, and, by induction, $\PilmuTerm [\lmuNF'] b $ is in normal form; since $a$ is fresh, that process does not input over $a$, so $ \SPilmuTerm [l x . \lmuNF'] a $ is normal form.

 \item[{$\lmuNF = \muterm `a . [`b] \lmuNF' ~ (`a \not= `b \Or `a \in \lmuNF', \lmuNF' \not= \muterm`g.[`d] \lmuNF'' )$}]
Then $ \SPilmuTerm [u `a . `b \lmuNF'] a = \PilmuTerm [u `a . `b \lmuNF'] a $, this follows immediately by induction.

 \item[$\lmuNF = \Sub \lmuNF' x := M ~ (\hv(\lmuNF') \not= x ) $]
Then $ \SPilmuTerm [s \lmuNF' x := M ] a = \PilmuTerm [s \lmuNF' x := M ] a $.
By induction, $\PilmuTerm [\lmuNF'] a $ is in normal form; since $x$ is not the head-variable of $\lmuNF$, the process $\PilmuTerm [\lmuNF'] a $ has no reachable input over $x$, so no synchronisation is possible over $x$; also, no synchronisation is possible inside $\PilmuTerm [M] w $, as above.

 \item[$\lmuNF = \Sub \lmuNF' `a := M{`.}`g ~ (\hn(\lmuNF') \not= `a ) $]
Then $ \SPilmuTerm [c {\lmuNF'} `a := M . `g ] a = \PilmuTerm [C {\lmuNF'} `a := M . `g ] a $.
By induction, $\PilmuTerm [\lmuNF'] a $ is in normal form.
Note that $`a$ is only a reachable output in $\PilmuTerm [\lmuNF'] a $ if $\lmuNF'$ is an abstraction and $a = `a$; this is impossible, since we choose $a$ fresh.
As above, no synchronisation is possible inside $\PilmuTerm [M] w $.
\qed

 \end{description}
 \end{Proof}

Notice that $ \SPilmuTerm [u `a . `b {u `g . `d \lmuNF}] a = \PilmuTerm [u `a . `b {u `g . `d \lmuNF}] a $, which is in normal form, so some reducible terms in $\lmux$ are mapped to processes in normal form; this does not contradict the above result, of course.

We can now show the following termination results:
 \begin{theorem} [Termination] \label{lambda head termination}
 \begin{enumerate}
 \firstitem If $M \rtcredxh[\nf] N $, then $ \PilmuTerm [M] a \convergesPi $.
 \item If $M \rtcredbmu[\hnf] N $, then $ \PilmuTerm [M] a \convergesPi $.

 \end{enumerate}
 \end{theorem}

 \begin{Proof}
 \begin{enumerate}

 \firstitem
By Lem.~\ref{head pi normal form}, if $N$ is in explicit head-normal from, then $ \PilmuTerm [N] a $ is in normal form, and by Thm~\ref{head reduction simulation}, $ \PilmuTerm [M] a \rtcredPi \proc{P} $ with $\proc{P} \wbisim \PilmuTerm [N] a $.
Since in the proof of Thm~\ref{head reduction simulation}, $\wbisimG$ only removes processes in normal form, this implies that \proc{P} is in normal form.
 \item
By Lem.~\ref{head reduction}, there exists $L$ in {\HNF} such that $ M \rtcredh[\nf] L $;
by Prop.~\ref{explicit versus head}, there exists $\lmuNF$ such that $ M \rtcredxh[\nf] \lmuNF $; by the previous part, $ \PilmuTerm [M] a \convergesPi $.
\qed

 \end{enumerate}
 \end{Proof}

Notice also that this result is stronger than the formulation of the termination result for Milner's interpretation in \cite{Sangiorgi-Walker-Book'01}, since it models reduction to head-normal form, not just lazy normal form.
Since terms that have a normal form have a head-normal form as well, Theorem~\ref{lambda head termination} immediately leads to:
 \begin{corollary} \label{lambda termination}
If $M \convergesbmu $, then $ \PiLmuSem [M] a \convergesPi $.
 \end{corollary}

}

\Paper{ \section{Weak reduction for $\lmu$ and $\lmux$} 
 \label{weak reduction} }
\CLAC{ \section{Weak \Paper{reduction and }equivalences for $\lmu$ and $\lmux$}
\label{first full abstraction section} 
 \label{weak reduction}
}

\label{weak equivalences}


\Paper{It seems widely accepted that bisimilarity-like equivalences have become the standard when studying interpretations of $`l$-calculi into the $`p$-calculus.
This creates a point of concern with respect to full abstraction.}
Since $`D`D$ and $`W`W$ (where \CLAC{$`D = `lx.xx$ and }$`W = `ly.yyy$\Paper{; we will use $`W$ again below}) are closed terms that do not interact with any context, they are contextually equivalent; any well-defined interpretation of these terms into the $`p$-calculus, be it input based or output based, will therefore map those to processes that are weakly bisimilar to $\Zero$, and therefore to weakly bisimilar processes.
Abstraction, on the other hand, enables interaction with a context, and therefore the interpretation of $`lz.`D`D$ will \emph{not} be weakly bisimilar to $\Zero$.
%
We therefore cannot hope to model standard $`b`m$-equality in the $`p$-calculus in a fully-abstract way; rather, we need to consider a notion of reduction that considers \emph{all} abstractions meaningful; therefore, the only kind of reduction on $`l$-calculi that can naturally be encoded into the $`p$-calculus is \emph{weak} reduction.

\Paper{

\DDinpi

 \begin{example} \label{DD is a zero process}
Consider the reduction of $`D`D$ that was given in Example~\ref{reductions example}; by Theorem~\ref{head reduction simulation}, we have for example the reduction of $\SPilmuTerm [`D `D] a \wbisim \SPilmuTerm [\Sub {\Sub xy y := x } x := `ly.yy ] a $ as shown in Figure~\ref{DDinpi},
which shows that the interpretation of $`D`D$ reduces without creating output over $a$ -- it always occurs inside a sub-process of the shape
 \[ \begin{array}{rcl}
\PiExContsub c := y . a
 \end{array} \]
and does not input, since the head-variable is always bound, so $\SPilmuTerm [`D `D] a $ is weakly bisimilar to $\Zero$ (see also Lem.~\ref{zero lemma}).
Therefore,
 \[ \begin{array}{lclcl}
\PilmuTerm[`lz.`D`D] a
	&\ByDef&
\PilmuTerm[l z . {`D`D}] a
	&\wbisim& \\ &&
\New zb ( \Zero \Par \Out a <z,b> )
	&\wbisim& \\ &&
\PilmuTerm[l z . {`W`W}] a
	&\ByDef&
\PilmuTerm[`lz.`W`W] a
 \end{array} \]
So, for full abstraction, we are forced to consider $`lz.`D`D$ and $`lz.`W`W$ equivalent, and therefore, we need to consider \emph{weak} equivalences on terms.
 \end{example}

Generally, the concept of \emph{weak} reduction refers to the variant of calculi that eliminate the contextual rule
 \[ \begin{array}{rcl}
M \red N &\Then& `lx.M \red `lx.N
 \end{array} \]
For the $`l$-calculus, then a closed normal form is an abstraction.
Note that, in the context of $\lmu$, this is no longer the case, since also context switches might occur inside the term.

We will now introduce the correct notions in $\lmu$.
}

 \begin{definition}
We define the notion $\redwbmu$ of \emph{weak $`b`m$-reduction} as in Def.~\ref{lmu reduction}, the notion $\redwh$ of \emph{weak head reduction}\footnote{This notion is also known as \emph{lazy} reduction; for the sake of keeping our terminology consistent, we prefer to call it weak head reduction.} on $\lmu$ as in Def.~\ref{head reduction definition}, and the notion $\redwxh$ of \emph{weak explicit head reduction} on $\lmux$ as in Def.~\ref{explicit head reduction}, by (also) eliminating the rules:
 \[ \begin{array}{rcl}
\Sub (`ly.M) x := N &\red & `l y.(\Sub M x := N ) \\
\Sub (`lx.M) `a := N{`.}`g & \red & `lx.(\Sub M `a := N{`.}`g ) \\
M \red N &\Then&
`lx.M \red `lx.N
 \end{array} \]
 \end{definition}

We define the notion of weak head-normal forms, the normal forms with respect to weak head-reduction:

 \begin{definition}[Weak head-normal forms for $\lmu$] \label{weak head reduction definition}
 \begin{enumerate}

 \firstitem
The $\lmu$ \emph{weak head-normal forms} (\WHNF) are defined through the grammar:
 \[ \begin {array}{rrl@{\quad}l}
\lmuwHNF & ::= &
`lx. M
	\\ &\mid &
xM_1\dots M_n & (n \geq 0)
	\\ &\mid &
\muterm`a. [`b] \lmuwHNF
	& (`a \not= `b \textrm{ or } `a \in \fn(\lmuwHNF),
		\Paper{\textrm{ and }}\lmuwHNF \not= \muterm `g.[`d] \lmuwHNF' )
 \end {array} \]

 \item
We say that $M$ \emph{has a \WHNF} if there exists $\lmuwHNF$ such that $M \rtcredwh \lmuwHNF$.

\Comment{
 \item
We define \emph{values} as those \WHNF s defined through:
 \[ \begin {array}{r\CLAC{@{~}}r\CLAC{@{~}}ll}
\lmuV & ::= & `lx. M
	\\ &\mid &
\muterm`a. [`b] \lmuV & (`a \not= `b \textrm{ or } `a \not\in \lmuV, \\ &&& ~
	\textrm{ and }\lmuV \not= \muterm `g.[`d] \lmuV' )
 \end {array} \]
}

 \end{enumerate}
 \end{definition}
\Paper{As before, the normal forms of weak head reduction are the \WHNF s.}

The main difference between \HNF s and \WHNF s is in the case of abstraction: where the definition of {\HNF} only allows for the abstraction over a \HNF, for \WHNF s any term can be the body.
Moreover, notice that both $\LTerm<l z . {`D`D}>$ and $\LTerm<l z . {`W`W}>$ are in \WHNF.

Since ${\redwxh} \subseteq {\redxh}$, we can show the equivalent of Lem~\ref{head reduction} and Thm.~\ref{rtc soundness} also for \emph{weak explicit head reduction}:

 \Paper{
 \begin{proposition} \label{weak head reduction}
If $ M \rtcredbmu N $ with $N$ in {\WHNF}, then there exists $\lmuwxHNF$ such that $ M \rtcredwh[\nf] \lmuwxHNF $ and $ \lmuwxHNF \rtcredbmu N $ without using $\redwh$.
 \end{proposition}

 \begin{corollary} \label{weak soundness}
}
\CLAC{
 \begin{theorem} [cf.~\cite{Bakel-Vigliotti-IFIPTCS'12}] \label{weak soundness}
}
 \begin{enumerate}
 \firstitem
If $ M \rtcredwxh N $, then $\SPilmuTerm [M] a \wbisim \SPilmuTerm [N] a $.
 \item
If $ M \rtcredbmu N $ with $N$ in {\WHNF}, then there exists $\lmuwxHNF$ such that $ M \rtcredwxh[\nf] \lmuwxHNF $ and $ \lmuwxHNF \rtcredx N $ without using $\redwxh$.
 \end{enumerate}
\CLAC{
 \end{theorem}
}
 \Paper{
 \end{corollary}
}

We also define weak explicit head-normal forms.

 \begin{definition}[Weak explicit head-normal forms\Paper{ for $\lmu$}] \label{weak explicit head reduction definition}

 \begin{enumerate}

 \firstitem
The $\lmux$ \emph{weak explicit head-normal forms \CLAC{\\ }(\WEHNF)} are defined through\Paper{ the grammar}:
 \[ \def\arraystretch{1.1} \begin {array}{rrl@{\qquad}l}
\lmuwxHNF & ::= &
`lx. M \, \Vect{\exsub y := N } \, {\Vexcontsub `s := Q . `t } 
	\\ &\mid &
xM_1\dots M_n \, \Vect{\exsub y := N } \, {\Vexcontsub `s := Q . `t } & (n \geq 0, ~ x \notele \Vect{y})
	\\ &\mid &
\muterm `a . [`b] \lmuwxHNF \, \Vect{\exsub y := N } \, {\Vexcontsub `s := Q . `t }
&	\\ \multicolumn{4}{r}
	{(`b \notele \Vect{`s}, ~ `a \not= `b \textrm{ or } `a \in \fn(\lmuwxHNF), 
	\textrm{ and } \lmuwxHNF \not= `m`g.[`d] \lmuwxHNF' ) }
 \end {array} \]

 \item
We say that $M \ele \lmux$ \emph{has an \WEHNF} if there exists $\lmuwxHNF$ such that $M \rtcredwxh \lmuwxHNF$.

\Comment{
 \item
We define \emph{values} as those weak head-normals defined through:
 \[ \begin {array}{rrll}
\lmuV & ::= & `lx. M
	\\ &\mid &
\muterm`a. [`b] \lmuV & (`a \not= `b \textrm{ or } `a \not\in \lmuV, \\ &&& ~
	\textrm{ and }\lmuV \not= \muterm `g.[`d] \lmuV' )
 \end {array} \]
}

 \end{enumerate}
 \end{definition}

 \begin{remark} \label{substitution remark}
In the context of reduction (normal and weak), when starting from pure terms, the substitution operation can be left inside terms in normal form, as in
 \[ \begin{array}{rcl}
(`lx.yM)NL &\redxh& \Sub yM x := N \, L .
 \end{array} \]
However, since, by Barendregt's convention, $x$ does not appear free in $L$, the latter term is operationally equivalent to $\Sub yML x := N $; in fact, these two are equivalent under $\equivwh$ (see Def.~\ref{equivwh definition}), and also congruent when interpreted as processes.
\Paper{
 \[ \begin{array}{lclcl}
\SPilmuTerm[a {S yM x := N } L] a
	& \ByDef &
\PilmuTerm[A {S yM x := N } L] a
	& \congruent & \\ &&
\PilmuTerm[S {A yM L} x := N] a
	& \ByDef & \\ &&
\SPilmuTerm[S {a yM L} x := N] a
 \end{array} \]
}
Since in weak reduction the reduction $\Sub (`lx.M) y := N $ for $`lx . (\Sub M y := N ) $ is not allowed, also this substitution can be considered to stay at the outside.
Therefore, without loss of generality, for readability and ease of definition we will use a notation for terms that places all explicit substitutions on the outside.\footnote{This is exactly the approach of Krivine's machine, where explicit substitutions are called \emph{closures} that form an environment in which a term is evaluated.}
So actual terms can have substitutions inside, but they are written as if they appear outside.
To ease notation, we will use $\Ssub$ for a set of substitutions of the shape $ \exsub x := N $ or $ \excontsub `a := N . `g $ when the exact contents of the substitutions is not relevant; we write $x \ele \Ssub $ if $\exsub x := N \ele \Ssub $ and similarly for $`a \ele \Ssub $.

 \end{remark}

We can show that the interpretation of a term without {\WHNF} \Paper{gives a process that }is weakly bisimilar to $\Zero$.

 \begin{lemma} \label{zero lemma}
If $M$ has no {\WEHNF} (so $M$ also has no \WHNF), then $\PilmuTerm[M] a \wbisim \Zero $.
 \end{lemma}

 \begin{Proof}
If $M$ has no {\WEHNF}, then $M$ has no leading abstractions and all terms generated by reduction have a weak explicit head redex.
If $M = \muterm `a . [`b] N $, then $\PilmuTerm[M] a \ByDef \PilmuTerm[u `a . `b N] a \wbisim \Zero $, so also $ \PilmuTerm[N] `b \wbisim \Zero $; therefore we can assume $M$ itself does not start with a context switch.

We reason by coinduction on the explicit weak head reduction sequence from $M$ and analyse the cases of weak explicit head reduction.
For example,
 \[ \def\arraystretch{1.2} \begin{array}{lcl}
\PilmuTerm [\VSub {(`lx.P_1)P_2 \dots P_n} y := Q \, {\Vexcontsub `a := R . `b }] a ~ \ByDef
	\\ \dquad
\NewA \Vect{c} (\PilmuTerm [l x . P_1] c_1 \Par \Comment{ {} \\ \dquad \dquad } {\VPiExContSub c_{i-1} := P_i . c_i } \Par {\VPiExSub y := Q } \Par \, {\VPiExContSub `a := R . `b })
 \end{array} \]
where $c_{n-1} = a$\CLAC{. }\Paper{ and
 \[ \def\arraystretch{1.1} \begin{array}{rcl}
\PiExContSub c_{i-1} := P_i . c_i &=& \PiExContFsub c_{i-1} := P_i . c_i \\
\PiExSub y_j := Q_j &=& \PiExsub y_j := Q_j \\
\PiExContSub `a_k := R_k . `b_k &=& \PiExContFsub `a_k := R_k . `b_k
 \end{array} \]}
Since a synchronisation over $c_1$ is possible, the process is not in normal form.
Observe that all outputs are over bound names or under guard, and since the result of the reduction has no head variable, no input is exposed. 
So \Paper{reduction of $\PilmuTerm[M] a $ cannot exhibit an input or an output, so }$\PilmuTerm[M] a \wbisim \Zero $.%
\qed
 \end{Proof}
\Paper{\DDnorename

The reduction of $\SPilmuTerm [a `D `D] a $ is given in Fig.~\ref{DDnogarb}, which shows that the interpretation of $`D`D$ reduces without creating output over $a$; notice that the individual steps of the above reduction in $\redxh$ in Example~\ref{redxh examples} are respected in Fig.~\ref{DDnogarb}.

As a direct consequence of this result,\Paper{ as for Milner's and Sangiorgi's interpretations,} our interpretation is not extensional, since
$ \PilmuTerm[`D`D] a \wbisim \Zero $, whereas $ \PilmuTerm[`lx.`D`Dx] a \ByDef \PilmuTerm[l x . {`D`Dx}] a \not\wbisim \Zero $.
}

We can show the following property.

 \begin{lemma} \label{redwh to redwxh lemma} \label{redbmu to redwxh lemma}

%
Let $M$ and $N$ be pure $\lmu$-terms; then
$M \rtcredwh[\nf] N$ if and only if there exists $N'$, $\Ssub$ such that $M \rtcredwxh[\nf] N' \, \Ssub $, and $ N' \, \Ssub\rtcredxsub[\nf] N $.

%

 \end{lemma}



We will now define equivalences $\equivwbmu$ and $\equivwh$ between terms of $\lmu$, and $\equivwxh$ between terms of $\lmux$ (the last two are defined coinductively as bisimulations), that are based on weak reduction, and show that the last two equate the same pure $\lmu$-terms.
These notions all consider terms without {\WHNF} equivalent.
This is also the case for the approximation semantics we present in the next section.

First we define a weak equivalence generated by the reduction relation $\redwbmu$.

 \begin{definition} \label{wbmu equivalence}
We define $\equivwbmu$ as the smallest congruence that contains:
 \[ \begin{array}{rcl@{\quad}l}
M, N \emph{ have no \WHNF} &\Then& M \equivwbmu N \\
(`l x . M ) N & \equivwbmu & M [ N \For x ] \\
(\muterm `a . \Cmd ) N & \equivwbmu & \muterm`g. \Cmd [N{`.}`g \For `a] & (`g \textit{ fresh}) \\
\quad \muterm`a.[`b]\muterm`g.[`d]M & \equivwbmu & \muterm`a.([`d]M[`b \For `g]) \\
\muterm `a . [`a] M & \equivwbmu & M & (`a \notele M) 
\Comment{
M \equivwbmu N \And N \equivwbmu L &\Then& N \equivwbmu L \\
 \\
M &\equivwbmu& M \\
M \equivwbmu N & \Then &
\multicolumn{2}{l}{ \begin{cases}
N \equivwbmu M \\
`lx.M \equivwbmu `lx.N \\
LM \equivwbmu LN \\
ML \equivwbmu NL \\
\muterm `a . [`b] M \equivwbmu \muterm `a . [`b] N \\
 \end{cases} }
}
 \end{array} \]
 \end{definition}

\Paper{
Notice that $`D`D \equivwbmu `W`W$ and $`lz.`D`D \equivwbmu `lz.`W`W$, but $`D`D \not \eqbmu `W`W$; moreover, $\equivwbmu$ is closed under reduction.
In Section~\ref{approximation section} we will show that two terms are equivalent in $\equivwbmu$ if and only if they have the same set of weak approximants.
}

Since reduction is confluent, the following is immediate.

 \begin{proposition} \label{confluence equiv}
If $M \equivwbmu N$ and $M \rtcredwbmu \lmuwHNF$, then there exists $\lmuwHNF'$ such that $\lmuwHNF \equivwbmu \lmuwHNF'$ and $N \rtcredwbmu \lmuwHNF'$.
 \end{proposition}
\Paper{Notice that Prop.~\ref{confluence eq} is formulated with respect to $\eqbmu$, not $\equivwbmu$.}

The other two equivalences we consider are generated by \emph{weak head reduction} and \emph{weak explicit head reduction}.
We will show in Theorem~\ref{equivwh is equivwxh} that these coincide for pure, substitution-free terms.

 \begin{definition} [Weak head equivalence] \label{equivwh definition}
The relation $\equivwh$ is defined co-inductively as the largest symmetric relation such that: $M \equivwh N$ if and only if either $M$ and $N$ have both no \WHNF, or both $M \rtcredwh[\nf] M'$ and $ N \rtcredwh[\nf] N'$, and either:
 \begin {itemize}

 \item if $M' = xM_1\dots M_n$ $(n \geq 0)$, then $ N' = xN_1\dots N_n$ and $ M_i \equivwh N_i $ for all $\iotn$; or

 \item if $M' =`lx.M''$, then $N' = `lx.N''$ and $M'' \equivwh N''$; or

 \item if $M' = \muterm`a.[`b]M''$, then $N' = \muterm`a.[`b]N''$ (so $`a \not= `b$ or $`a \ele \fn(M'')$, $M'' \not= \muterm`g.[`d] R$, and similarly for $N''$), and $M'' \equivwh N''$.

 \end {itemize}
 \end{definition}
Notice that $`lz.`D`D \equivwh `lz.`W`W$ because $`D`D \equivwh `W`W$, since neither has a \WHNF.

\Paper{
We perhaps need to clarify the details of this definition.
The notion of weak head equivalence captures the fact that, once weak head reduction has finished, there are sub-terms that can be reduced further.
This process will generate (in principle) infinite terms and the equivalence expresses when this produces equal (infinite) terms.
However, it also equates terms that have no \WHNF.
As can be seen from Def.~\ref{weak head reduction definition}, a context switch $\muterm`a.[`b]N$ is in {\WHNF} only if $N$ is; so when we state in the third case that $M \rtcredwh[\nf] \muterm`a.[`b]M''$, by the fact that this reduction has terminated, we know that $M''$ \emph{is} in \WHNF.
}

We will now define a notion of weak explicit head equivalence, that, in approach, corresponds the weak head equivalence but for the fact that now explicit substitutions are part of terms.

 \begin{definition} [Weak explicit head-equivalence] \label{equivwxh definition}
The relation $\equivwxh$ is defined co-in\-ductively as the largest symmetric relation such that: $M \equivwxh N$ if and only if either $M$ and $N$ have both no $\redwxh$-normal form, or both $M \rtcredwxh[\nf] M' \, \Ssub $ and $ N \rtcredwxh[\nf] N' \, \Ssub' $, and either:
 \begin {itemize}

 \item if $M' = xM_1\dots M_n $ $(n \geq 0)$, then $ N' = xN_1\dots N_n $ (so $x \notele \Ssub$, $x \notele \Ssub'$) and $M_i \, \Ssub \equivwxh N_i \, \Ssub ' $ for all $\iotn$; or

 \item if $M' =`lx.M''$, then $N' = `lx.N''$ and $M'' \, \Ssub \equivwxh N'' \, \Ssub' $; or

 \item if $M' = \muterm`a.[`b]M''$, then $N' = \muterm`a.[`b]N''$ (so $`a \not= `b$ or $`a \ele \fn(M'')$, $M'' \not= \muterm`g.[`d] R$, so $`b \notele \Ssub$, $`b \notele \Ssub'$, and similarly for $N''$) and $M'' \, \Ssub \equivwxh N'' \, \Ssub' $.

 \end {itemize}
 \end{definition}
Notice that $\muterm`a.[`b]`D`D \equivwxh `D`D$.


The following results formulate the strong relation between $\equivwh$ and $\equivwxh$, and therefore between $\redwh$ and $\redwxh$.
We first show that pure terms that are equivalent under $\equivwxh$ are also so under $\equivwh$.


 \begin{lemma}
Let $M$ and $N$ be pure $\lmu$-terms; then $M \equivwh N $ if and only if there are $M'$, $N'$ such that $M' \rtcredxsub[\nf] M$ and $N' \rtcredxsub[\nf] N$, and $M' \equivwxh N' $.

 \begin{Proof}
 \begin{description}

 \item[only if]
By co-induction on the definition of $\equivwh$.
If $M \equivwh N $, then either:
 \begin {itemize} 

 \item $M \rtcredwh[\nf] xM_1\dots M_n$ and $N \rtcredwh[\nf] xN_1\dots N_n$ and $ M_i \equivwh N_i$, for all $\iotn$.
Then, by Lem.~\ref{redwh to redwxh lemma}, there are $\Vect{M'_i}$ such that \Paper{both}
 \[ \begin{array}{ccccc}
M &\rtcredwxh[\nf]& xM'_1\dots M'_n ~ \Ssub &\rtcredxsub& xM_1\dots M_n
\\
N &\rtcredwxh[\nf]& xN'_1\dots N'_n ~ \Ssub' &\rtcredxsub& xN_1\dots N_n
 \end{array} \]
But then $M'_i \, \Ssub \rtcredxsub[\nf] M_i$ and $N'_i \, \Ssub' \rtcredxsub[\nf] N_i$, for all $\iotn$; then by induction, $M'_i \, \Ssub \equivwxh N'_i \, \Ssub' $ for all $\iotn$.
But then $M \equivwxh N$.




 \end {itemize}
The other cases are similar.

 \item[if]
By co-induction on the definition of $\equivwxh$.
If there are $M'$, $N'$ such that $M' \rtcredxsub[\nf] M$ and $N' \rtcredxsub[\nf] N$, and $M' \equivwxh N' $, then 
either:

 \begin {itemize}

 \item $M' \rtcredwxh[\nf] xM'_1\dots M'_n ~ \Ssub $, $N' \rtcredwxh[\nf] xN'_1\dots N'_n ~ \Ssub' $ and $ \Vect{M'_i \, \Ssub \equivwxh N'_i \, \Ssub' } $.
Let, for all $\iotn$, $M'_i \, \Ssub \rtcredxsub[\nf] M_i$ and $N'_i \, \Ssub \rtcredxsub[\nf] N_i$ then by induction, $\Vect{M_i \equivwh N_i}$.
Notice that we have $M' \rtcredwxh[\nf] xM'_1\dots M'_n ~ \Ssub \rtcredxsub[\nf]$ $xM_1\dots M_n$.
Let $M' = M'' \, \Ssub'' $, so $ M'' \, \Ssub'' \rtcredwxh[\nf] xM'_1\dots M'_n ~ \Ssub'\,\Ssub'' $, where $\Ssub = \Ssub'\,\Ssub'' $.
Let $M'' \, \Ssub'' \rtcredxsub[\nf] M $, then by Lem.~\ref{redwh to redwxh lemma}, we also have $M \rtcredwxh[\nf] xM''_1\dots M''_n ~ \Ssub' \rtcredwh[\nf] xM_1\dots M_n $.
Then, again by Lem.~\ref{redwh to redwxh lemma}, $M \rtcredwh[\nf] xM_1\dots M_n$; likewise, we have $N \rtcredwh[\nf] xN_1\dots N_n$.
But then $M \equivwh N$.




 \end {itemize}
The other cases are similar.\qed

 \end{description}
 \end{Proof}

 \end{lemma}

Notice that this lemma in fact shows:

 \begin{theorem} \label{equivwh is equivwxh}
Let $M,N \ele \lmu$, then $M \equivwxh N \Iff M \equivwh N$.
 \end{theorem}

\Comment{

 \begin{lemma} \label{equivwxh implies equivwh}
For $M,N \ele \lmu$: $M \equivwxh N \Then M \equivwh N$.
 \end{lemma}

 \begin{Proof}
We define $ {\equivR} \subseteq \lmu^2 $ by: $ M \equivR N $ if and only if $ M,N$ have no {\WHNF} or there exist $M',N'\ele \lmux $ such that $ M' \equivwxh N' $, $ M' \rtcredxsub M $, and $ N' \rtcredxsub N $.
We show that, for $M,N \ele \lmu$: $M \equivwxh N \Then M \equivR N \Then M \equivwh N$.
\CLAC{ \leftmargini 0pt}
 \begin{description}

 \item [$M \equivR N \Then M \equivwh N$]
Let $M',N'\ele \lmux$ be such that $M' \equivwxh N'$, $M' \rtcredxsub M$, and $N' \rtcredxsub N $.
We reason by co-induction on the definition of $\equivwxh$.

 \begin {itemize} \itemsep 3pt

 \item $M$ and $N$ have both no \WHNF; then also $M \equivwh N$.

 \item Assume $M' \rtcredwxh[\nf] xM'_1\dots M'_n ~ \Ssub $ and $N' \rtcredwxh[\nf] xN'_1\dots N'_n ~ \Ssub' $ such that $ M'_i \, \Ssub \equivwxh N'_i \, \Ssub' $ for all $\iotn$.

Let $M'_i \, \Ssub \rtcredxsub[\nf] M_i$ and $N'_i \, \Ssub \rtcredxsub[\nf] N_i$ (mark that $\rtcredxsub[\nf]$ just removes the explicit substitutions), then by definition of $\equivR$ we have $M_i \equivR N_i$, so by induction, $M_i \equivwh N_i$, for all $\iotn$.
Then also $xM'_1\dots M'_n ~ \Ssub \rtcredxsub[\nf] xM_1\dots M_n$.

Assume that $M'$ has some substitutions, so $M' = M'' \, \Ssub'' $, then also $ M'' \, \Ssub'' \rtcredwxh[\nf] xM'_1\dots M'_n ~ \Ssub'\,\Ssub'' $, where $\Ssub = \Ssub'\,\Ssub'' $.

Let $M$ be the result of removing all substitutions in $M''$, so such that $M'' \, \Ssub'' \rtcredxsub[\nf] M $, then by Lem.~\ref{redwh to redwxh lemma}, $M \rtcredwh[\nf] xM_1\dots M_n$; likewise, we can show that $N \rtcredwh[\nf] xN_1\dots N_n$.
So we have shown that $M_i \equivwh N_i$ for all $\iotn$, and we obtain $M \equivwh N$.

\Paper{
 \item Assume {$M' \rtcredwxh[\nf] `lx.M'_0 \, \Ssub $ and $N' \rtcredwxh[\nf] `lx.N'_0 \, \Ssub' $ such that $M'_0 \, \Ssub \equivwxh N'_0 \, \Ssub' $}.
Let $M'_0 \, \Ssub \rtcredxsub[\nf] M_0$ and $N'_0 \, \Ssub \rtcredxsub[\nf] N_0$,
then $M_0 \equivR N_0$ by definition, so by induction, $M_0 \equivwh N_0$.
Then also $`lx.M'_0 \, \Ssub \rtcredxsub[\nf] `lx.M_0$.

Let $M' = M'' \, \Ssub'' $, then also $ M'' \, \Ssub'' \rtcredwxh[\nf] `lx.M'_0 ~ \Ssub'\,\Ssub'' $, where $\Ssub = \Ssub'\,\Ssub'' $.
Take $M$ such that $M'' \, \Ssub'' \rtcredxsub[\nf] M $, then by Lem.~\ref{redwh to redwxh lemma}, $M \rtcredwh[\nf] `lx.M_0$; likewise, we can show that $N \rtcredwh[\nf] `lx.N_0$.
But then $M \equivwh N$.

 \item Assume {$M' \rtcredwxh[\nf] \muterm`a.[`b]M_0$ and $N' \rtcredwxh[\nf] \muterm`a.[`b]N_0$, such that $M_0 \equivh N_0 $ and both have a \WHNF.}
Let $M_0 \, \Ssub \rtcredxsub[\nf] M'_0$ and $N_0 \, \Ssub \rtcredxsub[\nf] N'_0$,
then $M'_0 \equivR N'_0$, so by induction, $M_0 \equivwh N_0$.
Then also $\muterm`a.[`b]M_0' \, \Ssub \rtcredxsub[\nf] \muterm`a.[`b]M_0$.
Let $M' = M'' \, \Ssub'' $, then also $ M'' \, \Ssub'' \rtcredwxh[\nf] \muterm`a.[`b]M_0 ~ \Ssub'\,\Ssub'' $, where $\Ssub = \Ssub'\,\Ssub'' $.
Take $M$ such that $M'' \, \Ssub'' \rtcredxsub[\nf] M $, then by Lem.~\ref{redwh to redwxh lemma}, $M \rtcredwh[\nf] \muterm`a.[`b]M_0$; likewise, we can show that $N \rtcredwh[\nf] \muterm`a.[`b]N_0$.
But then $M \equivwh N$.
}

 \item $M' \rtcredwxh[\nf] xM'_1\dots M'_n $ or $M' \rtcredwxh[\nf] \muterm`a.[`b]M_0$: similar

 \end {itemize}

So $M \equivR N$ implies $M \equivwh N$.

 \item [$M,N \ele \lmu \Then ( M \equivwxh N \Then M \equivR N )$]
Take $M' = M$, and $N' = N$.
\qed
 \end{description}
 \end{Proof}


We can also show the converse of the previous result, which is:

 \begin{lemma} \label{equivwh implies equivwxh}
For $M,N \ele \lmu$: $M \equivwh N \Then M \equivwxh N$.
 \end{lemma}

 \begin{Proof}
\qed
 \end{Proof}

We can now easily show:

 \begin{theorem} \label{equivwh is equivwxh}
Let $M,N \ele \lmu$, then $M \equivwxh N$ if and only if $M \equivwh N$.
 \end{theorem}

 \Paper{\begin{Proof}
By Lem.~\ref{equivwxh implies equivwh} and~\ref{equivwh implies equivwxh}.
\qed
 \end{Proof}}
}

\CLAC{ \section{Full abstraction for the logical interpretation} \label{Full abstraction} \label{approximation section}

In this section we will show our main result, that the logical encoding is fully abstract with respect to weak equivalence between pure $\lmu$-terms.
To achieve this, we show in Thm.~\ref{FA equivxh equivC} that $\PilmuTerm[M] a \wbisim \PilmuTerm[N] a $ \emph{iff} $M \equivwxh N$.
We are thus left with the obligation to show that $M \equivwxh N$ \emph{iff} $M \equivwbmu N$.
In Thm.~\ref{equivwh is equivwxh} we have shown that $M \equivwxh N$ \emph{iff} $M \equivwh N$, for pure terms; to achieve $M \equivwh N$ \emph{iff} $M \equivwbmu N$, we go through a notion of \emph{weak approximation}; based on Wadsworth's approach \cite{Wadsworth'76}, we define $\equivwA$ that expresses that terms have the same weak approximants and show that $ M \equivwh N $ \emph{iff} $ M \equivwA N $ \emph{iff} $ M \equivwbmu N $.

}

We can \Paper{also }show \Paper{the following results that state }that if the interpretation of $M$ produces an output, then $M$ reduces by head reduction to an abstraction; similarly, if the interpretation of $M$ produces an input, then $M$ reduces by head reduction to a term with a head variable.

 \begin{lemma} \label{pi reduction characterisation}
 \begin{enumerate} \itemsep 0pt

 \firstitem
If $\PilmuTerm [M] a \Outson a $, then there exist $x, N$ and $\Ssub$ such that $ \PilmuTerm [M] a \wbisim \PilmuTerm [`lx.N\, \Ssub] a $, and $ M \rtcredwxh[\nf] `lx.N \, \Ssub $.

 \item
If $\PilmuTerm [M] a \Outson c $, with $a \not= c$, then there exist $`a, c, x, N$ and $\Ssub$ such that $ \PilmuTerm [M] a \wbisim \PilmuTerm [\muterm`a.[c]`lx.N\, \Ssub] a $, and $ M \rtcredwxh[\nf] \muterm`a.[c]`lx.N \, \Ssub $.


 \item
If $\PilmuTerm [M] a \Inson x $, then there exist $\Vect{z_j}, x, \Vect{N_i}, c$ and $\Ssub$ with $x \notele \Vect{z_j}$, $m \geq 0$, and $n \geq 0$ such that
 \begin{itemize}
 \item $\PilmuTerm [M] a \wbisim \PilmuTerm [`lz_1\dots z_m.xN_1\dots N_n \, \Ssub] c $;
 \item $ M \rtcredwxh[\nf] `lz_1\dots z_m.xN_1\dots N_n \, \Ssub $ if $a = c$;
 \item $ M \rtcredwxh[\nf] \muterm`a.[c]`lz_1\dots z_m.xN_1\dots N_n[a \For `a] \, \Ssub $, if $a \not= c$.
 \end{itemize}

 \end{enumerate}
 \end{lemma}

 \begin{Proof}
Straightforward.
\qed
 \end{Proof}


\Comment{
 \[ \begin{array}{lcl}
\SPilmuTerm [(`lx.x)(`ly.z)] a
	&=& \\
\PilmuTerm [a {l x . x} `ly.z ] a
	&=& \\
\PilmuTerm [s {v x} x := `ly.z ] a
	&=& \\
\New w ( \BEq w=a \Par \PilmuTerm [l y . z ] w )
	&=& \\
\PilmuTerm [l y . {v z} ] a
 \end{array} \]
}

As to the reverse, we can show:

 \begin{lemma} \label{bmu reduction characterisation}
 \begin{enumerate} \itemsep 0pt

 \firstitem
If $ M \rtcredwxh[\nf] `lx.N \, \Ssub $, then $ \PilmuTerm [M] a \Outson a $.

 \item
If $ M \rtcredwxh[\nf] \muterm`a.[`b]`lx.N \, \Ssub $, then $ \PilmuTerm [M] a \Outson `b $.

 \item
$\PilmuTerm [M] a \Inson x $ if $ M \rtcredwxh[\nf] xN_1\dots N_n \, \Ssub $ or $ M \rtcredwxh[\nf] \muterm`a.[`b]xN_1\dots N_n \, \Ssub $.

 \end{enumerate}
 \end{lemma}

 \begin{Proof}
Straightforward.
\qed
 \end{Proof}

}

\Paper{ \section{Weak approximation for \texorpdfstring{$\lmu$}{}}
\label{approximation section}

The notions of \emph{approximant} and \emph{approximation} were first introduced by Wadsworth for the {\LC}~\cite{Wadsworth'76}, where they are used in order to better express the relation between equivalence of meaning in Scott's models and the usual notions of conversion and reduction.
Wadsworth defines approximation of terms through the replacement of any parts of a term remaining to be evaluated (\emph{i.e.}~$`b$-redexes) by $\bot$.
Repeatedly applying this process over a reduction sequence starting with $M$ gives a set of approximants, each giving some - in general incomplete - information about the result of reducing $M$.
Once this reduction produces $ `lx.yN_1\dots N_n $, all remaining redexes occur in $N_1, \ldots, N_n$, which then in turn will be approximated.

Following this approach, Wadsworth \cite{Wadsworth'76} defines $\SetAppr(M)$ (similar to Def.~\ref{approximation} below) as the set of approximants of the $`l$-term $M$, which forms a meet semi-lattice.
In \cite{Wadsworth'78}, the connection is established between approximation and semantics, by showing
 \[ \begin{array}{rcl}
\Sem{M}_{\Dinfty} \, p &=& \bigsqcup \, \Set { \Sem{A}_{\Dinfty} \, p \mid A \ele \SetAppr(M) }.
 \end{array} \]
So, essentially, approximants are partially evaluated expressions in which the locations of incomplete evaluation (\emph{i.e.}~where reduction \emph{may} still take place) are explicitly marked by the element $\bottom$; thus, they \emph{approximate} the result of computations. Intuitively, an approximant can be seen as a `snapshot' of a computation, where we focus on that part of the resulting program which will no longer change, which corresponds to the (observable) \emph{output}.
}

\CLAC{Essentially following \cite{Wadsworth'76}, w}\Paper{W}e now define a \emph{weak approximation semantics} for $\lmu$.
Approximation for $\lmu$ has been studied by others as well \cite{Saurin'10,deLiguoro'13}; however, seen that we are mainly interested in \emph{weak} reduction here, we will define \emph{weak} approximants, which are normally not considered.

 \begin{definition}[Weak approximation for $\lmu$] \label{weak approximation lmu} \label{approximation}

 \begin{enumerate} \itemsep 0pt

 \firstitem
The set of $\lmu$'s \emph{weak approximants} $\SetwApprlmu$ \Paper{with respect to $\redbmu$ }is defined through the grammar:\Paper{\footnote{For `normal' approximants, the case $`lx.\ftAppr$ normally demands that $\ftAppr \not= \bot$, as motived by the relation with $\Dinfty$.}}
\Paper{ \[ \begin{array}{rrl@{\quad}l}
\wAppr & ::= & \bot \\
	&\mid & x\wAppr^1\dots \wAppr^n & (n \geq 0) \\
	&\mid & `lx.\wAppr \\
	&\mid & \muterm`a.[`b]\wAppr & (`a \not= `b \textrm{ or } `a \in \wAppr,
	\wAppr \not= \muterm `g.[`d] \wAppr', ~ \wAppr \not= \bot )
 \end{array} \]
}\CLAC{
 \[ \begin{array}{rrl@{\quad}l}
\wAppr & ::= & 
\bot \mid `lx.\wAppr \mid x\wAppr^1\dots \wAppr^n & (n \geq 0)
\\
	&\mid & \muterm`a.[`b]\wAppr & (`a \not= `b \textrm{ or } `a \in \wAppr, ~
	\wAppr \not=	\muterm `g.[`d] \wAppr', ~ \wAppr \not= \bot )
 \end{array} \]
}

 \item
The relation ${\dirapp} \subseteq \SetwApprlmu \times \lmu $ is \Paper{defined as }the smallest preorder that is the compatible extension of $\bot \dirapp M$.\Paper{\footnote{Notice that if $\ftAppr_1 \dirapp M_1$, and $\ftAppr_2 \dirapp M_2$, then $\ftAppr_1\ftAppr_2$ need not be an approximant; it is one if $\ftAppr_1 = x\ftAppr_1^1\dots \ftAppr_1^n$, perhaps prefixed with a number of context switches of the shape $\muterm`a.[`b]$.}
 \[ \begin {array}{rcl}
 \bottom &\dirapp& M \\
M \dirapp M' & \Then & `l x . M \dirapp `l x . M' \\
M \dirapp M' & \Then & \muterm `g.[`d] M \dirapp \muterm `g.[`d] M' \\
M_1 \dirapp M'_1 \And M_2 \dirapp M'_2 & \Then & M_1M_2 \dirapp M_1'M_2'
 \end {array} \] }


 \item \label{appr def}
\Paper{ The set of \emph{weak approximants} of $M$, $\SetwApprlmu(M)$, is defined as:\footnote{Notice that we use $\redbmu$ here, not $\redwbmu$; the approximants are weak, not the reduction.}%
 \[ \begin{array}{rcl}
\Set{ \wAppr \ele \SetwApprlmu \mid \Exists N \ele \lmu \Pred[ M \rtcredbmu N \And \wAppr \dirapp N] } .
 \end{array} \]}
\CLAC{ $ \begin{array}{@{}rcl}
\SetwApprlmu(M) & \ByDef & \Set{ \wAppr \ele \SetwApprlmu \mid \Exists N \ele \lmu \Pred[ M \rtcredbmu N \And \wAppr \dirapp N] } .
 \end{array} $}

 \item
\emph{Weak approximation equivalence} is defined through:
 $ \begin{array}{rcl@{}}
M \equivwA N &\ByDef& \SetwApprlmu(M) = \SetwApprlmu(N)
 \end{array} $.

 \end{enumerate}
 \end{definition}
Notice that, in part~\ref{appr def}, the approximants are weak, not the reduction.

\Comment{ Notice that
$ \begin{array}[t]{ccccc}
\SetwApprlmu(`lz.`D`D)
	&=&
\Set{\bot, `lz.\bot}
	&=&
\SetwApprlmu(`lz.`W`W)
	\\
\SetwApprlmu(\muterm `a . [`b] `D`D)
	&=&
\Set{\bot}
	&=&
\SetwApprlmu(`D`D)
 \end{array} $ \\ }

The relationship between the approximation relation and reduction is characterised by\Paper{ the following result}:
 \begin{lemma} \label{approximation lemma redbmu}
 \begin{enumerate}

 \firstitem If $\wAppr \dirapp M$ and $M \rtcredbmu N$, then $\wAppr \dirapp N$.

 \item If $ \wAppr \ele \SetwApprlmu(N) $ and $ M \rtcredbmu N $, then also $ \wAppr \ele \SetwApprlmu(M) $.

 \item If $\wAppr \ele \SetwApprlmu(M)$ and $M \redbmu N$, then there exists $L$ such that $N \rtcredbmu L$ and $\wAppr \dirapp L$.

 \item $M$ is a {\WHNF} if and only if there exists $\wAppr \not= \bot $ such that $ \wAppr \dirapp M $.

 \end{enumerate}
 \end{lemma}

\Paper{ \begin{Proof}
Easy. \qed
 \end{Proof}}

\Paper{
We could also have defined the set of approximants of a term coinductively:

 \begin{definition} \label{alternative approximation}
We define $\SetwApprlmuAlt(M)$ coinductively by:

 \begin {itemize}

 \item If $ \wAppr \dirapp M $, then $ \wAppr \ele \SetwApprlmuAlt(M)$.

 \item
if $M \rtcredwh xM_1\dots M_n$ $(n \geq 0)$, then
$ \SetwApprlmuAlt(M) = \Set { x\wAppr^1\dots \wAppr^n \mid \wAppr^i \ele \SetwApprlmuAlt(M_i) , \iotn } $.

 \item if $M \rtcredwh `lx.N$, then
$ \SetwApprlmuAlt(M) = \Set { `lx.\wAppr \mid \wAppr \ele \SetwApprlmuAlt(N) } $.

 \item if $M \rtcredwh \muterm`a.[`b]N$, with $`a \not= `b$ or $`a \ele \fn(N)$, $N \not= \muterm`g.[`d] L$, then
$ \SetwApprlmuAlt(M) = \Set { \muterm`a.[`b]\wAppr \mid \wAppr \ele \SetwApprlmuAlt(N) } $.

 \end {itemize}
 \end{definition}

We can show that these definitions coincide:

 \begin{lemma} \label{coincide lemma}
$\SetwApprlmuAlt(M) = \SetwApprlmu(M)$.
 \end{lemma}

 \begin{Proof}
 \begin{description}
 \item[$\subseteq$]
If $\wAppr \ele \SetwApprlmuAlt(M)$, then by Def.~\ref{alternative approximation} either:

 \begin{description}

 \item [$ \wAppr \dirapp M $] Immediate.

 \item [$ \wAppr = x\wAppr^1\dots \wAppr^n $]
Then $ M \rtcredwh xM_1\dots M_n $, with $ \wAppr^i \ele \SetwApprlmuAlt(M_i) $.
By coinduction, also $ \wAppr^i \ele \SetwApprlmu(M_i) $; then, by Def.~\ref{weak approximation lmu}, there exist $M'_i$ such that $ M_i \rtcredbmu M'_i $ and $ \wAppr^i \dirapp M'_i $.
Then, since ${\rtcredwh} \subseteq {\rtcredbmu}$, in particular $ M \rtcredbmu xM'_1\dots M'_n$ and also $ \wAppr \dirapp xM'_1\dots M'_n $, so $ \wAppr \ele \SetwApprlmu(M) $.



 \end{description}
The other cases are similar.

 \item[$\supseteq$]
If $ \wAppr \ele \SetwApprlmu(M) $, then by Def.~\ref{weak approximation lmu}, there exists $N$ such that $ M \rtcredbmu N $ and $ \wAppr^i \dirapp N $.
Now either:

 \begin{description}

 \item [$ \wAppr \dirapp M $]  Trivial.

 \item [$ \wAppr = x\wAppr^1\dots \wAppr^n $]
Since $ x\wAppr^1\dots \wAppr^n \dirapp N $, $ N = xN_1\dots N_n $, and $ \wAppr^i \dirapp N_i $.
Then by Def.~\ref{alternative approximation} $ \wAppr^i \ele \SetwApprlmuAlt(N_i) $, and by induction, $ \wAppr^i \ele \SetwApprlmu(N_i) $.
By Lem.~\ref{weak head reduction}, there exist $M_i$ $(\iotn)$ such that $M \rtcredwh xM_1 \dots M_n \rtcredbmu xN_1\dots N_n $; since ${\rtcredwh} \subseteq {\rtcredbmu}$, this implies that $M_i \rtcredbmu N_i$; then by Lem.~\ref{approximation lemma redbmu}, 
$ \wAppr^i \ele \SetwApprlmu(M_i) $.
Then, by Def.~\ref{alternative approximation}, $ \wAppr \ele \SetwApprlmuAlt(M) $.



 \end{description}
The other cases are similar.
 \end{description}
 \end{Proof}
As a result, below we will use whichever definition is convenient.

Notice that the first and latter cases in Def.~\ref{alternative approximation} overlap for terms that are already in head-normal form. In fact, we can show:

 \begin{lemma}
If $\wAppr \ele \SetwApprlmu(M) $ and $ \wAppr' \dirapp \wAppr $, then $ \wAppr' \ele \SetwApprlmu(M) $.
 \end{lemma}
 \begin{Proof}
By Def.~\ref{weak approximation lmu} and transitivity of $\dirapp$.\qed
 \end{Proof}
By Lem.~\ref{coincide lemma}, this result also holds for $ \SetwApprlmuAlt(`.) $.

}

As is standard in other settings, interpreting a $\lmu$-term $M$ through its set of weak approximants $\SetwApprlmu(M)$ gives a semantics.

 \begin{theorem} [Weak approximation semantics] \label{approx seman lmux}
If $M \eqbmu N$, then $M \equivwA N$%
\Comment{$\SetApprlmu(M) = \SetwApprlmu(N) $}.
 \end{theorem}

 \begin{Proof}
\CLAC{Using Prop.~\ref{confluence eq} and Lem.~\ref{approximation lemma redbmu}.\qed}
\Paper{
\setbox100=\hbox{\hspace*{\leftmargini}\textit{Proof. }}
$ 
\begin {array}[t]{@{}lclcl} 
M \eqbmu N \And \wAppr \ele \SetwApprlmu(M)
	& \Then & \\
M \eqbmu N \And \Exists L \Pred[ M \rtcredbmu L \And \wAppr \dirapp L ]
	& \Then & (\textit{\ref{confluence eq}}) \\
\Exists L,K \Pred[ L \rtcredbmu K \And N \rtcredbmu K \And \wAppr \dirapp L ]
	& \Then & (\textit{\ref{approximation lemma redbmu}}) \\
\Exists K \Pred[ N \rtcredbmu K \And \wAppr \dirapp K ]
	& \Then &
\wAppr \ele \SetwApprlmu(N) \qed
 \end{array} $
}
\Comment{
$ \kern-6mm \begin {array}[t]{@{}lllll} \kern6mm
M \eqbmu N \And \wAppr \ele \SetwApprlmu(M)
	& \Then &
\kern -10mm
M \eqbmu N \And \Exists L \Pred[ M \rtcredbmu L \And \wAppr \dirapp L ]
	& \Then (\textit{\ref{confluence eq}}) \\
\Exists L,K \Pred[ L \rtcredbmu K \And N \rtcredbmu K \And \wAppr \dirapp L ]
	& \Then (\textit{\ref{approximation lemma redbmu}}) &
\Exists K \Pred[ N \rtcredbmu K \And \wAppr \dirapp L ]
	& \Then & \\
\wAppr \ele \SetwApprlmu(N) \qed
 \end{array} $
}
 \end{Proof}

The reverse implication of this result does not hold, since terms without {\WHNF} (which have only $\bot$ as approximant) are not all related by reduction.
But we can show the following full abstraction result:

 \begin{theorem} [Full abstraction of $\equivwbmu$ versus $\equivwA$]
\label{full abstr approx seman Lmux}
$M \equivwbmu N$ if and only if $M \equivwA N$.
 \end{theorem}

 \begin{Proof}


 \begin {description}

 \item[if]
By co-induction on the definition of the set of weak approximants.
\CLAC{ \leftmarginii 0pt}
\Paper{%
If $\SetwApprlmu(M) = \Set{\bottom} = \SetwApprlmu(N)$, then both $M$ and $N$ have no {\WHNF}, so $M \equivwbmu N$.
Otherwise, either:

 \begin{description} \itemindent -1em

 \item[$x\wAppr^1\dots \wAppr^n \ele \SetwApprlmu(M) \And x\wAppr^1\dots \wAppr^n \ele \SetwApprlmu(N)$] \CLAC{ ~ }

Then by Def.~\ref{alternative approximation}
$M \rtcredwh xM_1\dots M_n$ for some $M_i$ ($\iotn$) and $\wAppr^i \ele \SetwApprlmu(M_i)$.
Likewise, there exist $N_i$ ($\iotn$) such that $N \rtcredwh xN_1\dots N_n$ and $\wAppr^i \ele \SetwApprlmu(N_i)$.
So $\SetwApprlmu(M_i) = \SetwApprlmu(N_i)$ and by induction $M_i \equivwbmu N_i$, for $\iotn$.
Since $\equivwbmu$ is a congruence, also $xM_1\dots M_n \equivwbmu xN_1\dots N_n$; since $\equivwbmu$ is closed under reduction $\redwbmu$, it is also under $\redwh$, and we have $M \equivwbmu N$.

\Comment{%
AAAAAAAAAAAAAAAAAAAAAAAAAAAAAAAAAA

 \item[$x\wAppr^1\dots \wAppr^n \ele \SetwApprlmu(M) \And x\wAppr^1\dots \wAppr^n \ele \SetwApprlmu(N)$] \CLAC{ ~ }

If $x\wAppr^1\dots \wAppr^n \ele \SetwApprlmu(M)$, then there exist $L$ such that $M \rtcredbmu L$ and $x\wAppr^1\dots \wAppr^n \dirapp L$, so $L = xL_1\dots L_n$ with $\wAppr^i \dirapp L_i$.

Then, by Lem.~\ref{redbmu to redwxh lemma}, $M \rtcredwh xM_1\dots M_n$ with, for $\iotn$, $M_i \rtcredwbmu L_i$; then $\wAppr^i \ele \SetwApprlmu(M_i)$.

Likewise, there exist $L'$ such that $N \rtcredbmu L'$ and $x\wAppr^1\dots \wAppr^n \dirapp L'$, so $L' = xL'_1\dots L'_n$ with $\wAppr^i \dirapp L'_i$, and $N \rtcredwh xN_1\dots N_n$ with, for $\iotn$, $N_i \rtcredwbmu L'_i$; then $\wAppr^i \ele \SetwApprlmu(N_i)$.

So $\SetwApprlmu(M_i) = \SetwApprlmu(N_i)$, so by induction $M_i \equivwbmu N_i$, for $\iotn$.
But then $xM_1\dots M_n \equivwbmu xN_1\dots N_n$; since $\equivwbmu$ is closed under reduction, also $M \equivwbmu N$.


}

 \end {description}
The other cases are similar.
}

 \item[only if]
As the proof of Theorem~\ref{approx seman lmux}, but using Proposition~\ref{confluence equiv} rather than~\ref{confluence eq}.\qed

 \end {description}
 \end{Proof}

We can also show that weak head equivalence and weak approximation equivalence coincide:
 \begin{theorem} \label{eqA is eqh}
$M \equivwh N $ if and only if $M \equivwA N$.
 \end{theorem}

 \begin{Proof}
Straightforward, by coinduction.\qed

\Comment{
 \begin{description} \itemsep 4pt

 \item[if : $ \SetwApprlmu(M) = \SetwApprlmu(N) \Then M \equivwh N $]
By co-induction on the definition of the set of weak approximants.

 \begin{description}

 \item [$ \SetwApprlmu(M) = \Set{\bottom} = \SetwApprlmu(N) $]
Then both $M$ and $N$ have no \WHNF, so $ M \equivwh N $.

 \item [$ \wAppr = x\wAppr^1\dots \wAppr^n $]
Then $ M \rtcredwh xM_1\dots M_n $, and $ \wAppr^i \ele \SetwApprlmu(M_i) $, for $\iotn$.
Since $ \SetwApprlmu(M) = \SetwApprlmu(N) $, also $ N \rtcredwh xN_1\dots N_n $, with $ \wAppr^i \ele \SetwApprlmu(N_i) $, so $ \SetwApprlmu(M_i) = \SetwApprlmu(N_i) $.
Then, by coinduction, $ M_i \equivwh N_i $, so $ M \equivwh N $.


]
 \end{description}
The other cases are similar.

 \item[only if : $M \equivwh N \Then M \equivwA N $]
By co-induction on the definition of $\equivwh$.
 \begin{description}

 \item [$M$ and $N$ have no {\WHNF}]
Then $ \SetwApprlmu(M) = \Set{\bottom} = \SetwApprlmu(N) $.

 \item[$ M \rtcredwh xM_1\dots M_n $]
Then also $N \rtcredwh xN_1 \dots N_n$, and $ M_i \equivwh N_i $ for $\iotn$, and by coinduction, $ M \equivwA N $, so $ \SetwApprlmu(M_i) = \SetwApprlmu(N_i)$.
Then, by Def.~\ref{alternative approximation}, we have $ \SetwApprlmu(M) = \SetwApprlmu(N)$.



 \end{description}
The other cases are similar.

 \end{description}
}
 \end{Proof}

We can \Paper{also }define $\Sem{M}_{\SetwApprlmu} = \sqcup \, \Set{ \wAppr \mid \wAppr \ele \SetwApprlmu(M) } $, with $\sqcup$ the least-upper bound with respect to $\dirapp$; then $\Sem{`.}_{\SetwApprlmu}$ corresponds to the ($\lmu$ variant of) L\'evy-Longo trees.
Combined with the results shown in the previous section, we now also have the following result that states that all equivalences coincide:
 \begin{corollary} \label{all lmu weak equivalences}
Let $M,N \ele \lmu$, then $M \equivwxh N \Iff M \equivwh N \Iff M \equivwA N \Iff M \equivwbmu N $.
 \end{corollary}



\Paper{ \section{Full abstraction for the logical interpretation} \label{Full abstraction} }

We now come to the main result of this paper, where we show a full abstraction result for our logical interpretation.
First we show the relation between weak explicit head equivalence and weak bisimilarity.

 \begin{theorem} [Full abstraction of $\wbisim$ versus $\equivwxh$]
\label{FA equivxh equivC}
For any $M,N \ele \lmux$:
$\PilmuTerm[M] a \wbisim \PilmuTerm[N] a $ if and only if $M \equivwxh N$.
 \end{theorem}

{
 \begin{Proof}

\CLAC{ \leftmargini 0pt 
}

 \begin{description} \itemsep 4pt 

 \item[if]
By co-induction on the definition of $\equivwxh$.
Let $M \equivwxh N$, then either $M$ and $N$ have both no $\redwxh$-normal form, so, by Lem.~\ref{zero lemma}, their interpretations are both weakly bisimilar to the process $\Zero$; or
both $M \rtcredwxh[\nf] M' \, \Ssub $ and $ N \rtcredwxh[\nf] N' \, \Ssub' $ (let $\Ssub = \Vexsub y := P \, \Vexcontsub `a := Q . `b $, and $\Ssub' = \Vexsub y := P' \, \Vexcontsub `a := Q' . `b $), and either:

 \begin{description}
 \CLAC{ \itemindent -5mm}

 \item [$M' = xM_1\dots M_n $ $(n \geq 0)$, $ N = xN_1\dots N_n $ and $M_i \, \Ssub \equivwxh N_i \, \Ssub ' $, for all $\iotn$] ~

We have $\PilmuTerm[M] a \wbisim \PilmuTerm[xM_1\dots M_n \, \Ssub] a $ and $\PilmuTerm[N] a \wbisim \PilmuTerm[xN_1\dots N_n \, \Ssub'] a $ by Corollary~\ref{weak soundness}.
Notice that
 \[ \begin{array}{lcl}
\PilmuTerm [xM_1\dots M_n \, \Ssub] a &=&
\New \Vect{cy`a} (\PilmuTerm [v x] c_1 \Par {\VPiExContSub c_i := M_i . c_{i+1} } \Par {} \PiLmuSem[\Ssub] )
 \end{array} \]
where $c_n = a$ and
 \[ \begin{array}{rcl}
\PiLmuSem[\Ssub] &=& {\VPiExSub y := P } \Par {\VPiExContSub `a := Q . `b } \\
\PiExContSub c_i := M_i . c_{i+1} &=&
{ \PiExContFsub c_i := M_i . c_{i+1} } \\
\PiExSub y_j := P_j &=& \PiExsub y_j := P_j \\
\PiExContSub `a_k := Q_k . `b_k &=&
 {\PiExContFsub `a_k := Q_k . `b_k }
 \end{array} \]
and similar for $\PilmuTerm [xN_1\dots N_n \, \Ssub'] a $.
By induction,
 \[ \begin{array}{rclccll}
\New \Vect{y`a} (\PilmuTerm[M_i] w \Par {} \PiLmuSem[\Ssub]) & \ByDef &
\PilmuTerm[M_i \, \Ssub] w & \wbisim &
\PilmuTerm[N_i \, \Ssub'] w & \ByDef &
\New \Vect{y`a} (\PilmuTerm[N_i] w \Par {} \PiLmuSem[\Ssub'])
 \end{array} \]
%
Since $\wbisim$ is a congruence, also
 \[ \begin{array}{ccl}
\PiExContFsub c_i := M_i . c_{i+1} \Par \PiLmuSem[\Ssub]
	& \wbisim & 
\PiExContFsub c_i := N_i . c_{i+1} \Par \PiLmuSem[\Ssub']
 \end{array} \]
for all $\iotn$, so also
 $ \PilmuTerm [xM_1\dots M_n \, \Ssub] a \wbisim \PilmuTerm [xN_1\dots N_n \, \Ssub'] a
$ but then also $\PilmuTerm[M] a \wbisim \PilmuTerm[N] a $.

\Paper{
 \item [ $M' = `lx.M''$, $N' = `lx.N''$, and $M'' \, \Ssub \equivwxh N'' \, \Ssub' $]
By Corollary~\ref{weak soundness}, we have $\PilmuTerm[M] a \wbisim \PilmuTerm[`lx.M'' \, \Ssub ] a $ and $\PilmuTerm[N] a \wbisim \PilmuTerm[`lx.N'' \, \Ssub' ] a $.
Notice that
 \[ \begin{array}{rcl}
\PilmuTerm [`lx.M'' \, \Ssub ] a &\ByDef&
\New \Vect{y`a} ( \PilmuTerm [l x . M''] a \Par {} \PiLmuSem[\Ssub] )
	\\ [1mm]
\PilmuTerm [`lx.N'' \, \Ssub' ] a &\ByDef&
\New \Vect{y`a} ( \PilmuTerm [l x . N''] a \Par {} \PiLmuSem[\Ssub'] )
 \end{array} \]
with $\Ssub$ and $\Ssub'$ as in the previous part and $a$ not in $\Ssub$ or $\Ssub'$.
By induction,
 \[ \begin{array}{rclclcl}
\New \Vect{y`a} ( \PilmuTerm [M''] b \Par {} \PiLmuSem[\Ssub] )
	& \ByDef &
\PilmuTerm[M'' \, \Ssub] b
	& \wbisim & \CLAC{ \\ }
\PilmuTerm[N''\, \Ssub'] b
	& \ByDef &
\New \Vect{y`a} ( \PilmuTerm [N''] b \Par {} \PiLmuSem[\Ssub] )
 \end{array} \]
Since $\wbisim$ is a congruence, also
$\PilmuTerm[M] a \wbisim \PilmuTerm[N] a $.

 \item [{$M' = \muterm`g.[`d]M''$, $N' = \muterm`g.[`d]N''$}]
Then $M''$ and $N''$ themselves are in normal form and $M'' \, \Ssub \equivwxh N'' \, \Ssub' $, with $\Ssub$, $\Ssub'$ as above.
By Corollary~\ref{weak soundness}, $\PilmuTerm[M] a \wbisim \PilmuTerm[\muterm`g.[`d]M'' \, \Ssub ] a $ and $\PilmuTerm[N] a \wbisim \PilmuTerm[\muterm`g.[`d]N'' \, \Ssub' ] a $.
Notice that
 \[ \begin{array}{\CLAC{l@{~}c@{~}l}\Paper{rcl}cl}
\PilmuTerm [{\muterm`g.[`d].M''} \, \Ssub ] a
	&\ByDef& \CLAC{ \\ }
\New \Vect{y`a} ( \PilmuTerm [{ M''[a \For `g]}] `d \Par {} \PiLmuSem[\Ssub] )
	&\ByDef&
\PilmuTerm[{M''[a \For `g]} \, \Ssub] `d
	\\ [2mm]
\PilmuTerm [{\muterm`g.[`d].N''} \, \Ssub'] a
	&\ByDef& \CLAC{ \\ }
\New \Vect{y`a} ( \PilmuTerm [{ N''[a \For `g]}] `d \Par {} \PiLmuSem[\Ssub'] )
	&\ByDef&
\PilmuTerm[{N''[a \For `g]} \, \Ssub'] `d
 \end{array} \]
By induction, $\PilmuTerm[{M''[a \For `g]} \, \Ssub] `d \wbisim \PilmuTerm[{N''[a \For `g]} \, \Ssub'] `d $; 
since $\wbisim$ is a congruence, also
$\PilmuTerm[M] a \wbisim \PilmuTerm[N] a $.

}

\CLAC{\item [{$M' = `lx.M''$ or $M' = \muterm`g.[`d]M''$}] Similar.}

 \end{description}


 \item[only if]
We distinguish the following cases.

 \begin{enumerate}

 \item $ \PilmuTerm[M] a $ can never input nor output; then $ \PilmuTerm[M] a \wbisim \Zero \wbisim \PilmuTerm[N] a $.
Assume $M$ has a weak head-normal form, then by Lem.~\ref{bmu reduction characterisation}, $ \PilmuTerm[M] a $ is not weakly bisimilar to $\Zero$; therefore, $M$ and $N$ both have no weak head-normal form.

 \item $\PilmuTerm[M] a \Outson c $, then by Lem.~\ref{pi reduction characterisation}, $\PilmuTerm[M] a \wbisim \New xb ( \PilmuTerm[M'] b \Par \Out c <x,b> \Par \PiLmuSem[\Ssub] ) $, and $ M \rtcredwxh `lx.M' \, \Ssub $.
Since $\PilmuTerm[M] a \wbisim \PilmuTerm[N] a $,
also $\PilmuTerm[N] a \Outson c $, so $ \PilmuTerm[N] a \wbisim \New xb ( \PilmuTerm[N'] b \Par \Out c <x,b> \Par \PiLmuSem[\Ssub'] ) $ and $ N \rtcredwxh `lx.N' \, \Ssub' $.
Then also $ \PilmuTerm[M'] b \Par \PiLmuSem[\Ssub] \wbisim \PilmuTerm[N'] b \Par \PiLmuSem[\Ssub'] $, so $\PilmuTerm[M' \, \Ssub] a \wbisim \PilmuTerm[N' \, \Ssub'] a $
and by induction, $M' \, \Ssub \equivwxh N' \, \Ssub'$; so also $M \equivwxh N$ by definition.

 \item
If $\PilmuTerm [M] a \not\!\Outson c $, but $\PilmuTerm [M] a \Inson x $, then by Lem.~\ref{pi reduction characterisation}, $ \PilmuTerm[M] a \wbisim \PilmuTerm [xM_1\dots M_n \, \Ssub] a' $ and $ M \rtcredwxh xM_1\dots M_n \, \Ssub$.
We have
 \[ \begin{array}{lcl}
\PilmuTerm [xM_1\dots M_n \, \Ssub] a' &=&
\New \Vect{cy`a} (\PilmuTerm [v x] c_1 \Par {\VPiExContSub c_i := M_i . c_{i+1} } \Par \PiLmuSem[\Ssub])
 \end{array} \]
\Paper{where $c_n = a'$ and
 \[ \begin{array}{rcl}
\PiLmuSem[\Ssub] &=& {\VPiExSub y := P } \Par {\VPiExContSub `a := Q . `b } \\
\PiExContSub c_i := M_i . c_{i+1} &=& { \PiExContFsub c_i := M_i . c_{i+1} } \\
\PiExSub y_j := P_j &=& \PiExsub y_j := P_j \\
\PiExContSub `a_k := Q_k . `b_k &=& { \PiExContFsub `a_k := Q_k . `b_k }
 \end{array} \]}
\CLAC{with $\PiLmuSem[\Ssub] $,
$ \PiExContSub c_i := M_i . c_{i+1} $,
$ \PiExSub y_j := P_j $, and
$ \PiExContSub `a_k := Q_k . `b_k $ are defined as above.}

Since $\PilmuTerm[M] a \wbisim \PilmuTerm[N] a $, again by Lem.~\ref{pi reduction characterisation}, $\PilmuTerm[N] a \wbisim \PilmuTerm [xN_1\dots N_n \, \Ssub'] a'' $ and $ N \rtcredwxh xN_1\dots N_n \, \Ssub' $.
Notice that
 \[ \begin{array}{lcl}
\PilmuTerm [xN_1\dots N_n \, \Ssub'] a'' & = &
\New \Vect{cy`a} (\PilmuTerm [v x] c_1 \Par {\VPiExContSub c_i := N_i . c_{i+1} } \Par {} \PiLmuSem[\Ssub'])
 \end{array} \]
\Paper{
where $c_n = a''$ and
 \[ \begin{array}{rcl}
\PiLmuSem[\Ssub'] &=& {\VPiExSub y := P' } \Par {\VPiExContSub `a := Q' . `b } \\
\PiExContSub c_i := N_i . c_{i+1} &=& \PiExContFsub c_i := N_i . c_{i+1} \\
\PiExSub y_j := P'_j &=& \PiExsub y_j := P'_j \\
\PiExContSub `a_k := Q'_k . `b_k &=& { \PiExContFsub `a_k := Q'_k . `b_k }
 \end{array} \]
}
\CLAC{with $\PiLmuSem[\Ssub'] $, $\PiExContSub c_i := N_i . c_{i+1} $, $ \PiExSub y_j := P'_j $, and $ \PiExContSub `a_k := Q'_k . `b_k $ similar to above.}
Then we have
 \[ \begin{array}{rcl}
\PilmuTerm [xM_1\dots M_n \, \Ssub] a' &\wbisim& \PilmuTerm [xN_1\dots N_n \, \Ssub'] a'' ,
\end{array} \]
so $a' = a''$ and $\PilmuTerm [M_i' \, \Ssub ] w \wbisim \PilmuTerm [N_i' \, \Ssub'] w $; then by induction, $M_i' \, \Ssub \equivwxh N_i' \, \Ssub' $, and \Paper{then also }$M\equivwxh N $.
\qed

 \end{enumerate}
 \end{description}
 \end{Proof}
}

We now obtain our main result:

 \begin{theorem} [Full abstraction] \label{main result}
Let $M,N \ele \lmu$, then $ \PilmuTerm[M] a \wbisim \PilmuTerm[N] a $ if and only if $
 M \equivwbmu N $.

\Paper{\begin{Proof}
By Corollary~\ref{all lmu weak equivalences} and Theorem~\ref{FA equivxh equivC}.
\qed
 \end{Proof}
}

 \end{theorem}

\Paper{
Since ${\eqbmu} \subseteq {\equivwbmu}$, the following is immediate.

 \begin{corollary} \label{semantics}
If $M \eqbmu N$, then $ \PilmuTerm[M] a \wbisim \PilmuTerm[N] a $.
 \end{corollary}
which states that our interpretation gives a semantics for $\lmu$.

 \subsection*{Conclusions and Future Work}
We have found a new, simple and intuitive interpretation of $\lmu$-terms in $`p$ that respects head reduction with explicit substitution.
For this interpretation, we have shown that termination is preserved, and that it is sound and complete.
We have shown that, for our context assignment system that uses the type constructor $\arrow$ for $`p$ and is based on classical logic, typeable $\lmu$-terms are interpreted by our interpretation as typeable $`p$-processes, preserving the types.
}


\CLAC{ \section*{Conclusions and future work}
We have studied the output based, logic-inspired interpretation of untyped $\lmu$ with explicit substitution into the $`p$-calculus and shown that this interpretation is fully abstract with respect to weak equivalence between terms and weak bisimilarity between processes.

We have defined the 
weak equivalences $\equivwbmu$, $\equivwh$, $\equivwxh$, and $\equivwA$ on $\lmu$ terms, and shown that these all coincide.
We then proved that $M \equivwxh N \Iff \PilmuTerm[M] a \wbisim \PilmuTerm[N] a $, which, combined with our other results, essentially shows that $\PilmuTerm[`.] `. $ respects equality between L\'evy-Longo trees for $\lmu$.}

We will investigate the relation between our interpretation and the {\sc cps}-translation of Lafont, Reus, and Streicher \cite{Lafont-Reus-Streicher'93}.

\Paper{ 
\bibliography {references} 
}
 \CLAC{ \input {biblio} }




 \end {document}